\documentclass[apj]{emulateapj}

\usepackage{color}

\newcommand{\ba}{{\bf a}}
\newcommand{\bb}{{\bf b}}

\newcommand{\bj}{{\bf j}}

\newcommand{\br}{{\bf r}}
\newcommand{\bu}{{\bf u}}

\newcommand{\bx}{{\bf x}}

\newcommand{\bA}{{\bf A}}
\newcommand{\bB}{{\bf B}}

\newcommand{\bE}{{\bf E}}
\newcommand{\bJ}{{\bf J}}

\newcommand{\bU}{{\bf U}}
\newcommand{\bR}{{\bf R}}

\newcommand{\obj}{{\overline{{\hbox{\boldmath $\j$}}}}}

\newcommand{\OL}{\overline}
\newcommand{\lb}{\label}
\newcommand{\be}{\begin{equation}}
\newcommand{\ee}{\end{equation}}
\newcommand{\ber}{\begin{eqnarray}}
\newcommand{\eer}{\end{eqnarray}}
\newcommand{\bers}{\begin{eqnarray*}}
\newcommand{\eers}{\end{eqnarray*}}

\newcommand{\boJ}{{\bf {\cal J}}}

\newcommand{\boepsilon}{\hbox{\boldmath $\varepsilon$}}
\newcommand{\botau}{\hbox{\boldmath $\tau$}}
\newcommand{\boeta}{\hbox{\boldmath $\eta$}}
\newcommand{\borho}{\hbox{\boldmath $\rho$}}
\newcommand{\boxi}{\hbox{\boldmath $\xi$}}
\newcommand{\bomega}{\hbox{\boldmath $\omega$}}
\newcommand{\bdot}{\hbox{\boldmath $\cdot$}}
\newcommand{\bdots}{\hbox{\boldmath $:$}}
\newcommand{\btimes}{\hbox{\boldmath $\times$}}
\newcommand{\grad}{\hbox{\boldmath $\nabla$}}
\newcommand{\Bell}{\hbox{\boldmath $\ell$}}
\newcommand{\bzed}{\hbox{\boldmath $0$}}

\shorttitle{Fast Magnetic Reconnection}
\shortauthors{Eyink, Lazarian \& Vishniac}

\begin{document}


\title{Fast Magnetic Reconnection and Spontaneous Stochasticity}


\author{Gregory L. Eyink
}
\affil{Department of Applied Mathematics \& Statistics and Department of Physics \& Astronomy, \\
The Johns University University, Baltimore, MD 21218, USA}
\email{eyink@jhu.edu}

\author{A. Lazarian}
\affil{Department of Astronomy, University of Wisconsin, 475 North 
Charter Street, Madison, WI 53706, USA}

\and

\author{E. T. Vishniac}
\affil{Department of Physics and Astronomy, McMaster University, 1280 Main Street West, 
Hamilton, ON L8S 4M1, Canada} 




\begin{abstract}
Magnetic field-lines in astrophysical plasmas are expected to be frozen-in at 
scales larger than the ion gyroradius.  The rapid reconnection of magnetic flux 
structures with dimensions vastly larger than the gyroradius requires a breakdown 
in the standard Alfv\'en flux-freezing law. We attribute this breakdown to ubiquitous 
MHD plasma turbulence with power-law scaling ranges of velocity and magnetic 
energy spectra.  Lagrangian particle trajectories in such environments become 
``spontaneously stochastic'', so that infinitely-many magnetic field-lines are advected 
to each point and must be averaged to obtain the resultant  magnetic field. The relative 
distance between initial magnetic field lines which arrive to the same final point 
depends upon the properties of two-particle turbulent dispersion. We develop 
predictions based on the phenomenological Goldreich \& Sridhar theory of strong 
MHD turbulence and on weak MHD turbulence theory. We recover the predictions 
of the Lazarian \& Vishniac theory for the reconnection rate of large-scale magnetic  
structures. Lazarian \& Vishniac  also invoked ``spontaneous stochasticity'', but of 
the field-lines rather than of the Lagrangian trajectories.  More recent theories of 
fast magnetic reconnection appeal to microscopic plasma processes that lead to 
additional terms in the generalized Ohm's law, such as the collisionless Hall term.  
We estimate quantitatively the effect of  such processes on the inertial-range 
turbulence dynamics and find them to be negligible in most astrophysical environments. 
For example, the predictions of the Lazarian-Vishniac theory are unchanged in 
Hall MHD turbulence with an extended inertial range, whenever the ion skin depth 
$\delta_i$ is much smaller than the turbulent integral length or injection-scale $L_i.$
\end{abstract}


\keywords{
galaxies: magnetic fields -- methods: theoretical -- MHD -- turbulence\\
$$\,$$}



$$\,$$

$$\,$$

$$\,$$

\section{Introduction}

It is generally believed that magnetic field embedded in a highly conductive 
fluid preserves its topology for all time due to magnetic fields being frozen-in \citep{Alfven42,Parker79}.  
Although ionized astrophysical objects, like stars and galactic disks, are almost perfectly 
conducting, they show indications of changes in topology, ``magnetic reconnection'', 
on dynamical time scales \citep{Parker70, Lovelace76, PriestForbes02}.  
Reconnection can be observed directly in the solar corona \citep{Innesetal97, 
YokoyamaShibata95, Masudaetal94}, but can also be inferred from the 
existence of large-scale dynamo activity inside stellar interiors \citep{Parker93, Ossendrijver03}.  
Solar flares usually \citep{Sturrock66} and $\gamma$-ray bursts often 
\citep{ZhangYan11,Foxetal05, Galamaetal98} are associated with magnetic reconnection.
Much recent work has concentrated on showing how reconnection 
can be rapid in plasmas with very small collisional rates 
\citep{Shayetal98, Drake01, Drakeetal06, Daughtonetal06}, which 
substantially constrains astrophysical applications of the corresponding 
reconnection models. 

Reconnection occurs rapidly in computer simulations due 
to the high values of resistivity (or numerical resistivity) that are employed 
at the low resolutions currently achievable. Therefore, if there are situations 
where magnetic fields reconnect slowly, numerical simulations do 
not adequately reproduce astrophysical reality. This means that if collisionless 
reconnection is the only way to make reconnection rapid, then 
the numerical simulations of many astrophysical processes, including those in 
interstellar media (ISM), which is collisional, are in error. At the same time, it is not possible 
just to claim that the reconnection must always be rapid empirically, 
as solar flares require periods of flux accumulation time, which correspond to slow 
reconnection.

To understand the difference between reconnection in astrophysical situations and 
in numerical simulations, one should recall that the dimensionless combination that
controls the reconnection rate is the Lundquist number\footnote{The magnetic
Reynolds number, which is the ratio of the magnetic field decay time to the eddy
turnover time, is defined using the injection velocity $v_l$ as a characteristic
speed instead of the Alfv\'en speed $v_A$, which is taken in the Lundquist
number.}, defined as $S = L_xv_A / \lambda$, where $L_x$ is the length of the
reconnection layer, $v_A$ is the Alfv\'en velocity, and $\lambda=\eta c^2/4\pi$ 
is Ohmic diffusivity. Because of the huge astrophysical 
length-scales $L_x$ involved, the astrophysical Lundquist numbers are also huge,
e.g. for the ISM they are about $10^{16}$, while present-day MHD simulations
correspond to $S<10^4$. As the simulation costs scale as $L_x^4$, where $L_x$
is the size of the box, it is feasible neither at present nor in the foreseeable future 
to have simulations with sufficiently high Lundquist numbers.

Plasma flows at such high Lundquist numbers are generically turbulent, since laminar flows
are then prey to numerous linear and finite-amplitude instabilities. Indeed, turbulence
is ubiquitous in astrophysical plasmas. This is sometimes driven turbulence due to an 
external energy source, such as supernova in the ISM \citep{NormanFerrara96,Ferriere01}, 
merger events and AGN outflows in the intercluster medium (ICM) 
\citep{Subramanianetal06,EnsslinVogt06,Chandran05}, 
and baroclinic forcing behind shock waves in interstellar clouds. In other cases, the turbulence is 
spontaneous, with available energy released by a rich array of instabilities, such as MRI 
in accretion disks \citep{BalbusHawley98}, kink instability of twisted 
flux tubes in the solar corona \citep{GalsgaardNordlund97a,GerrardHood03}, etc. 
Whatever its origin, the signatures of plasma turbulence are 
seen throughout the universe. Turbulent cascade of energy leads to long ``inertial ranges''
with power-law spectra that are widely observed, e.g. in the solar wind \citep{Leamonetal98,
Baleetal05}, in the ISM \citep{Armstrongetal94,ChepurnovLazarian10}, and in the ICM 
\citep{Schueckeretal04,VogtEnsslin05}. Often inertial-range spectra cannot be directly observed,
but only large-scale turbulent fluctuations, e.g. through their Doppler broadening of line 
spectra. Nevertheless, the power-law ranges are universal features of high-Reynolds-number
turbulence which, even when not seen, can be inferred to be present from enhanced 
rates of dissipation and mixing \citep{Eyink08}. Any theory of astrophysical reconnection must 
take into account the pre-existing turbulent environment. Even if the plasma flow is initially laminar, 
kinetic energy release by reconnection due to some slower plasma process may generate 
vigorous turbulent motion.

But can turbulence itself accelerate reconnection? \cite{LazarianVishniac99} 
[henceforth LV99] proposed a model of fast reconnection in the presence of 
sub-Alfv\'enic turbulence in magnetized plasmas. 
They identified stochastic wandering of magnetic field-lines 
as the most critical property of MHD turbulence which permits fast reconnection.  
As we discuss more below, this line-wandering widens the outflow region 
and alleviates the controlling constraint of mass conservation. The LV99 model 
has been successfully tested recently in \cite{Kowaletal09} (see also higher resolution results 
in \cite{Lazarianetal10}). 
The model is radically different from its predecessors which also appealed 
to the effects of turbulence (see more comparisons in section 5.4). For instance, unlike \cite{Speiser70} and 
\cite{Jacobson84} the model does not appeal to changes of microscopic properties of plasma. 
The nearest progenitor to LV99 was the work of  \cite{MatthaeusLamkin85,MatthaeusLamkin86}, 
who studied the problem numerically in 2D MHD and who suggested that magnetic reconnection 
may be fast due to a number of  turbulence effects, e.g. multiple X points and turbulent EMF. However,  
\cite{MatthaeusLamkin85,MatthaeusLamkin86} did not observe  the important role of magnetic field-line 
wandering,  and did not obtain a quantitative prediction for the reconnection rate, as did LV99.

The success of LV99 in identifying an MHD turbulence mechanism for fast reconnection leads, 
however, to a conflict with certain conventional beliefs. As the predicted reconnection velocity 
is independent of magnetic diffusivity $\lambda,$ the LV99 theory implies that field-line topology 
should change in MHD plasmas at a finite rate even in the limit of infinite Lundquist number.
This contradicts the accepted wisdom that magnetic field-lines should be ``nearly'' frozen-in to 
very high-conductivity MHD plasmas. It is implicit in the LV99 theory that the standard Alfv\'en 
Theorem on magnetic-flux conservation must be violated for $\lambda\rightarrow 0$
\citep{VishniacLazarian99}. 

In a separate, more recent development,  \cite{Eyink11} has critically examined the concept 
of frozen-in field-lines for turbulent MHD plasmas. See also \cite{EyinkAluie06,Eyink07,Eyink09}.
A first contribution of that work was to establish exact flux-conservation properties  for {\it resistive} 
MHD solutions. The magnetic field at a point was shown to result from averaging over an infinite 
ensemble of line-vectors stochastically advected to that point by the fluid velocity perturbed 
with a Brownian motion proportional to $\sqrt{\lambda}$. For smooth, laminar flows the contribution 
of the random Brownian term  disappears as $\lambda\rightarrow 0$ and standard flux-freezing 
is recovered. However, it was shown in \cite{Eyink11} that flux-conservation may remain stochastic 
even at infinite conductivity for rough, turbulent velocities. This remarkable behavior is due to 
``spontaneous stochasticity'' of Lagrangian fluid particles undergoing turbulent two-particle or 
Richardson diffusion: see \cite{Bernardetal98,GawedzkiVergassola00,EvandenEijnden00,
EvandenEijnden01,Chavesetal03}, and, for a review, \cite{Kupiainen03}. The work of Eyink 
removes the objection to LV99 based on the Alfv\'en theorem for high conductivity plasmas, 
since that law is fundamentally altered in a turbulent flow and poses no necessary obstacle 
to fast reconnection. Indeed, there are very close relations between the  work of Eyink and LV99. 
For example, it was pointed out in \cite{EyinkAluie06} that the  ``stochastic wandering'' 
of field-lines in a rough magnetic field which was invoked in LV99 is exactly analogous 
to the ``spontaneous stochasticity" phenomenon for Lagrangian particle trajectories 
guided by a rough, turbulent velocity field. 

In this work we shall establish even deeper connections between the work of Eyink and LV99.
In particular, we show that the detailed, quantitative predictions of the LV99 theory can be rederived 
by applying the stochastic flux-freezing laws of \cite{Eyink11} to the problem of magnetic reconnection.
In this derivation, the LV99 theory emerges as the natural turbulent analogue of the Sweet-Parker 
laminar solution, with resistive diffusion of field-lines through the reconnection layer replaced by 
turbulent Richardson diffusion. To estimate the latter, we employ the same phenomenological
turbulence model as did LV99, the \cite{GoldreichSridhar95} theory [hereafter, GS95], although
alternative theories would lead to similar results. The new derivation demonstrates the dynamical 
consistency of the LV99 theory and also eliminates the need for some difficult arguments developed 
in LV99 concerning the ``motion'' of  already-reconnected field-lines. 

A second main purpose of this paper is to explain in a clear physical manner the concept of 
``spontaneous stochasticity'', both of Lagrangian trajectories and of magnetic field-lines, and the 
related notion of stochastic flux-freezing. ``Spontaneous stochasticity'' has some rather dramatic 
implications, such as the breakdown of Laplacian determinism for classical dynamics. Nevertheless, 
the phenomenon is not difficult to understand in intuitive terms. We shall also try to explain 
the concept of stochastic flux-freezing in high (kinetic and magnetic) Reynolds-number
MHD turbulence in the simplest possible manner, explicating some points that were left implicit
in \cite{Eyink11}. 
An essential point which has been emphasized many times before \citep{Newcomb58,
Vasyliunas72, Alfven76} is that the notion of flux-freezing is largely 
conventional. While it is a very useful intuitive device,  the frozenness of field-lines 
is a ``meta-physical'' principle which cannot be subjected to direct experimental test. Any definition 
of field-line motion that is consistent with the hydromagnetic equations is equally acceptable.  
We believe that  the stochastic approach to flux-freezing will provide a powerful heuristic tool 
in astrophysics. In particular, there should be important applications whenever the plasma fluid 
velocity field has an extended inertial-range and the stochastic motion of field-lines is therefore 
due to turbulent advection rather than to resistivity. This paper presents an illustrative ``case study'' 
of an application to the problem of large-scale turbulent reconnection, but many other uses can 
be envisaged. As another example, \cite{EyinkNeto10}, \cite{Eyink10,Eyink11} discuss the critical 
role of stochastic flux-freezing to the mechanism of the turbulent kinematic dynamo for  
magnetic Prandtl numbers either small or unity.
 
The detailed contents of this paper are as follows: In \S 2 we briefly recall the justification 
for an MHD description of astrophysical plasmas.  In \S 3 we discuss the scaling laws of MHD turbulence 
and Lagrangian particle dynamics in MHD turbulence, and introduce the concepts of stochastic flux 
freezing \& spontaneous stochasticity. We next discuss the generalized Ohm's law in \S 4. Using 
the concept of spontaneous stochasticity, we rederive the predictions of LV99 theory of turbulent 
reconnection in \S 5 and consider various limiting cases. We provide a discussion of our results 
in \S 6 and a final summary in \S 7. 

\section{Validity of a Magnetohydrodynamic Description}

An MHD (or nearly MHD) description of a plasma can be justified based either on collisionality 
or on strong magnetization. This issue has been well-studied and the subject of many discussions, 
e.g. \cite{Kulsrud83}, so we shall here just briefly review the standard understanding. There are three 
characteristic length-scales of importance: the ion gyroradius $\rho_i,$ the ion mean-free-path length 
$\ell_{mfp,i}$ arising from Coulomb collisions, and the scale $L$ of large-scale variation of magnetic 
and velocity fields. Astrophysical plasmas are in many cases ``strongly collisional''
in the sense that $\ell_{mfp,i}\ll \rho_i,$ e.g. the interiors of stars and accretion disks. In such cases,
a standard Chapman-Enskog expansion yields a fluid description of the plasma \citep{Braginsky65}, 
either a two-fluid model for scales between $\ell_{mfp,i}$ and the ion skin-depth $\delta_i=
\rho_i/\sqrt{\beta_i}$ or simple MHD at scales much larger than $\delta_i$. However, 
astrophysical plasmas are often at most only ``weakly collisional'' with $\ell_{mfp,i}\gg \rho_i,$  
especially when they are very hot and diffuse. This can be seen from the simple relation 
\be \frac{\ell_{mfp,i}}{\rho_i}\propto \frac{\Lambda}{\ln\Lambda}\frac{v_A}{c}, \lb{lmfp-rho} \ee
which follows from the standard formula for the Coulomb collision frequency (e.g. see 
\cite{Fitzpatrick11}, eq.(1.25)). Here $\Lambda=4\pi n\lambda_D^3$ is the plasma parameter,
or number of particles within the Debye screening sphere. When astrophysical plasmas are very 
weakly coupled (hot and rarified), then $\Lambda$ is extremely large. Of course, $v_A<c$ 
as well, but the ratio $v_A/c$ is generally only moderately small in well-magnetized plasmas. 
Typical values for some weakly coupled cases are shown in the following table: 

\vspace{10pt}

\hspace{-27pt} \scalebox{.78}{
\begin{tabular}{llll}
\multicolumn{4}{c}{{\bf Table 1}} \\
\multicolumn{4}{c}{Representative Parameters for Some 
Weakly-Coupled Astrophysical Plasmas} \\
\hline
\hline
Parameter & warm ionized & post-CME & solar wind at  \\
& ISM${\,\!}^a$ & current sheets${\,\!}^b$ & magnetosphere${\,\!}^c$ \\
\hline
density $n,\,cm^{-3}$ & .5 & $7\times 10^7$ & 10 \\
temperature $T,\, eV$ & .7 & $10^3$ &10 \\
plasma parameter $\Lambda$ & $4\times 10^9$ & $2\times 10^{10}$ & $5\times 10^{10}$ \\
ion thermal velocity $v_{th,i},\,cm/s$ & $10^6$ & $3\times 10^7$ & $5\times 10^6$ \\
ion mean-free-path $\ell_{mfp,i},\,cm$ & $6\times 10^{11}$ & $10^{10}$ & $7\times 10^{12}$ \\
magnetic diffusivity $\lambda,\, cm^2/s$ & $10^7$ & $8\times 10^2$ & $6\times 10^5$ \\
\hline
magnetic field $B,\, G$ & $10^{-6}$ & 1 & $10^{-4}$ \\
plasma beta $\beta$ & 14 & 3 & 1 \\
Alfv\'en speed $v_A,\,cm/s$ & $3\times 10^5$ & $3\times 10^7$ & $7\times 10^6$ \\
ion gyroradius $\rho_i,\,cm$ & $10^8$ & $3\times 10^3$ & $6\times 10^6$ \\
\hline
large-scale velocity $U,\,cm/s$ & $10^6$ & $4\times 10^6$ & $5\times 10^6$ \\
large length scale $L,\,cm$ & $10^{20}$ & $5\times 10^{10}$ & $10^8$ \\
Lundquist number $S_L=\frac{v_A L}{\lambda}$ & $3\times 10^{18}$ & $2\times 10^{15}$ & $10^9$ \\
resistive length${\,\!}^*$ $\ell_\eta^\perp,\, cm$ & $5\times 10^5$ & 1 & 20\\ 
\hline
& & & \\
\multicolumn{4}{l}{\footnotesize{${\,\!}^a$\cite{NormanFerrara96,Ferriere01} \,\,\,\, ${\,\!}^b$\cite{Bemporad08} 
\,\,\,\, ${\,\!}^c$\cite{Zimbardoetal10}}}\\
\multicolumn{4}{l}{\footnotesize{*This nominal resistive scale is calculated from $\ell_\eta^\perp \simeq L (v_A/U)
S_{L}^{-3/4}$, assuming GS95 turbulence holds}}\\
\multicolumn{4}{l}{\footnotesize{down to that scale (see eq.(\ref{A2})), and should not be taken literally 
when $\ell_\eta^\perp<\rho_i.$}}\\
\end{tabular}
}

\vspace{10pt}

\noindent  
The primary expansion to justify a hydrodynamic description of these weakly coupled but 
well-magnetized plasmas is an expansion in the small ion gyroradius. This yields 
at all length scales larger than $\rho_i$ a description by ``kinetic MHD equations'', nearly 
identical to standard MHD. The most significant difference is that the pressure tensor in the momentum 
equation is anisotropic, with the two components $p_\|$ and $p_\perp$ of the pressure parallel 
and perpendicular to the local magnetic field direction determined from a 1-dimensional kinetic 
equation \citep{Kulsrud83}.  The magnetic field solves an ideal induction equation, 
if one ignores all collisional effects, although electron-ion collisions produce in reality 
a non-zero resistivity $\eta$. However, in very many cases (cf. the ISM and the magnetosphere 
in Table 1) the resistive length-scale $\ell_\eta^\perp$ is much smaller than $\rho_i$ and also than 
$\rho_e\approx \frac{1}{43}\rho_i$. Magnetic field-lines are, at least formally, well ``frozen-in'' on 
these scales. Plasmas that are not strongly collisional further divide into two cases: ``collisionless'' 
plasmas for which $\ell_{mfp,i}\gg L,$ the largest scales of interest, and ``weakly collisional'' plasmas for which 
$L\gg \ell_{mfp,i}.$ In the latter case the``kinetic MHD'' description can be further reduced in complexity 
at  scales greater than $\ell_{mfp,i}$ by including the Coulomb collision operator and making again a
Chapman-Enskog expansion. This reproduces a fully hydrodynamic MHD description at those scales, 
with anisotropic transport behavior associated to the well-magnetized limit.  Among our examples 
in Table 1 above, the warm ionized ISM is ``weakly collisional'', while post-CME current sheets 
and the solar wind impinging on the magnetosphere are (nearly) ``collisionless.''

Additional important simplifications occur if the following assumptions are satisfied:  turbulent fluctuations 
are small compared to the mean magnetic field,  have length-scales parallel to the mean field much larger than 
perpendicular length-scales, and have frequencies low compared to the ion cyclotron frequency. These are 
standard assumptions of the Goldreich-Sridhar (GS95) theory of MHD turbulence, discussed in the following
section, but are often valid much more generally in astrophysical plasmas. They are the basis of the 
``gyrokinetic approximation'' which was developed for fusion plasmas but more recently extensively 
applied in astrophysics \citep{Schekochihinetal07,Schekochihinetal09}. At length-scales greater than the 
ion gyroradius $\rho_i$, which mainly concern us in this work, a remarkable reduction occurs. The incompressible
shear-Alfv\'en wave modes have an autonomous dynamics unaffected by the compressive modes and described 
by the simple ``reduced MHD'' (RMHD) equations. This fact is of fundamental importance for the theory 
developed in the present work, permitting us to base our analysis on an incompressible MHD fluid model. 
Compressible fast magnetosonic waves are subject to strong damping both from collisionless 
effects and in shocks,  and, also, are not regenerated by the incompressible dynamics (see below). 
The other compressible plasma modes (slow and entropy waves) are ``passively'' transported  by the 
shear-Alfv\'en waves. For ``collisionless'' plasmas the compressible modes  are described by gyrokinetic 
equations for a distribution function $g$ in phase space  and dissipated by collisionless damping. 
For ``weakly collisional'' plasmas at scales greater than $\ell_{mfp,i}$ the slow and entropy waves are 
described by the compressible MHD equations and damped primarily by (field-parallel) viscosity and 
thermal  diffusion. 

Although we are generally interested in this work in reconnection phenomena at length scales
much larger than $\rho_i,$ there are some situations where the scales of interest are comparable to the ion 
gyroradius, e.g. in the magnetosphere (see Table 1).  At scales smaller than $\rho_i$ but larger than 
$\rho_e$, the plasma is described by an ion kinetic equation and a system of ``electron reduced MHD'' 
(ERMHD) equations for kinetic Alfv\'en waves \citep{Schekochihinetal07,Schekochihinetal09}. 
This system exhibits the ``Hall effect'', with distinct ion and electron mean flow velocities and magnetic 
field-lines frozen-in to the electron fluid. The ERMHD equations (or the more general ``electron MHD'' or 
EMHD equations) produce the typical features of ``Hall reconnection'' such as quadrupolar magnetic 
fields in the reconnection zone \citep{UzdenskyKuslrud06}.\footnote{Because the Hall 
MHD equations have played a prominent role in magnetic reconnection research of the past decade 
\citep{Shayetal98, Shayetal99,Wangetal00,Birnetal01,Drake01,Malakitetal09,Cassaketal10},
it is worth remarking that those equations are essentially never applicable in astrophysical
environments. A derivation of Hall MHD based on collisionality requires that the ion skin-depth $\delta_i$ 
must satisfy the conditions $\delta_i\gg L\gg \ell_{mfp,i}.$ The second inequality is needed so that a two-fluid 
description is valid at the
scales $L$ of interest, while the first inequality is needed so that the Hall term remains 
significant at those scales.  However, substituting $\delta_i=\rho_i/\sqrt{\beta_i}$ into (\ref{lmfp-rho})
yields the result 
$$\frac{\ell_{mfp,i}}{\delta_i}\propto \frac{\Lambda}{\ln\Lambda}\frac{v_{th,i}}{c}. $$
The ratio $v_{th,i}/c$ is generally small in astrophysical plasmas, but the plasma parameter 
$\Lambda$ is usually large by even much, much more (see Table 1). Thus, it is usually the case that 
$\ell_{mfp,i}\gg \delta_i,$ unless the ion temperature is extremely low. A collisionless derivation of Hall 
MHD from gyrokinetics requires also a restrictive condition of cold ions (\cite{Schekochihinetal09}, 
Appendix E).  Thus, Hall MHD is literally valid only for cold, dense plasmas like those produced in 
some laboratory experiments, such as the MRX reconnection experiment \citep{Yamada99,Yamadaetal10}.}
At length scales smaller than $\rho_e,$ kinetic equations are 
required to describe both the ions and the electrons. It is at these scales that the magnetic flux
finally ``unfreezes'' from the electron fluid, due to effects such as Ohmic resistivity, electron inertia,
finite electron gyroradius, etc. However, as we shall discuss at length in this work, these weak effects 
are vastly accelerated by turbulent advection and manifested, in surprising ways, at far larger length scales.

We end this short review with just a few additional references. The effects of compressible modes 
have been extensively studied in numerical simulations of isothermal compressible MHD by 
\cite{ChoLazarian02,ChoLazarian03} (see also \cite{KowalLazarian10}). 
Separating MHD fluctuations into Alfv\'en, slow and fast modes, they showed
that the energy transfer from Alfv\'enic to compressible modes scales $\propto (\delta v)^2/(v_A^2+c_s^2)$, 
where $\delta v$ is the velocity perturbation at the given scale and $v_A$ and $c_s$ are Alfv\'en and 
sound speeds, respectively. As a result, for many astrophysically important cases the energy 
exchange is suppressed to a high degree. The Alfv\'en-wave cascade then decouples from the compressible 
degrees of freedom and the shear-Alfv\'en modes in the compressible regime exhibit scalings and 
anisotropy very similar to those in the incompressible regime. These results give further justification 
in our present study to adopting a simple model of an incompressible MHD plasma. As we shall 
see, it is the shear-Alfv\'en modes that are responsible for  all the principal effects discussed below. 
Finally, \cite{ChoLazarian04} have carried out a numerical study of three-dimensional EMHD 
turbulence. They observed a critically-balanced, anisotropic EMHD cascade quantitatively different, 
but qualitatively the same, as the GS95 strong MHD turbulence discussed in the following section.  
Although we focus in this work on reconnection phenomena at scales much larger than $\rho_i,$ 
similar ideas may apply at scales less than $\rho_i$ in the EMHD regime.

\section{Magnetohydrodynamic Turbulence}


Magnetized turbulence is a tough and complex problem (see \cite{Biskamp03} and 
references therein). A broad outlook on the astrophysical implications of the turbulence 
can be found in a review by \cite{ElmegreenScalo04}, while the effects of turbulence 
on molecular clouds and star formation are reviewed in \cite{McKeeOstriker07} and 
\cite{Ballesteros-Paredesetal06}. However, the issues of turbulence spectrum and its 
anisotropies, we feel, are frequently given less attention than  they deserve.  


In spite of its complexity, the turbulent cascade is self-similar  and universal 
(see \cite{MoninYaglom74}) over its inertial range. The physical variables are
proportional to simple powers of the eddy sizes over a large range of
sizes, leading to scaling laws expressing the dependence of certain
non-dimensional combinations of physical variables on the eddy
size. Robust scaling relations can predict turbulent properties on the
whole range of scales, including those that no large-scale numerical
simulation can hope to resolve. These scaling relations are extremely
important for obtaining an insight into processes on the small scales. 
In the interstellar medium, the inertial range of fluctuations is very appreciable, i.e.
from hundreds of kilometers to dozens of parsecs (see \cite{Armstrongetal94} for the data 
at small scale, and \cite{ChepurnovLazarian10} for its extension to pc scales) and therefore 
the details of the large-scale driving and of the small-scale dissipation become 
unimportant for the properties of turbulence at the long range of intermediate scales.

The presence of a magnetic field makes MHD turbulence anisotropic 
(\cite{MontgomeryTurner81, Matthaeusetal83, Shebalinetal83, Higdon84, 
GoldreichSridhar95}, see \cite{Oughtonetal03} for a review). The
relative importance of hydrodynamic and magnetic forces changes with scale, so the 
anisotropy of MHD turbulence does too. If the turbulence spectrum 
and its anisotropy are known, then many astrophysical results, e.g. the dynamics 
of dust, scattering and acceleration of energetic particles,  thermal conduction, 
can be obtained.

\subsection{The Goldreich-Sridhar picture of strong and weak MHD turbulence}

The standard theory for Alfv\'{e}nic turbulence is currently the one suggested by 
\cite{GoldreichSridhar95} [henceforth GS95]. The cornerstone of the theory is the 
so-called critical balance, which is the rough equality between the timescale 
of the eddy-turnovers perpendicular to the local magnetic field and the periods of waves 
associated with the eddies\footnote{This equality is frequently expressed following GS95 
convention in terms of parallel and perpendicular wavenumbers. However, as the wavenumbers 
are defined in the global magnetic field frame of reference, this may be misleading. Critical 
balance is a condition satisfied only in the frame related to the local magnetic field. The original 
GS95 work also uses closure relations which are true only in the global frame of reference. 
The importance of the local system of reference was implicit in \cite{Kraichnan65} but 
was first discussed within the GS95 model in subsequent publications (LV99, 
\cite{ChoVishniac00, MaronGoldreich01, Choetal02, ChoLazarian02}).}. 
The predictions of the GS95 model are in rough agreement with numerical simulations 
\citep{ChoVishniac00, MaronGoldreich01, Choetal02, BeresnyakLazarian06}, 
although some disagreement in terms of the measured spectral slope was noted. This 
disagreement produced a flow of papers with suggestions to improve the GS95 model 
by including additional effects like dynamical alignment \citep{Boldyrev05, Boldyrev06}, 
polarization intermittency \citep{BeresnyakLazarian06}, non-locality \citep{Gogoberidze07}. 
More recent studies in \cite{BeresnyakLazarian09, BeresnyakLazarian10} 
and \cite{Beresnyak11} indicate that all numerical simulations performed to date may not 
have sufficiently extended inertial range to get the actual spectral slope and therefore worries 
about the ``inconsistency'' of the GS95 model may be premature.

We shall add parenthetically that in a number of applications the empirical so-called 
composite 2D/slab model of magnetic fluctuations is used (see \cite{Bieberetal94}). 
In the latter model, which is also known as {\it two-component model}, 
it is assumed that interplanetary fluctuations can be described as a 
superposition of fluctuations with wave vectors parallel to the ambient large-scale 
magnetic field (so-called {\it slab modes}) and perpendicular to the mean field (so-called 
{\it two-dimensional modes}). This model results in a {\it Maltese cross} structure of magnetic 
correlations. This model was developed to account for the solar wind observations, which it does 
well by adjusting the intensity of two components (see, e.g., \cite{Matthaeusetal90}). 
From the theoretical point of view, as well as from the point of numerical testing, this standard model 
is rather vulnerable. Indeed, 2D fluctuations are consistent with the theory of weak Alfv\'{e}nic turbulence 
(see \cite{NgBhattacharjee96, LazarianVishniac99, Galtieretal00}), but this analytical 
theory predicts strengthening of the cascade with decrease of the turbulence scale. 
Therefore 2D fluctuations can describe Alfv\'{e}nic turbulence only over a limited range of scales. 
In addition, slab modes do not arise naturally in MHD numerical simulations with large-scale 
driving (see \cite{ChoLazarian02, ChoLazarian03}). We do not discuss this model of MHD turbulence
further, as it does not have physical motivation and is not supported by numerical modeling. It may be treated 
as a parameterization of a particular type of magnetic perturbation dominated by the peculiarities of driving. 
On the contrary, we consider turbulence which has an extended universal inertial range over which the 
effects of driving and dissipation are negligible.

Below we shall adopt the GS95 model of turbulence to describe the Alfv\'{e}nic part of MHD turbulent 
fluctuations. Note that the Alfv\'{e}nic perturbations are most important for magnetic field wandering 
which, according to LV99, is the process that enables fast reconnection. The GS95 model can be 
generalized to compressible turbulence and even for supersonic motions numerical calculations 
show that the Alfv\'{e}nic perturbations exhibit GS95 scaling \citep{ChoLazarian02, ChoLazarian03, 
KowalLazarian10}.  In what follows we consider MHD turbulence where the flows of energy 
in the opposite directions are balanced. When this is not true, i.e. when the turbulence has non-zero 
cross-helicity, the properties of turbulence depart substantially from the GS95 model
\footnote{Among the existing theories of imbalanced turbulence (see \cite{LithwickGoldreich07, 
BeresnyakLazarian08, Chandran08, PerezBoldyrev09}, all, but 
\cite{BeresnyakLazarian08} seem to contradict to numerical testing in 
\cite{BeresnyakLazarian09, BeresnyakLazarian10}. 
Solar wind presents a system with imbalanced turbulence. In compressible media 
the imbalance decreases due to reflecting of waves from pre-existing density fluctuations and 
due to the development of parametric instabililites.}. 
Similarly, we shall not discuss MHD turbulence at high magnetic Prandtl numbers,
when the field-parallel viscosity is much larger than resistivity (see 
\cite{Choetal02, Choetal03, Lazarianetal04}. 
This viscosity-dominated regime has important consequences for turbulence in partially ionized gas 
\citep{Lazarianetal04}, while our present study deals 
mostly with conducting fluids and fully ionized plasmas. Our considerations do apply for large 
magnetic Prandtl numbers at scales greater than the viscous length, if the kinetic Reynolds 
number is  also large. 

The nature of Alfv\'{e}nic cascade is expressed through the critical balance condition in the GS95 model 
of strong turbulence, namely,
\begin{equation} 
\ell_{\|}^{-1}v_A\sim \ell_{\bot}^{-1}\delta u_\ell,
\label{crit}
\end{equation}
where $\delta u_\ell$ is the eddy velocity, while $\ell_{\|}$ and $\ell_{\bot}$ are, respectively, 
eddy scales parallel and perpendicular to the local direction of magnetic field. The critical balance 
condition states that the parallel size of an eddy is determined by the distance Alfv\'{e}nic perturbation 
can propagate during an eddy turnover. The qualifier ``local'' is important\footnote{To stress the 
difference between local and global systems here we do not use the language of $k$-vectors. 
Wavevectors parallel and perpendicular to magnetic fields can be used, if only the wavevectors 
are understood in terms of a wavelet transform defined with the local reference system rather than 
ordinary Fourier transform defined with the mean field system.}, as no universal relations exist if 
eddies are treated in respect to the global mean magnetic field (LV99, \cite{ChoVishniac00, 
MaronGoldreich01, LithwickGoldreich01, Choetal02}). Combining (\ref{crit}) 
with the Kolmogorov cascade notion, i.e. that the energy transfer rate is 
$\delta u^2_{\ell}/(\ell_{\bot}/\delta u_{\ell})=const$ one gets $\delta u_\ell\sim \ell_\bot^{1/3}$
and $\ell_{\|}\sim \ell_{\bot}^{2/3}$. 
If the turbulence injection scale is $L_{i},$ then $\ell_{\|}\propto 
L_{i}^{1/3}\ell_{\bot}^{2/3}$,  which shows that the eddies get very 
anisotropic for small $\ell_{\bot}$. Recent measurements of anisotropy in the solar wind 
are consistent with these predictions \citep{Podesta10,Wicksetal10,Wicksetal11}. 

Critical balance is a feature of strong turbulence, which is the case when the 
turbulence is injected isotropically with velocity amplitude $u_L=v_A$. 
If the turbulence is injected at velocities $u_L\ll v_A$ (or anisotropically with $L_{i,\|}\ll L_{i,\bot}$),
then the cascade is weak, with $\ell_{\bot}$ of the eddies decreasing but $\ell_{\|}=L_i$ unchanged 
by resonant wave interactions \citep{MontgomeryMatthaeus95, LazarianVishniac99, 
Galtieretal00}. In other words, as a result of the weak cascade the eddies get thinner, but 
preserve the same length along the local magnetic field. The energy cascade rate in the weak 
MHD turbulence regime is \citep{Kraichnan65, NgBhattacharjee96, LazarianVishniac99} 
\begin{equation}
\varepsilon\approx \tau_{corr} \delta u_\ell^4/\ell_\bot^2 \approx
\delta u_\ell^4 L_i/v_A \ell_\bot^2, 
\label{cascade}
\end{equation}
where $\tau_{corr} = L_i/v_A$ is the decorrelation time due to large-scale Alfv\'en waves.
This implies that the scaling of weak turbulence is 
\begin{equation}
\delta u_\ell \sim u_L (\ell_{\bot}/L_i)^{1/2}.
\label{weak}
\end{equation}
For weak turbulence initially the LHS of Eq.~(\ref{crit}) is larger than the RHS,  but as $\ell_{\bot}$ 
decreases this eventually makes Eq.~(\ref{crit}) satisfied. 

Comparing Eqs. (\ref{weak}) 
and (\ref{crit}) for $\ell_\|=L_i$, one can see that the 
transition to the strong MHD turbulence happens at the scale $L_{trans}=L_i M_A^2$
and the velocity at this scale is $v_{trans}=u_L M_A$, with $M_A=u_L/v_A\ll 1$ the Alfv\'enic 
Mach number of the turbulence \citep{LazarianVishniac99, Lazarian06}. Thus, weak turbulence 
has a limited,  i.e. $[L_i, L_i M_A^2]$ inertial interval 
before it gets into the regime of strong turbulence at smaller scales. Note   that weak and strong are
not the characteristics of the amplitude of turbulent perturbations, but the strength of non-linear interactions 
(see more discussion in \cite{Choetal03}) and very weak Alfv\'{e}nic perturbations can 
correspond to a strong Alfv\'{e}nic cascade.  

While GS95 assumed that the turbulent energy is isotropically injected with amplitude $u_L=v_A$ 
at the scale $L_i$, LV99  provided general relations for the turbulent scaling at small scales for the case 
that the injection velocity $u_{L}$ is less or equal to $v_A$. 
Combining Eqs. (\ref{crit}), (\ref{cascade}) at $\ell_\bot=L_i,$ and the constant flux condition, 
we get the relations between the parallel and perpendicular scales of eddies in the strong 
GS95 cascade range,  which can 
be written in terms of $\ell_{\|}$ and $\ell_{\bot}$ (LV99):
\begin{equation}
\ell_{\|}\approx L_i \left(\frac{\ell_{\bot}}{L_i}\right)^{2/3} M_A^{-4/3}
\label{Lambda}
\end{equation}
\begin{equation}
\delta u_{\ell}\approx u_{L} \left(\frac{\ell_{\bot}}{L_i}\right)^{1/3} M_A^{1/3}.
\label{vl}
\end{equation} 
As we discuss later, the present day debates of whether GS95 approach should be augmented 
by additional concepts like ``dynamical alignment'', ``polarization'', ``non-locality'' \citep{Boldyrev06, 
BeresnyakLazarian06, BeresnyakLazarian09, Gogoberidze07} do not change 
the nature of the reconnection of the weakly turbulent magnetic field. 

The above relations were exploited in LV99 to estimate magnetic 
field lateral diffusion. Moving a distance $s$ along field-lines, one finds that a 
pair of lines initially a distance $\ell_\bot^{(0)}$ apart at $s=0$ separate at the rate
\begin{equation}
\frac{d}{ds}\ell_\bot \simeq \frac{\delta b_\ell}{B_0}\simeq \frac{\delta u_\ell}{v_A}.
\label{diffusion}
\end{equation}
Substituting the scaling given by Eq.~(\ref{vl}),  one can solve to obtain 
\be  \ell_\bot^2 \simeq (s^3/L_i) M_A^4 \lb{B-rich} \ee
when $L_i>s\gg \ell_\bot^{(0)}.$  This result applies only until $s=L_i$ when 
$\ell_\bot(s)\simeq L_{trans},$ the transitional scale between weak and 
strong turbulence. Since $\ell_\|=L_i$ is constant in weak turbulence, there 
are no eddies with parallel length-scale greater than $L_i.$ Thus, as $s$
continues to increase, the lines encounter at every increment of $s$ by $L_i$ 
a new turbulent eddy and undergo a further separation by $L_{trans}.$ 
The result is a diffusive random-walk with 
\be \ell_\bot^2  \simeq L_{trans}^2 (s/L_i) \simeq  s L_i M_A^4 \lb{B-diff} \ee
for $s>L_i.$  The results (\ref{B-rich}),(\ref{B-diff}) are exact analogues 
for magnetic field-lines of the 2-particle diffusion of Lagrangian particle trajectories 
discussed in the next section and can be obtained more rigorously by the methods 
employed there. Since this discussion is somewhat out of logical order, we defer 
it to Appendix A.

A remarkable feature of (\ref{B-rich}),(\ref{B-diff}) is their complete
lack of dependence on the initial separation $\ell_\perp^{(0)}$ of the 
lines. As we shall see below, it is this property which leads in the LV99 model
to fast magnetic reconnection in MHD turbulence independent of the microscopic 
resistivity.  The shear-Alfv\'enic component of turbulence is the most 
important for magnetic field-line wandering. \cite{ChoLazarian02, ChoLazarian03} 
demonstrated the correspondence of the Alfv\'enic component of compressible 
MHD turbulence, obtained by decomposition into fundamental modes, 
to the Alfv\'enic turbulence in incompressible media. Thus we shall use the relations 
above to treat magnetic fields in realistically compressible astrophysical fluids.

\subsection{Lagrangian particle dynamics in MHD turbulence}

It will be important for what follows to develop a theoretical understanding 
of the dynamics of Lagrangian particles in MHD turbulence. Here one means 
by ``particles'' not the microscopic charged constituents of  the plasma (electrons 
and ions), but instead macroscopically small and microscopically large parcels of 
plasma fluid. The Lagrangian perspective is particularly important to understand 
turbulent processes and there has been an explosion of research in this area over
the last decade (see \cite{Falkovichetal01, SalazarCollins09, ToschiBodenschatz09}).  
There has been relatively little work done on the Lagrangian 
properties of MHD turbulence, however, where the effects of Alfv\'en waves on fluid 
particle motion can be significant. We shall here develop a theory of Lagrangian 
MHD turbulence at the level of GS95 phenomenology by a semi-quantitative use
of the equations of motion. 

We begin with the simplest case of 1-particle or Taylor diffusion, which concerns 
the displacement $\delta\bx (t)=\bx(t)-\bx_0$ of a Lagrangian fluid particle from 
its initial position as it moves according to the advection equation
\be \frac{d}{dt}\bx(t)=\bU(t) \ee
with $\bU(t)=\bu(\bx(t),t)$ the Lagrangian particle velocity. At early 
times $t\ll \tau_L$ or short compared with the eddy-turnover time $\tau_L=u_L/L_i$, 
motion is ballistic with $\langle x^2(t)\rangle\sim u_L^2t^2.$ It was shown by 
\cite{Taylor21}
that the motion is diffusive 
$\langle \delta x_i(t)\delta x_j(t)\rangle \sim 2 D_{ij} t$ 
at times $t\gg \tau_L$, with the eddy-diffusivity   
\be D_{ij}=\frac{1}{2}\int_{-\infty}^{+\infty} dt\,\langle U_i(t) U_j(0)\rangle. 
\lb{taylor} \ee
These results carry over unchanged to MHD turbulence. For the weak magnetic field
case with $M_A\geq 1$ (including hydrodynamic turbulence with $M_A=\infty),$ 
$\langle U(t)U(0)\rangle\sim u_L^2 e^{-t/\tau_L}$ 
\citep{Mordantetal01, BusseMuller08}. Thus, 
\be D\sim u_L^2\tau_L \sim u_L L_i,  \lb{strong-D1} \ee
exactly as for hydrodynamics of neutral fluids. When $M_A\simeq 1,$ one expects 
slight differences from the hydrodynamic case due to anisotropy of MHD turbulence. 
For example, the velocity components $U_\|$ and $U_\perp$ ought to have times scales 
$\tau_{L\,\|}>\tau_{L\,\perp},$ since the parallel components are pseudo-Alfv\'enic
modes with essentially passive scalar dynamics, due to reduced nonlinear interactions. 
Thus, one expects that $D_\|>D_\perp,$ as indeed observed in simulations of 
\cite{BusseMuller08}.  For $M_A\ll 1,$ there are more profound effects due to the 
strong field. Weak MHD turbulence becomes relevant, for which the nonlinear time-scale 
is $\tau_L=M_A^{-2}\omega_A^{-1}$ with $\omega_A=v_A/L_i$ the 
forcing-scale Alfv\'en wave frequency. (See \cite{Kraichnan65, SridharGoldreich94,Galtieretal00}; 
cf. also eq.(\ref{tau-ell}) below for $\ell_\perp=L_i$). Furthermore, fluid particles in the 
Lagrangian frame experience fast shear-Alfv\'enic oscillations in velocity, so that 
\be \langle U_\perp(t)U_\perp(0)\rangle\sim u_L^2 {\rm Re}\left(e^{i\omega_At-t/\tau_L}\right). \ee
Thus, (\ref{taylor}) gives
\be D_\perp\sim u_L^2 \frac{\tau_L}{(\omega_A\tau_L)^2+1} . \lb{1-eddy-diff-eq} \ee
Since $\omega_A\tau_L=M_A^{-2}\gg 1,$ 
\be D_\perp \sim u_L^2 \frac{1}{\omega_A^2\tau_L} \sim u_L L M_A^3. \lb{weak-D1} \ee
There is a large reduction in the turbulent 1-particle diffusivity due to inefficient 
advection by Alfv\'en wave modes. Physically, the velocity of a particle transported 
by Alfv\'en wave turbulence experiences rapid oscillations in sign that lead to large
cancellations in the net displacement.  

We note as an aside that the Taylor 1-particle diffusivity given by (\ref{taylor}) 
coincides with the effective turbulent diffusivity of an advected scalar field (e.g. 
temperature),  whenever the ``eddy-diffusivity'' concept is valid. In order for such 
a description of the turbulence effects to be accurate, the scalar fields must vary slowly 
on spatial scales of order $L_i$ and, in that case, they experience an augmentation 
of their molecular diffusivity by a ``turbulent diffusivity''  precisely given by (\ref{taylor}). 
Our results (\ref{strong-D1}),(\ref{weak-D1}) thus reproduce  those of \cite{Lazarian06} 
for the eddy thermal diffusivity in MHD turbulence. Of course, in general, 
when there is no separation between the scale of variation of the scalar field and the 
scale $L_i$ of the turbulence, then the gradient-transport assumption breaks down 
(e.g. \cite{TennekesLumley72}, \S 2.3). In such cases, eddy-diffusivity models can produce 
erroneous results and may suffice only for order-of-magnitude estimates of scalar transport.  

Two-particle turbulent diffusion or Richardson diffusion concerns instead the separation 
$\Bell(t)=\bx(t)-\bx'(t)$ between a pair of Lagrangian fluid particles. It was proposed 
by \cite{Richardson26} that this separation grows in turbulent flow according to the 
formula
\be \frac{d}{dt}\langle \ell_i(t)\ell_j(t)\rangle=\langle
D_{ij}(\Bell)\rangle \lb{richardson} \ee
with a scale-dependent eddy-diffusivity $D(\ell).$ In hydrodynamic turbulence
Richardson deduced that $D(\ell)\sim \varepsilon^{1/3}\ell^{4/3}$ (see also
\cite{Obukhov41}) and thus $\ell^2(t)\sim \varepsilon t^3. $ An analytical 
formula for the 2-particle eddy-diffusivity was later derived by 
\cite{Batchelor50, Kraichnan66}:
\be D_{ij}(\Bell)=\int_{-\infty}^0 dt \langle \delta U_i(\Bell,0)
\delta U_j(\Bell,t)\rangle \ee
with $ \delta U_i(\Bell,t)\equiv U_i(\bx+\Bell,t)-U_i(\bx,t)$ the relative velocity
at time $t$ of a pair of fluid particles which were at positions $\bx$ and $\bx+\Bell$
at time 0. This formula is the analogue for 2-particle eddy-diffusivity of Taylor's formula 
(\ref{taylor}) for 1-particle diffusivity and it is valid in MHD turbulence just as in 
hydrodynamic turbulence. We can assume similarly as before that 
\be \langle \delta U_\perp(\ell,0)\delta U_\perp(\ell,t)\rangle 
    \sim \delta u^2_\ell {\rm Re}\left(e^{i\omega_{A\,\ell}t-|t|/\tau_\ell}\right), \ee
where $\omega_{A,\ell}$ is the Alfv\'en wave-frequency and $\tau_\ell$ is the nonlinear 
interaction time both at length-scale $\ell.$   The result for the 2-particle diffusivity 
analogous to (\ref{1-eddy-diff-eq}) is 
\be D_\perp(\ell) \sim \delta u^2_\ell \frac{\tau_\ell}{(\omega_{A\,\ell}\tau_\ell)^2+1} . 
\lb{2-eddy-diff-eq} \ee
We consider here only the case $M_A<1,$ 
for which weak turbulence holds when $\ell_\perp>L_{trans}$ and strong turbulence when 
$\ell_\perp<L_{trans}.$ We shall also treat particle dispersion only in the direction perpendicular 
to the background magnetic field, since this is what is required later for the application to 
reconnection. See  \cite{BusseMuller08, Eyink11} for some discussion of particle 
dispersion in the field-parallel direction. 

We first consider the strong turbulence regime when $\ell_\perp(t)<L_{trans}.$ 
In this case, $\tau_\ell =\varepsilon^{-1/3}\ell_\perp^{2/3}$ and 
$\omega_{A \,\ell}=v_A/\ell_\|,$ where $\ell_\|$ is the parallel 
scale of the turbulent eddy with perpendicular scale $\ell_\perp(t)$ 
(and not the particle-separation $\ell_\|(t)$ in the field-parallel direction).  
Because of the critical balance condition in strong MHD turbulence, 
\be  \omega_{A\,\ell} \tau_\ell=(const.)\ee
In that case 
\be \frac{\tau_\ell}{(\omega_{A\,\ell}\tau_\ell)^2+1}= f \tau_\ell \ee
with $f=[(\omega_{A\,\ell}\tau_\ell)^2+1]^{-1}<1$ a constant factor which 
represents the reduced efficiency of particle transport due to wave oscillations. 
Eq.(\ref{2-eddy-diff-eq}) gives
\be D_\perp(\ell) \sim f \delta u^2_\ell \tau_\ell \sim f
 \varepsilon^{1/3}\ell^{4/3}_\perp \ee
with the same scaling as the original \cite{Richardson26} 2-particle diffusivity.  
Thus, 
\be \ell^2_\perp (t) \sim f^3 \varepsilon t^3,  \lb{strong-2part-law} \ee
showing the same $t^3$-law as in the hydrodynamic case, with only a reduced 
coefficient. 

We now consider the weak turbulence regime with $\ell_\perp(t)>L_{trans}.$ 
The pair of particles are advected apart by a turbulent eddy with corresponding 
perpendicular length-scale but with $\ell_\|=L_i$ for all eddies, so that 
$\omega_{A\,\ell}=\omega_A=v_A/L_i,$ independent of $\ell_\perp.$ 
The nonlinear interaction time $\tau_\ell$ is estimated from 
\be  \frac{\delta u^2_\ell}{\tau_\ell}=\varepsilon =\frac{u_L^4}{v_AL_i}\ee
together with (\ref{weak}) to give
\be \tau_\ell = (\ell_\perp/v_A) M_A^{-2}. \lb{tau-ell} \ee
Since
\be  \omega_A\tau_\ell=\ell_\perp/L_{trans}\gg 1, \ee
it follows from (\ref{2-eddy-diff-eq}) that 
\be D_\perp(\ell) \sim \frac{\delta u^2_\ell}{\omega_A^2\tau_\ell}
    \sim \frac{\varepsilon}{\omega_A^2}= u_LL_iM_A^{3}. \ee
Note that $D_\perp(\ell)$ is identical with the 1-particle diffusivity $D_\perp$ 
given by (\ref{weak-D1}) and, in particular, is independent of $\ell_\perp.$
It thus follows that in the weak turbulence regime 
\be \ell^2_\perp(t) \sim u_LL_iM_A^3 t \lb{weak-2part-law} \ee
and 2-particle diffusion is identical to 1-particle diffusion. It should be 
possible to verify this interesting prediction by analytical methods of weak 
turbulence theory for Lagrangian particle motion (see \cite{Balk02}).
 The crossover between the dispersion laws (\ref{strong-2part-law}) and 
(\ref{weak-2part-law}) occurs at the time $t_c\sim L_i/v_A=M_A t_L.$

\section{Flux-Freezing and Spontaneous Stochasticity}

There is an intimate connection between magnetic reconnection and flux-freezing. 
Indeed, some authors consider that reconnection is precisely any topology-changing 
evolution of the magnetic field due to nonexistence of a smooth, flux-preserving velocity 
field.  For example, \cite{Greene93} proposed to define reconnection ``as evolution in which 
it is not possible to preserve the global identification of some field lines.'' See also 
\cite{HornigSchindler96}.  For this reason, a necessary prelude to our treatment  of
turbulent magnetic reconnection is a careful discussion of how flux-freezing operates 
in MHD turbulence. 

\subsection{Stochastic flux-freezing} 

Magnetic field lines do not move. Magnetic fields evolve under the Maxwell equations, of course, 
but single field lines do not possess an identity over time. As has long been understood, 
magnetic field-line motion is a ``metaphysical'' concept, defined by convention and not testable 
by experiment (cf. \cite{Newcomb58, Vasyliunas72, Alfven76}). Any line-velocity which 
leads to agreement with the Maxwell equations for the magnetic field evolution is equally acceptable, 
e.g. the bulk plasma velocity $\bu$ in ideal MHD, the ${\bf E}\btimes \bB$ drift velocity in low energy 
plasmas, etc. Flux-freezing is such a powerful heuristic tool, however, that it is often forgotten that the 
concept of magnetic line-motion rests on certain conventions and is not a direct, physical reality. In 
particular, real physical plasmas always possess a Spitzer plasma resistivity (along with other possible 
forms of non-ideality, such as electron inertia, electron pressure anisotropy, etc.) The validity of flux-freezing 
in the standard sense depends upon careful consideration of  the limit when such non-ideal effects are 
taken to be small. 

Since field-lines do not move in reality, one is free to adopt any convention for their motion
which is consistent with the dynamical equations. Let us for the moment assume the validity
of the resistive induction equation
\be \partial_t\bB = \grad\btimes (\bu\btimes\bB) +\lambda\triangle\bB \lb{induct} \ee
with $\lambda=\eta c^2/4\pi$ the magnetic diffusivity for a simple scalar resistivity $\eta.$
\footnote{It may be that the flux-freezing constraint is broken in actual astrophysical 
plasmas by some other microscopic mechanism (e.g. electron inertia, finite electron gyroradius, etc.).
We shall consider such alternatives below (section 4.3) and show that they do not alter our final 
conclusions. In particular, we argue that the large-scale global reconnection rate in a turbulent 
plasma is independent of whatever microscopic plasma process produces field-line breaking. 
It is thus pedagogically useful to begin with the simplest case.}
A smooth, deterministic velocity field $\bu_*$ is flux-preserving if the magnetic field obeys 
\be \partial_t\bB = \grad\btimes (\bu_*\btimes\bB). \lb{flux-pres} \ee
Unfortunately, for the resistive MHD model in three space dimensions, there is generally no 
flux-preserving velocity whatsoever. For example, \cite{WilmotSmithetal05} provide a counterexample 
with a closed magnetic field-line through which the flux is changing in time, so that no deterministic 
flux-preserving velocity can exist. This seems to leave one without any consistent and meaningful approach 
to discuss field-line ``motion'' in the presence of Ohmic resistivity. 

On the other hand, magnetic line-motion in a resistive MHD plasma may be  very naturally regarded 
as {\it stochastic} \citep{Eyink09}. Let the plasma fluid velocity be $\bu(\bx,t),$ assumed here to be 
divergence-free for an incompressible plasma\footnote{The assumption of incompressibility 
is convenient to simplify the arguments and, also, realistic when considering the RMHD dynamics of
shear-Alfv\'en modes. As a matter of fact, the formulas (\ref{SDE})-(\ref{tildeB}) below all generalize to the 
compressible case. See \cite{Eyink11}.}. Then we may define stochastic Lagrangian trajectories 
$\widetilde{\bx}(\ba,t)$ by the initial-value problem 
\be \left\{ \begin{array}{l}
                {{d\widetilde{\bx}}\over{dt}}(\ba,t) =\bu(\widetilde{\bx}(\ba,t),t) 
                            +\sqrt{2\lambda} \widetilde{\boeta}(t), \cr 
                \widetilde{\bx}(\ba,t_0) = \ba, 
                \end{array} \right.  \lb{SDE}  \ee
where $\widetilde{\boeta}(t)$ is a three-dimensional vector white-noise process, delta-correlated
in time. For any initial magnetic field $\bB_0(\bx)$ one can form the random field
\be \widetilde{\bB}(\bx,t)= \left. \bB_0(\ba)\bdot\grad_a\widetilde{\bx}(\ba,t)\right|_{\widetilde{\ba}(\bx,t)} 
     \lb{virtB} \ee
where $\widetilde{\ba}(\bx,t)$ is the spatial inverse map to $\widetilde{\bx}(\ba,t).$ If the flow 
map were deterministic, then (\ref{virtB}) would be the standard {\it Lundquist formula} for the 
magnetic  field. What is true in the present case with $\lambda>0$ is that the ensemble-average field
\be \bB(\bx,t) = \overline{\,\,\widetilde{\bB}(\bx,t)\,\,} \lb{meanB} \ee
over all independent realizations of the white-noise is the exact solution of the resistive 
induction equation (\ref{induct}) with initial condition 
\be \bB(\bx,t_0)=\bB_0(\bx). \lb{ic} \ee In fact, (\ref{SDE})-(\ref{meanB}) are mathematically equivalent 
to (\ref{induct}) and (\ref{ic}). It should not be too surprising that averaging over Brownian motions will 
reproduce the Laplacian diffusion term $\lambda\triangle\bB$. 

\begin{figure}
\begin{center}
\includegraphics[width=160pt,height=360pt]{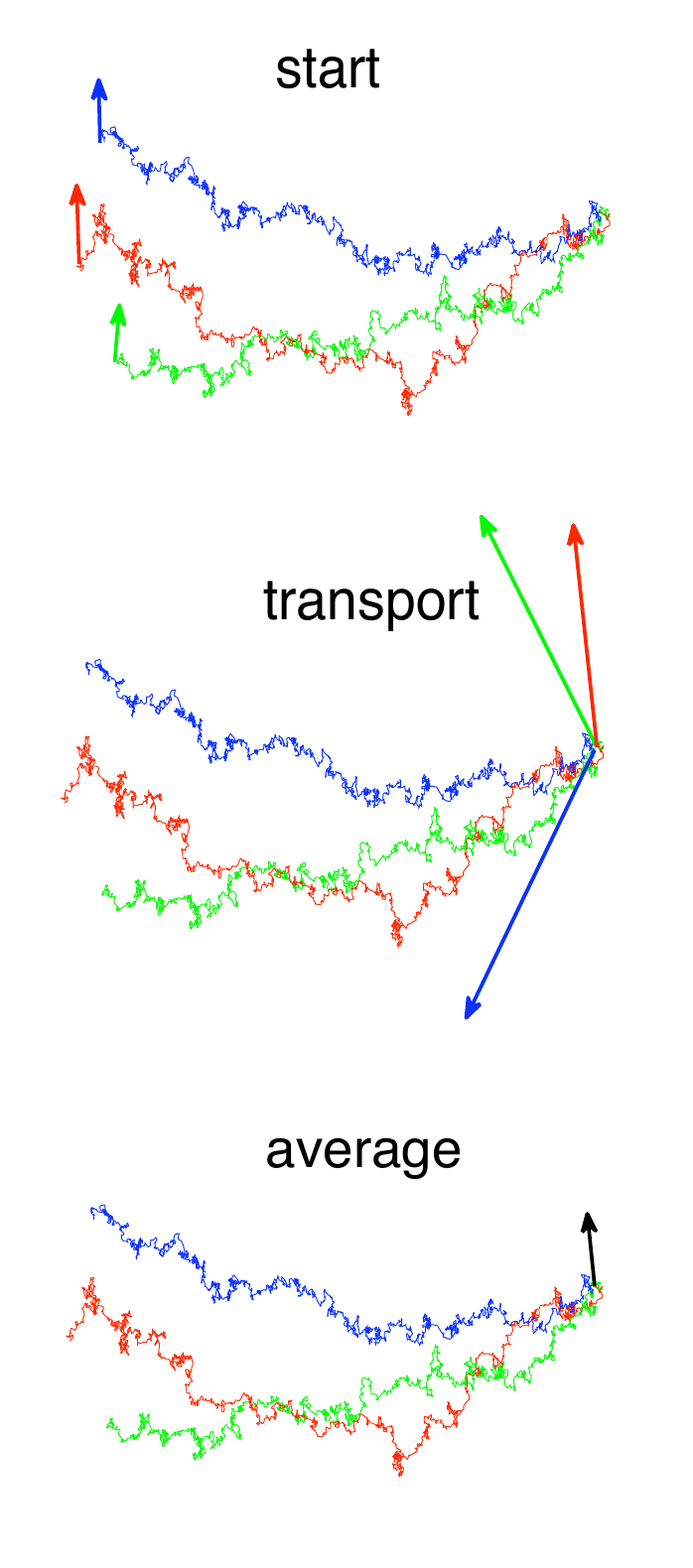}\\
\end{center}
\caption{{\it Illustration of the stochastic Lundquist formula.} 
Three stochastic Lagrangian trajectories running backward in time from a common 
point are shown in \textcolor{red}{red},  \textcolor{blue}{blue}, \textcolor{green}{green}.
Starting field vectors, represented by correspondingly colored arrows, are transported 
along the trajectories, stretched and rotated, to the common final point. These ``virtual field 
vectors'' are then averaged to give the resultant magnetic field at that point, indicated by 
the {\bf black} arrow.}
\end{figure}

The eqs.(\ref{SDE})-(\ref{meanB}) are equivalent to the following path-integral formula, 
derived in \cite{Eyink11}:  
\begin{eqnarray}
&& \bB(\bx,t) =
 \int_{\ba(t)=\bx} {\cal D}\ba \,\, \bB_0(\ba(t_0))\bdot \boJ(\ba,t)  \,\,\,\,\,\,\,\,\cr
&& \,\,\,\,\times 
\exp\left(-{{1}\over{4\lambda}}\int_{t_0}^t d\tau\, |\dot{\ba}(\tau)-\bu^\nu(\ba(\tau),\tau)|^2\right)
\,\,\,\,\,\,\,\,\, \lb{path-int} \end{eqnarray}
where $\bB$ is interpreted as a 3-dimensional row vector and $\boJ(\ba,\tau)$ is a 
$3\times 3$ matrix satisfying the following ODE along the trajectory $\ba(\tau)$:
\be \left\{ \begin{array}{l} 
              {{d}\over{d\tau}}\boJ(\ba,\tau)=\boJ(\ba,\tau)\grad_x\bu(\ba(\tau),\tau) \cr
              \boJ(\ba,t_0)={\bf I}. 
              \end{array} \right. \lb{Jeq} \ee
By applying $\grad_a$ to (\ref{SDE}) it is easy to see that $\boJ(\ba,\tau)=
\grad_a\widetilde{\bx}(\ba(t_0),\tau)$ and thus we may identify 
\be  \widetilde{\bB}(\ba(\tau),\tau)=\bB_0(\ba(t_0))\bdot \boJ(\ba,\tau).  \lb{tildeB} \ee
As often with such path-integration techniques, only the initial field $\bB_0(\bx)$ 
and the final field $\bB(\bx,t)$ have real physical meaning and the intermediate 
random fields $\widetilde{\bB}(\bx,t)$ are merely calculational devices. We shall refer
to the latter as the ``virtual'' magnetic fields, since they enter as non-real, intermediate 
states between two physical fields at different times. Equation (\ref{meanB}) [or (\ref{path-int})] 
shows that summing over all of the ``virtual'' fields at any instant reproduces the true, physical 
magnetic field. It follows from (\ref{virtB}) [or (\ref{tildeB})] that each ``virtual'' field $\widetilde{\bB}$ 
is topologically equivalent to the initial field $\bB_0$ and is simply stretched, deformed and advected 
by its own stochastic flow $\widetilde{\bx}.$ Field-lines for two different ``virtual'' fields that arise 
from statistically independent flows may pass directly through each other as they move,
since they correspond to completely different random realizations. The topology of the field-lines 
is only changed by the final averaging step, which corresponds to resistive gluing 
of all the ``virtual'' magnetic field vectors which arrive to the same point $\bx$ at time $t.$  
It  is this resistive averaging which reconnects the lines of the initial magnetic field $\bB_0$. 
See Fig.~1 for a pictoral representation of this process.

We shall refer to either the relations (\ref{SDE})-(\ref{meanB}) or the equivalent 
path-integral formula (\ref{path-int})-(\ref{Jeq}) as the {\it stochastic flux-freezing} (SFF)
relations. They provide a consistent approach to describe the ``motion'' of magnetic field-lines
in the presence of resistivity. Setting $\lambda=0$ one formally recovers the standard 
deterministic flux-freezing relations of ideal MHD, but this conclusion is not rigorous. 
A careful discussion is required of the limiting process involved.     
 
\subsection{Spontaneous stochasticity}  
 
Standard flux-freezing arguments are invoked in the limit of high conductivity or 
vanishingly small magnetic diffusivity.  It is usually assumed that flux-freezing in the 
conventional sense will hold better and better as  $\lambda\rightarrow 0.$ This is not true, 
however, when the plasma flow is turbulent. Instead, flux-freezing in this circumstance 
breaks down in a completely novel and unexpected way:  it remains stochastic 
\citep{Eyink07, Eyink11}. 
 
Consider first the  simpler situation of a smooth, laminar solution of the resistive MHD 
equations in the limit of very small $\lambda.$ Define the standard (deterministic) Lagrangian 
flow map by 
\be  \left\{ \begin{array}{l} 
                 {{d\bx}\over{dt}}(\ba,t)=\bu(\bx(\ba,t),t), \cr 
                 \bx(\ba,0)= \ba.  \end{array} \right.  \lb{ODE} \ee 
It is not hard to see under the stated assumptions that\footnote{A more precise
statement is that 
$ \langle|\widetilde{\bx}(\ba,t)-\bx(\ba,t)|^2\rangle\leq \frac{6\lambda}{K}(e^{Kt}-1) $
where $K$ is the ``Lipschitz constant'' of the velocity field $\bu.$ See \cite{FreidlinWentzell84},
Ch.2. In the presence of Lagrangian chaos this upper bound may be attained, with $K$ 
the leading Lyapunov exponent. Note that for $t\ll 1/K$ the upper bound reduces to $6\lambda t,$
the usual diffusive estimate. The important fact is that the bound vanishes 
as $\lambda\rightarrow 0.$}
\be  |\widetilde{\bx}(\ba,t)-\bx(\ba,t)|=O(\sqrt{\lambda t}) \lb{x-xtilde} \ee
for typical realizations of the ``virtual'' flows $\widetilde{\bx}.$ The real magnetic 
field is also given to very good accuracy by the standard Lundquist formula, with errors 
that vanish as $\lambda\rightarrow 0.$  Thus, each ``virtual'' field line is wiggling a small 
distance $\sim\sqrt{\lambda t}$ around a particular physical field-line for very small 
$\lambda$ (and not too large times $t$). For this reason, one does not need to make 
use of the concept of  ``virtual'' field lines for smooth, laminar solutions of the MHD 
equations. The lines of force of the real, physical magnetic field are themselves ``nearly'' 
frozen-in for very small resistivities and one  does not need to make an essential distinction 
between real and ``virtual''  fields. 
 
This is not the case for rough, turbulent solutions of the resistive MHD equations, 
even for vanishingly small $\lambda.$ By ``rough'' fields we mean here more precisely
that $\bu,\bB$ which solve the MHD equations have power-law energy spectra 
$\propto k^{-n},$ with $1<n<3,$ for a long range of wavenumbers $k$ with 
kinematic viscosity $\nu$ and magnetic diffusivity $\lambda$ both small. For 
example, in the GS95 theory of MHD turbulence the three-dimensional (anisotropic) 
energy spectra for both $\bu,\bB$ can be written in the form 
\be  E(k_\perp,k_\|)\sim v_A^2 M_A^{4/3} L_i^{-1/3} k_\perp^{-10/3}
          f\left({{k_\| L_i^{1/3}}\over {k_\perp^{2/3}M_A^{4/3}}}\right). \lb{3Dspectrum} \ee
where we added, compared to the original expressions, the dependence on $M_A$ following LV99. The
original GS95 theory was formulated for $M_A\equiv 1$, as we mentioned earlier. Note, that the above
expression is expected to be valid for large enough $k_\|$ and $k_\perp.$ The function $f$ is not specified in
GS95, but Cho, Lazarian \& Vishniac (2002) showed that it can be fitted by a Castaing function (see Castaing,
Gagne \& Hopfinger 1990), which, in its turn, outside the vicinity of $k_{\|}=0,$ can be approximated by an exponent. As we mentioned earlier, $k_{\|}$ and $k_{\bot}$ in the expression are not the wavevectors in the conventional sense as the corresponding scales should be measured in the local system of reference\footnote{The use of $k$ vectors in GS95 stems from the fact that the concept the local system of reference was not introduced till the later works.}  

One thus finds that $E(k_\perp)\propto 
k_\perp^{-5/3}$ and $E(k_\|)\propto k_\|^{-2},$ so that the fields predicted by GS95 
are ``rough''  in our sense. In this case, (\ref{x-xtilde}) is invalid except for very short times, 
due to the properties of 2-particle Richardson diffusion. For example, in the transverse 
direction perpendicular to the local magnetic field, (\ref{x-xtilde}) holds only for times less 
than $t_\lambda=O(\sqrt{\lambda/\varepsilon}).$ 
If one considers a pair of independent ``virtual'' flows $\widetilde{\bx},\widetilde{\bx}',$
then for times $t\gg t_\lambda$ in the root-mean-square sense
\be |\widetilde{\bx}_\perp(\ba,t)-\widetilde{\bx}_\perp'(\ba,t)| \sim (\varepsilon t^3)^{1/2}. 
     \lb{x-xtilde-Rich} \ee
See eq.(\ref{strong-2part-law}). This result is completely independent of magnetic 
diffusivity, with very remarkable consequences. As $\lambda,\nu\rightarrow 0,$ and thus also 
$t_\lambda\rightarrow 0,$ the two realizations stay far apart for {\it all} times $t$ and do not 
collapse to a single real particle trajectory. This phenomenon has been called {\it spontaneous 
stochasticity} because the distance between ``virtual'' particle trajectories stays finite (and random) 
in the limit, similar to the way that a spontaneous magnetization develops in a ferromagnet below 
the Curie temperature when an external magnetic field  is taken to zero \citep{Bernardetal98,
GawedzkiVergassola00,EvandenEijnden00,EvandenEijnden01,Chavesetal03,Kupiainen03}. 
In the limit $\lambda,\nu\rightarrow 0$ all of the ``virtual'' flows $\widetilde{\bx}(\ba,t)$ solve 
the deterministic initial-value problem (\ref{ODE}), which then possesses infinitely many 
solutions!  For more discussion of spontaneous stochasticity and for a review of experimental 
and numerical evidence of this remarkable phenomenon, see \cite{Eyink11}. 

It is an immediate consequence that in a ``rough''  turbulent velocity field, the conventional 
notion of flux-freezing must break down. The violations are not small. Taking the square 
of (\ref{x-xtilde}) as the conventional estimate for line-slippage, $\langle \ell^2_\perp
\rangle_{{\rm frozen}}\sim \lambda t,$ and comparing with the predicted amount (\ref{x-xtilde-Rich}), 
$\langle \ell_\perp^2\rangle\sim \varepsilon t^3,$ one finds that 
\be \frac{\langle\ell^2(t)\rangle}{\langle\ell^2(t)\rangle}_{{\rm frozen}}\sim M_A^2
    \left(\frac{t}{t_L}\right)^2 S_L, \ee
where $t_L=L_i/u_L$ is the eddy-turnover time and $S_L=v_AL_i/\lambda$ is the Lundquist
number based on the turbulence injection scale. For $S_L\gg 1$ typical of astrophysical 
systems, the violations of standard flux-freezing are enormous and increase with time.  It is 
also not true however that flux-freezing is completely violated. Stochastic flux-freezing in 
the sense of (\ref{SDE})-(\ref{meanB}) remains valid for the ``virtual'' particle trajectories 
in a plasma flow which corresponds to a high (kinetic and magnetic) Reynolds-number, 
turbulent MHD solution. 

There is another way to understand spontaneous stochasticity  by considering the 
real trajectories of actual plasma fluid elements, which obey (\ref{ODE}). 
Suppose that one considers two fluid-elements located initially at locations $\ba,\ba'$ which 
are  displaced from each other by a small amount $\borho=\ba'-\ba.$ At early 
times the separation of the particles is ballistic \citep{Batchelor50}. For example, GS95 theory implies that 
in the transverse direction 
\be  |\bx_\perp(\ba',t)-\bx_\perp(\ba,t)|\sim |\bu_\perp(\ba')-\bu_\perp(\ba)|t 
   \sim (\varepsilon \rho_\perp)^{1/3} t \lb{Batchelor} \ee
in the rms sense, for times $t<t_0=O\left( (\rho_\perp^2/\varepsilon)^{1/3}\right)$
and separations $\rho_\perp>\rho_\nu=O((\nu^3/\varepsilon)^{1/4}).$ However, 
for $t\gg t_0$ the Richardson-type result 
\be  |\bx_\perp(\ba',t)-\bx_\perp(\ba,t)| \sim (\varepsilon t^3)^{1/2}. 
     \lb{Richardson} \ee
again holds in the rms sense. Note that the initial particle separation $\borho$
is completely ``forgotten'' at sufficiently long times!  See Fig.~2. In the limit taking 
first $\nu\rightarrow 0$ and then $\rho\rightarrow 0,$ one obtains infinitely 
many solutions of the deterministic initial-value problem (\ref{ODE}). In fact, one can see by taking 
$\widetilde{\ba}'=\ba+\widetilde{\borho}$ for a random displacement $\widetilde{\borho}$
that the real fluid particle trajectories also become random in this limit. More precisely, 
consider an ensemble of random displacements inside the ball $|\widetilde{\borho}|
<\rho_0. $ Then one obtains a random ensemble of real particle trajectories by 
solving (\ref{ODE}) for $\bx(\ba+\widetilde{\borho},t)$ and taking the limits first 
$\nu\rightarrow 0$ and then $\rho_0\rightarrow 0.$ In fact, it can be shown using 
Richardson's diffusion approximation for 2-particle dispersion that this is the same as the 
random ensemble of ``virtual'' particle trajectories obtained by solving (\ref{SDE}) and taking 
the limits $\nu,\lambda\rightarrow 0$ together \citep{EvandenEijnden00}. Put another way, 
for every ``virtual'' particle trajectory $\widetilde{\bx}(\ba,t)$ there is a real fluid particle 
trajectory $\bx(\ba+\widetilde{\borho},t)$ such that $|\widetilde{\borho}|\rightarrow 0$
and $|\widetilde{\bx}(\ba,t)-\bx(\ba+\widetilde{\borho},t)|\rightarrow 0$ in the limit 
$\nu,\lambda\rightarrow 0.$ ``Virtual'' particle trajectories coincide 
with real fluid particle trajectories in this statistical sense. 

\begin{figure}
\begin{center}
\includegraphics[width=160pt,height=360pt]{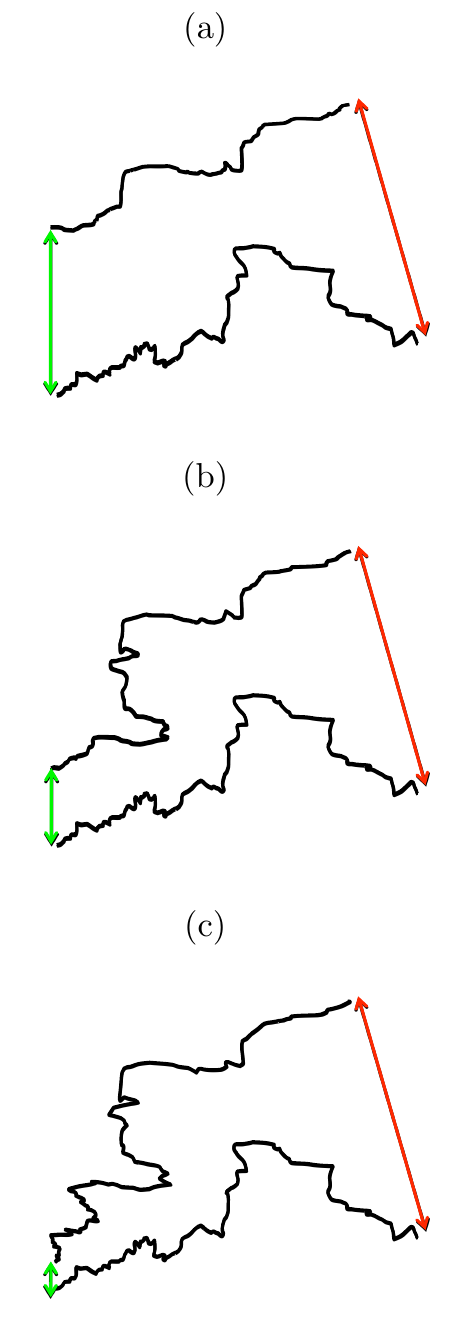}\\
\end{center}
\caption{{\it Illustration of spontaneous stochasticity.} 
Three pairs of Lagrangian trajectories are shown in panels (a),(b),(c), with initial separations 
at time $t_0$ between the two members of the pair indicated by \textcolor{green}{green} arrows and 
final separations at time $t$ indicated by \textcolor{red}{red} arrows. The initial separations are progressively 
decreased going from (a) to (b) to (c), but the final separations remain the same! Continuing 
in this manner, one obtains as a limit the situation
with more than one Lagrangian trajectory for the same initial point. }
\end{figure}

The breakdown of conventional flux-freezing due to ``roughness'' of the fields and spontaneous 
stochasticity of the Lagrangian flows means that magnetic line-topology is no longer preserved 
in time for the limit $\lambda\rightarrow 0.$ The infinite ensemble of ``virtual magnetic fields''
frozen-in to the stochastic flows all have exactly the same line-topology as the initial magnetic field. 
However, the average over the ensemble of  virtual field-lines that arrive to the same final point 
which gives the resultant physical magnetic field will, in general, change the topology. In particular, 
this permits fast topology change (magnetic reconnection) independent of the precise value of the 
resistivity. It is worth noting, however, that not all topological conservation laws are vitiated. Magnetic 
helicity conservation, for example, is particularly robust, even when velocity and magnetic fields are 
extremely ``rough''. It has been proved by \cite{Caflischetal97} [Theorem 4.2] that magnetic helicity 
is conserved in the limit as $\lambda\rightarrow 0$ even for such ``rough'' fields, as long as the scaling 
exponents of velocity and magnetic field 3rd-order structure functions remain positive. This 
result is consistent with very extreme singularities such as sparsely distributed tangential 
discontinuities (current and vortex sheets).  Thus, fast reconnection due to the mechanism 
of spontaneous stochasticity will generally conserve magnetic helicity.

One last important remark: everything that we have stated about Lagrangian 
particle trajectories applies just as well to the magnetic field-lines $\boxi(\bx,\sigma)$, 
defined by solving (at each fixed time $t$)
\be  \left\{ \begin{array}{l} 
                 {{d\boxi}\over{d\sigma}}(\bx,\sigma) =\bB(\boxi(\bx,\sigma),t) , \cr 
                 \boxi(\bx,0)= \bx.  \end{array} \right.  \lb{line-ODE} \ee
Here $\boxi(\bx,\sigma)$ is the field-line which passes through point $\bx$ 
(at time $t$). The parameter $\sigma$ is related to arc-length $s$ along the 
field-line by $ds=|\bB(\boxi(\bx,\sigma),t)|d\sigma.$ In GS95 theory, field lines 
which correspond to two nearby points $\bx,\bx'$ with $\br=\bx'-\bx$ will separate  
in the transverse direction as
\be |\boxi_\perp(\bx',s)-\boxi_\perp(\bx,s)|\sim (s^3/L_i)^{1/2}M_A^2 \ee
in the rms sense, for $s\gg s_0=O(v_A(r_\perp^2/\varepsilon)^{1/3})$ and for 
$r_\perp>r_\lambda=O\left((\lambda^3/\varepsilon)^{1/4}\right). $ See eq.(\ref{B-rich}). 
This is exactly the ``stochastic line-wandering'' which was invoked in the LV99 
theory of fast magnetic reconnection and, later, in the 
theory of thermal conduction in a turbulent MHD plasma (see \cite{NarayanMedvedev01} 
for $M_A=1$ and Lazarian (2006) for an arbitrary $M_A$). Just as for Lagrangian
trajectories, the initial separation $\br$ between field lines is ``forgotten'' after
proceeding a large enough distance along them. This is the reason that the 
aforementioned results for the thermal conductivity do not depend 
upon the electron Larmor radius $\rho_e.$ In fact, magnetic field-lines become 
stochastic in the limit $\lambda\rightarrow 0$ in precisely the same sense as 
do Lagrangian particle trajectories in the limit $\nu\rightarrow 0.$\footnote{In the limit of
large viscosities that we do not consider here the physics of field wandering is somewhat different
and requires a separate discussion (see Lazarian, Vishniac \& Cho 2004).}  Consider
a random bundle of field lines $\boxi(\bx+\widetilde{\br},s)$ for a stochastic 
displacement vector $\widetilde{\br}$ distributed over the ball $|\widetilde{\br}|<r_0$ 
of radius $r_0.$ By taking 
the limits first $\lambda\rightarrow 0$ and then $r_0\rightarrow 0,$ one obtains
an infinite ensemble of magnetic field-lines all passing through the same point
\bx!  

\subsection{The generalized Ohm's law} 

Our arguments for the essential stochasticity of  turbulent flux-freezing in 
sections 4.1-2 may have given the impression that the origin of the randomness is necessarily 
connected with Ohmic resistivity. If our conclusions really did depend upon the 
resistive MHD model (\ref{induct}) assumed there, then they could be criticized as 
physically unrealistic. Indeed, there are in actuality many non-ideal terms that appear 
in the magnetic equations of motion for an ionized plasma, which may be summarized 
in the generalized Ohm's law
\begin{eqnarray} 
&&{\bf E}=-\frac{1}{c}\bu\btimes\bB+\eta_\perp\bJ_\perp+\eta_\|\bJ_\| 
+\frac{\bJ\btimes \bB}{nec}\cr
&&   \,\,\,\,\,-\frac{\grad\bdot {\bf P}_e}{nec} 
+ \frac{m_e}{ne^2}\left(\frac{\partial\bJ}{\partial t} + \grad\bdot(\bu\bJ+\bJ\bu-\frac{1}{ne}\bJ\bJ)\right),
\lb{gen-ohm} \end{eqnarray}      
when quasineutrality and $m_e\ll m_i$ are assumed. See \cite{Vasyliunas75, 
PriestForbes00, Bhattacharjeeetal99}. The electric fields appearing on the righthand 
side are, respectively, the motional field, Ohmic fields associated to perpendicular and 
parallel resistivities, the Hall field, a contribution from the electron pressure tensor, and 
electron inertial contributions.  All of these terms 
have been invoked in various theories of fast magnetic reconnection.  However, we 
shall argue that in the turbulent environments that we consider none of these terms 
alters our fundamental conclusions. In the first place, the stochasticity of flux-freezing 
in turbulent plasmas is due not to resistivity (or to other terms in the generalized Ohm's 
law) but instead to the ``roughness'' of the MHD solutions in a long inertial range. 
In the second place, the precise microscopic plasma mechanism of ``line-breaking''
that acts at small scales (below the ion or electron gyroradii) is irrelevant to the inertial-range 
turbulence dynamics, which will be fundamentally the same for any such mechanism.

To demonstrate the first point, we show that flux-freezing is effectively stochastic in 
the turbulent  inertial-range even for the ideal induction equation
\be  \partial_t\bB =\grad\btimes(\bu\btimes\bB) \lb{ideal} \ee
with {\it all} non-ideal terms in (\ref{gen-ohm}) set to zero! Here we assume that $\bu,\bB$ are 
smooth at length-scales below some cut-off $<\ell_d$ but rough (turbulent) at larger scales. 
Notice that the above equation (\ref{ideal}) is realistic at leading-order for collisionless,  
magnetized plasmas at scales larger than the ion-gyroradius $\rho_i$ \citep{Kulsrud83},
while the smoothness scale $\ell_d$ set by field-perpendicular viscosity and resistivity is  
often of the same order or smaller than $\rho_i.$ Thus, our assumptions are quite realistic.
Because of the cut-off $\ell_d,$ flux-freezing in the standard sense must, in fact, be valid. 
How then can we claim that it becomes stochastic? To see this, consider the magnetic 
field observed at some finite space-resolution $\rho:$
\be \overline{\bB}_\rho(\bx,t)=\int d^3r \,\,G_\rho(\br) \bB(\bx+\br,t), \lb{coarse} \ee
where we have introduced a coarse-graining kernel $G_\rho$ to represent the smearing 
effect of the observation over a ball of radius $\rho$ around the space point $\bx$. We shall 
assume below that $\ell_d\ll \rho\ll L_i,$ the scale of the largest turbulent eddies. Thus, $\rho\gtrsim\rho_i$ 
satisfies these conditions. Applying the standard Lundquist formula for the frozen-in magnetic field, 
one obtains 
\be \overline{\bB}_\rho(\bx,t)=
\int 
\left.\frac{\bB(\ba,t_0)\bdot \grad_a \bx_{t,t_0}(\ba)}
{{\rm det}(\grad_a\bx_{t,t_0}(\ba))}
\right|_{\bx_{t,t_0}(\ba)=\bx+\br}G_\rho(\br). \,d^3r 
\lb{stoch-lund} \ee
The Lagrangian particle trajectories that appear in this formula start in the ball of radius $\rho$
around $\bx$ at time $t$ and then follow the flow velocity $\bu$ backward in time to $t_0,$
as illustrated in Fig.~3.  When $\rho$ lies in a GS95 turbulent inertial range, then the 
trajectories explosively separate to a perpendicular distance $\Delta x_\perp\sim 
(\varepsilon |t-t_0|^3)^{1/2},$ independent of $\rho$ at times $|t-t_0|\gg (\rho^2/\varepsilon)^{1/3}$.
\begin{figure}
\begin{center}
\includegraphics[width=250pt,height=175pt]{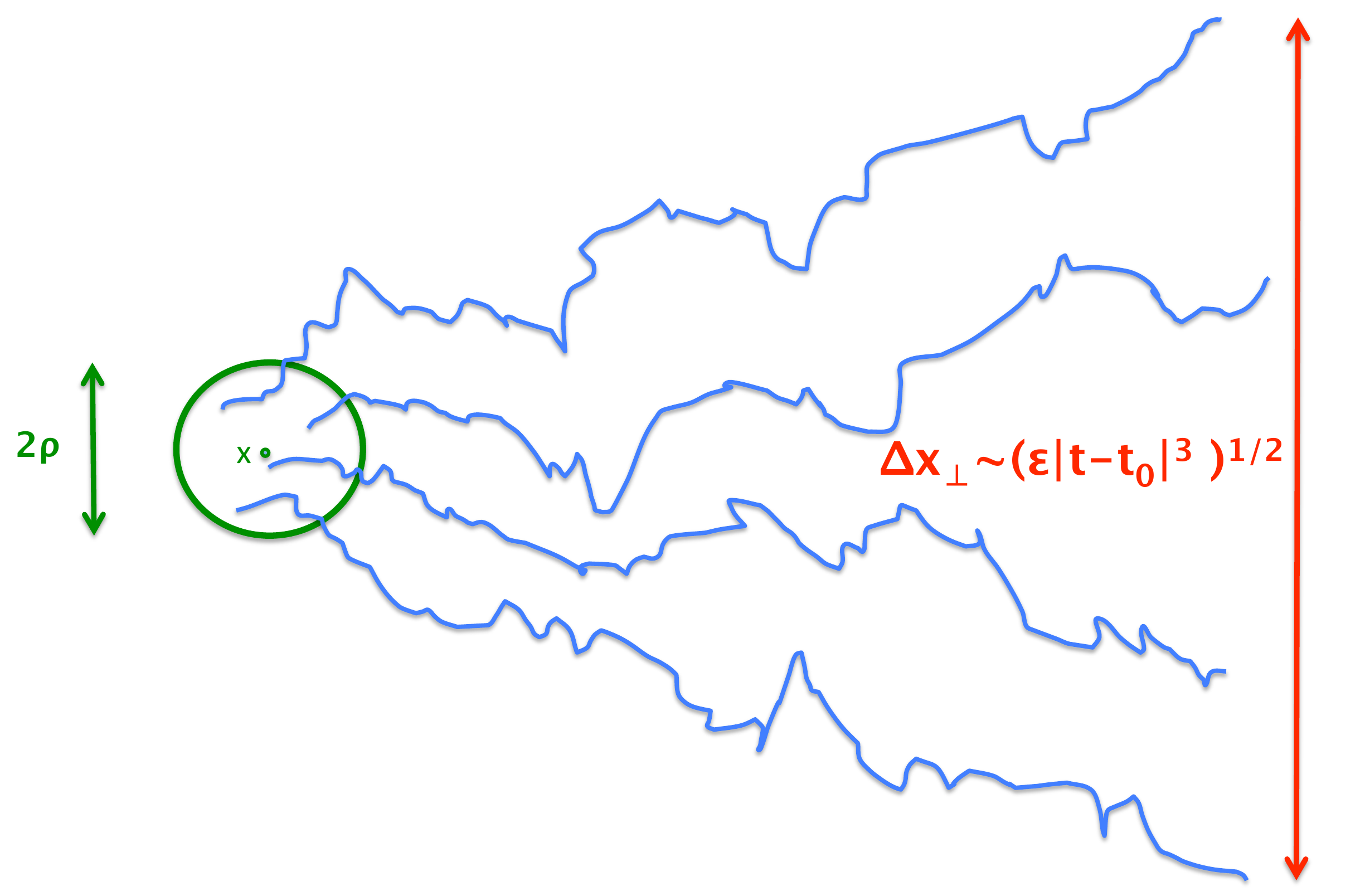}\\
\end{center}
\caption{Lagrangian trajectories that start in the ball of radius $\rho$ around space point $\bx$
at time $t$ move backward in time to $t_0,$ explosively separating to a field-perpendicular distance
$\Delta x_\perp\sim (\varepsilon |t-t_0|^3)^{1/2}$ which is independent of $\rho$ for $|t-t_0|\gg
(\rho^2/\varepsilon)^{1/3}.$} 
\end{figure}
The result is indistinguishable from the stochastic Lundquist formula (\ref{path-int})
which was derived in section 4.1 using the stochastic representation of Laplacian resistivity.  
In fact, in a formal mathematical limit taking first $\ell_d\rightarrow 0$, then $\rho\rightarrow 0,$ the 
Lagrangian trajectories in (\ref{stoch-lund}) remain stochastic and the two formulas coincide. 
This is a rigorous theorem for the Kazantsev-Kraichnan dynamo model, where it has been 
proved that the ensemble of stochastic Lagrangian trajectories as constructed above is 
precisely the same as that obtained for the $\lambda\rightarrow 0$ limit \citep{EvandenEijnden00}. 
Stochasticity of flux-freezing in not due intrinsically to resistivity or other microscopic plasma 
mechanisms that ``break'' field-lines but is, instead, a fundamental consequence of turbulent 
Richardson diffusion.

Higher-order terms in the generalized Ohm's law (\ref{gen-ohm}) that do not appear in the 
ideal equation (\ref{ideal}) will lead to melding and merging of field-lines at scales $<\rho_i.$ 
However, the above argument strongly suggests that these details of the microscopic plasma 
processes do not affect the dynamics at scales larger than $\rho_i.$ In some cases this 
can be shown more analytically by defining a suitable ``motion'' of field-lines consistent 
with the induction equation. For example, the formulation in section 4.1 based on addition 
of a Brownian motion to the Lagrangian particle dynamics, eq.(\ref{SDE}), can be carried 
over to certain instances of the generalized Ohm's law. See in particular \cite{Eyink09} for the 
Hall MHD equations. This approach is used in Appendix B to argue that neither the Hall effect 
nor Ohmic resistivity will have any significant influence on the inertial-range turbulence dynamics
at large enough scales. The Hall term, for example, does not affect the dynamics at scales 
much greater than $\delta_i=\rho_i/\sqrt{\beta_i},$ the ion skin depth. Unfortunately, it is 
difficult to extend this type of argument to all cases of the generalized Ohm's law, because 
it is not known how to define a ``motion'' of field-lines consistent with the induction equation 
for the general case.

On the other hand, there is a different argument which applies in general and leads to the same 
conclusion that flux-freezing must be intrinsically stochastic in turbulent plasmas. While the 
``motion'' of magnetic field-lines is a conventional and somewhat arbitrary concept, the motion 
of plasma is perfectly well-defined within the validity of an MHD description. Plasma fluid 
moves with the bulk velocity $\bu.$  Thus, field-lines may be tracked by ``tagging'' the lines 
with plasma fluid elements and then following these as Lagrangian fluid particles 
\citep{Newcomb58,Axford84}.  In the case of a smooth, laminar solution of the ideal MHD equations,  
this is unambiguous because of Alfv\'en's theorem: two plasma particles which start on a certain 
field-line must share a field-line for all times. One can then, by convention, consider 
this as the ``same'' field line as the initial one. This approach fails for a non-ideal Ohm's law, 
\be  {\bf E}+\frac{1}{c}\bu\btimes\bB=\bR, \ee
where $\bR$ represents all of the terms on the righthand side of (\ref{gen-ohm}) 
other than the motional term. Clearly, $\bR$ is just the electric field in the rest
frame of the plasma flowing with the bulk velocity $\bu.$  Emf due to these non-ideal terms 
leads to time-dependent magnetic flux in the rest frame, corresponding to a slippage of field 
lines. This vitiates the usual method to assign an identity to individual field-lines over time, 
because plasma elements shift their attachments to lines. Charged particles move along 
magnetic field-lines, but two plasma elements that start on one field-line will sit on 
distinct field-lines at later times.

\begin{figure}
\begin{center}
\includegraphics[width=240pt,height=360pt]{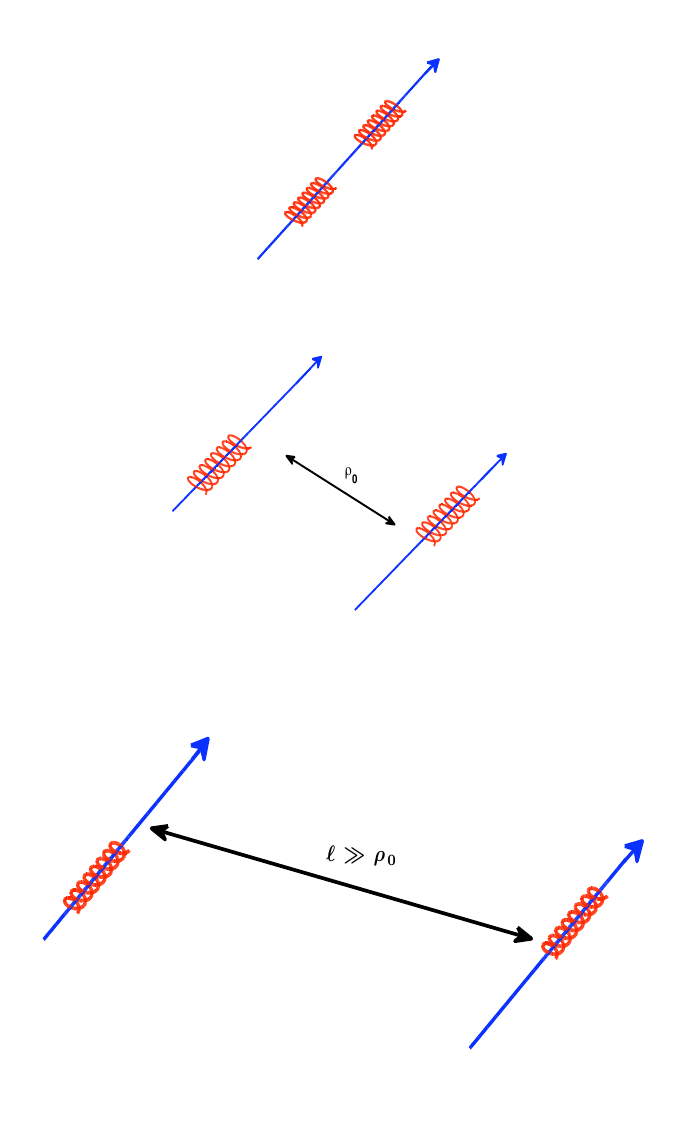}\\
\end{center}
\caption{{\it Advective acceleration of magnetic field-line separation.} {\it Top:} A pair 
of ions starting on the same field-line. {\it Middle:} Electric field in the frame of the plasma
changes the connections of the ions to field-lines separated by distance $\rho_0$.
{\it Bottom:} The ions are swept apart by fluid motions to a distance $\ell \gg \rho_0$
in a time $t_\ell$ that is independent of $\rho_0.$ The original field-line has effectively
``split''  into two separated by the macroscopic distance $\ell.$}
\end{figure}

Now consider a turbulent plasma where the non-ideal term is numerically ``small''
but the plasma has a turbulent inertial-range in which the velocity field $\bu$ is rough, with 
a power-law energy spectrum extending down to a smallest length-scale $\rho_0\approx \rho_i.$ 
The slight shifts in line-attachments are enormously amplified by explosive relative advection, 
as illustrated in Fig.~4. Consider a single magnetic field-line in this plasma and, along it,  two 
plasma fluid particles at  initial locations $\ba$ and $\ba'.$ Due to a combination of the non-ideal 
field $\bR$ and advection by sub-inertial-range eddies, the two plasma particles will end up on 
{\it distinct}  field-lines displaced a distance $|\bx(\ba',t)-\bx(\ba,t)|=\rho_0$ apart in a time $\tau_0$ 
which is  generally microscopically small compared with the eddy-turnover time $t_L.$ 
Because of (\ref{Richardson}), the two plasma elements will subsequently end up on field-lines displaced 
a macroscopic distance $(\varepsilon t^3)^{1/2}$ apart independent of $\rho_0$ in a time $t\gg  \tau_0$ 
\footnote{In the conventional view, for ideal plasmas with $\bR\equiv\bzed,$ the two elements would 
separate also to this distance but would remain on the same, highly-stretched field-line. This is
not the case with a nonideality $\bR\neq\bzed,$ however small.}. 
In effect, the original field-line has split apart into two lines separated by macroscopic distance
in a finite time $t.$ Of course, this same argument works if we take $\tilde{\ba}'=\ba+\tilde{\borho}$ for 
some small random displacement $\tilde{\borho}$ along the initial field-line with $|\tilde{\borho}|<\rho_0$ . 
We have already argued in section 4.2 that the random ensemble of Lagrangian trajectories 
$\bx(\ba+\tilde{\borho},t)$ will coincide for $t\gg \tau_0$ with the random ensemble $\tilde{\bx}(\ba,t)$ 
obtained from the perturbation by Brownian motion in eq.(19). The initial field-line will be stochastically
 ``frozen-in'' to this ensemble of random flows.

Turbulent advection accelerates the separation of field-lines so effectively that the microscopic 
plasma process of line-slippage, whatever its origin,  is rapidly ``forgotten''. The non-ideal 
terms in the generalized Ohm's law become insignificant in further separating lines which 
have exceeded the distance $\rho_0$ apart. 
When the plasma flow velocity $\bu$ is ``rough'', it is the motional 
term in the generalized Ohm's law which, after a very short time, dominates all of the other  terms 
in the separation of field-lines.      




\section{Large-Scale Magnetic Reconnection}

With the theory developed above, we now turn to our main topic of magnetic reconnection.  To be  
specific, we consider the steady-state reconnection of a pair of large-scale magnetic flux tubes which 
collide along a section of length $L_x.$ We assume incompressible plasma flow and anti-parallel 
magnetic flux-tubes of equal field-strength $|B_x|$. The situation could be generalized to allow 
for compressible plasma and asymmetric field-strengths \citep{CassakShay07}, flux-tubes 
intersecting at an angle or with a shared component of magnetic field (guide field) $B_z$  
\citep{Lintonetal01}, twisted flux-tubes with magnetic helicity \citep{ZweibelRhoads95}, etc. 
All of these are relevant for applications,  but we keep the set-up simple in order to make the new 
ideas clear.  The large-scale geometry we consider is thus precisely the two-dimensional configuration 
in classical Sweet-Parker theory. However, we are working in three-space dimensions and thus 
small-scale turbulence, in particular, will be fully 3D. 

\begin{figure}[!t]
 \begin{center}
\includegraphics[width=1.0 \columnwidth]{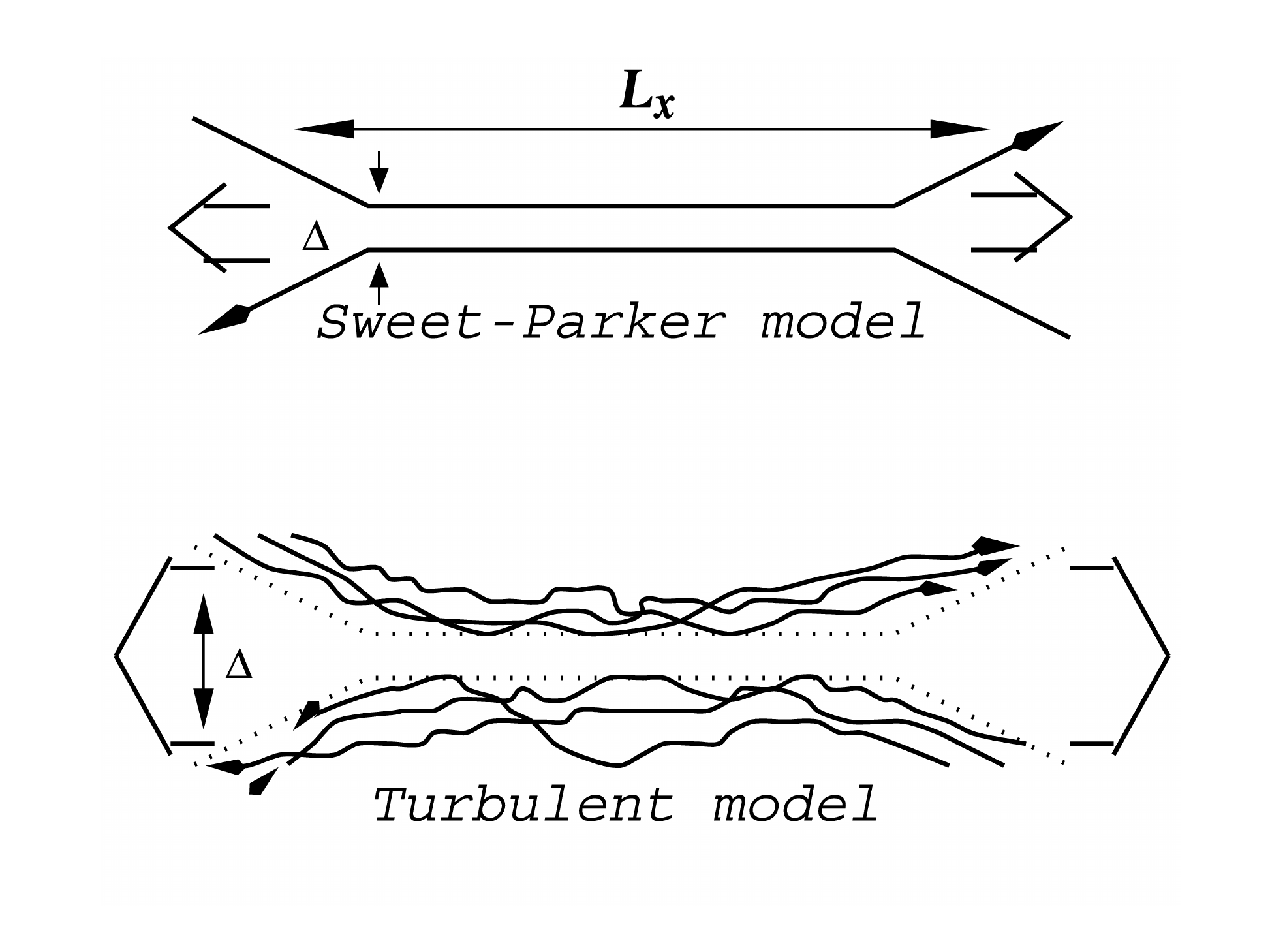}
\caption{{\it Upper plot}:
Sweet-Parker model of reconnection. The outflow
is limited by a thin slot $\Delta$, which is determined by Ohmic
diffusivity. The other scale is an astrophysical scale $L_x\gg \Delta$.
{\it Lower plot}: Reconnection of weakly stochastic magnetic field according to
LV99. The model that accounts for the stochasticity
of magnetic field lines. The outflow is limited by the wandering of
magnetic field lines, which depends on field line stochasticity.
From Lazarian et al. 2004.}
\label{fig_rec}
 \end{center}
\end{figure}

\subsection{A Rederivation of LV99 by Stochastic Flux-Freezing} \label{rederive}

The constraints on steady-state reconnection from the MHD balance equations are well-known.
For a reconnection layer of length $L_x$ and thickness $\Delta,$ the constraint of energy 
conservation is
\be {{1}\over{2}} \rho v_{out}^2v_{out}\Delta = {{1}\over{8\pi}}B^2_xv_{rec}L_x, \lb{energy} \ee
where $v_{rec}$ is the in-flow plasma speed or reconnection velocity and $v_{out}$ is the out-flow
velocity from the sides of the reconnection layer. Here it is assumed that the energy in the out-flow
jets is primarily kinetic, since the $B_y$ component of the outgoing field-lines at the outer edges
of the reconnection layer is still relatively small. The critical constraint of mass conservation is 
\be  v_{out} \Delta = v_{rec}L_x.  \lb{mass} \ee
Combining (\ref{energy}) and (\ref{mass}) yields
\be v_{out} = {{B_x}\over{\sqrt{4\pi \rho}}}=v_A, \lb{vout} \ee
so that the out-flow velocity is at the Alfv\'en speed of the in-coming magnetic fields.  The momentum 
equation enforces pressure balance along and across the reconnection layer and does not yield 
a new constraint in our setting. From (\ref{mass}) and (\ref{vout}) we see that 
\be   v_{rec}= v_A {{\Delta}\over{L_x}}. \lb{vrec} \ee
Our discussion to this point has been completely general and no particular mechanism of line-slippage
has been assumed. To proceed further, a specific reconnection mechanism must be invoked which 
permits one to estimate the thickness $\Delta$ of the layer. 

The classical Sweet-Parker model assumes resistive reconnection in a laminar MHD solution. 
There are then several ways to estimate the thickness $\Delta,$ which are all consistent.
The most common aproach is to use the steady-state Faraday's law, which implies the 
constancy in the vertical $y$-direction of the reconnection electric field $E_z$. Outside the 
reconnection layer $E_z=v_{rec}B_x/c,$ while inside $E_z=\eta J_z=\eta c B_x/4\pi \Delta.$ Equating 
these gives
\be v_{rec} = {{\lambda}\over{\Delta}}, \lb{vdiff} \ee
where $\lambda=\eta c^2/4\pi$ is the magnetic diffusivity. Combining (\ref{vrec}) and (\ref{vdiff}) 
gives the well-known Sweet-Parker results that 
\be   \Delta= L_x/\sqrt{S}, \,\,\,\,v_{rec} = v_A/\sqrt{S}, \lb{SP} \ee
where $S=v_A L_x/\lambda$ is the Lundquist number. Another way to obtain this result is 
based on standard flux-freezing ideas for a laminar plasma flow (Kulsrud, 2005). The mean-square 
vertical distance that a magnetic field-line can diffuse by resistivity in time $t$ is 
\be \langle y^2(t)\rangle \sim \lambda t. \lb{diff-dist} \ee
Each reconnected field-line is the result of resistive gluing of field-lines that diffuse vertically across 
the reconnection layer.  At the same time, the field lines are advected out of the sides of the 
reconnection layer of length $L_x$ at a velocity of order $v_A.$ Thus, the time that the lines can 
spend in the resistive layer is the Alfv\'en crossing time $t_A=L_x/v_A.$ Thus, field lines can only 
be merged that are separated by a distance 
\be \Delta = \sqrt{\langle y^2(t_A)\rangle} \sim \sqrt{\lambda t_A} = L_x/\sqrt{S}. \ee
We thus recover (\ref{SP}). This is an important dynamical consistency check on the Sweet-Parker model
of laminar resistive reconnection. 

For large Lundquist numbers, the Sweet-Parker reconnection speed (\ref{SP}) is too small to 
account for observed reconnection rates in most astrophysical settings. However, 
the Sweet-Parker laminar MHD solution is not appropriate for turbulent plasmas at large $S$.
For simplicity, we shall consider reconnection in the presence of a steady-state
background turbulence with an integral length-scale $L_i$ and rms velocity $u_L.$
We shall consider sub- and trans-Alfv\'enic turbulence with $u_L\leq v_A$ or $M_A\leq 1,$
under the assumption that the large-scale magnetic flux tubes are strong and well-defined 
in the presence of turbulence. In some situations the reconnection may be initially laminar and 
then turbulence initiated by instabilities triggered by small fluctuations from the environment. 
Initial fast reconnection by Hall effect  \citep{Shayetal98, Wangetal00,Birnetal01,Drake01}
or other collisionless processes may itself provide the energy release to generate small-scale 
MHD turbulence. We shall consider only the final stage of steady-state reconnection in the 
presence of turbulence with the given characteristic length and velocity scales.  

The key issue is to estimate $\Delta.$ We shall give two different methods of determining 
the width of the turbulent reconnection layer.

Our first estimate is based on stochastic flux-freezing/SFF (section 4) and 2-particle turbulent 
diffusivity (section 3.2). We have seen that GS95 theory implies that the turbulent 
separation of pairs of  ``virtual'' field-lines backward in time, in the $y$-direction perpendicular
to the mean magnetic field, is given by eqs.(\ref{strong-2part-law}) and (\ref{weak-2part-law}), or  
\be \langle y^2(t)\rangle \sim 
                     \left\{ \begin{array}{ll} 
                        \varepsilon t^3, & t_\eta\ll  t<t_c \cr 
                       u_LL_i M_A^3 t, & t>t_c, \cr
                       \end{array} \right. \lb{2part} \ee
for the strong and weak MHD turbulence regimes, respectively, with $t_c=L_i/v_A.$ 
Reconnected field-lines that emerge from the turbulent reconnection layer are 
constituted from field-lines obtained by following ``virtual'' field lines backward 
in time over the Alfv\'en crossing period $t_A=L_x/v_A.$ We thus obtain $\Delta=
\sqrt{\langle y^2(t_A)\rangle}$ as the width of the turbulent reconnection layer. 
There are two cases to consider, depending upon whether $t_A<t_c$ or $t_A>t_c.$
Using $\varepsilon \sim u_L^4/v_AL_i,$ we rewrite (\ref{2part}) as
 \be        \langle y^2(t)\rangle \sim \cases{ (L_x^3/L_i) M_A^4 (t/t_A)^3 &if $t<t_c$ \cr     
                                                                       L_iL_x M_A^4 (t/t_A),  &if $t>t_c$. \cr} \ee
Since $t_c \lessgtr t_A$ exactly when $L_i\lessgtr L_x,$ respectively, 
we get finally that
\be \Delta = L_x M_A^2 \min\left\{\left({{L_x}\over{L_i}}\right)^{1/2}, 
       \left({{L_i}\over{L_x}}\right)^{1/2}\right\}. \lb{Delta-LV99} \ee
This is precisely the result of \cite{LazarianVishniac99}. The present derivation avoids 
many complications in their original argument due to the fact that ``virtual'' field lines pass directly 
through each other.  LV99 had to consider the limit to the reconnection rate set by the speed with 
which reconnected magnetic field elements can pass through each other on the way out of the 
reconnection zone, employing a plausible but heuristic self-consistency argument.  This problem 
is avoided completely and rigorously by using SFF (which is mathematically equivalent to the
resistive MHD equations).

The thickness $\Delta$ of the turbulent reconnection layer was originally estimated in LV99
by a geometric argument, based upon the spontaneous stochasticity of the magnetic field-lines
themselves. The plasma in the reconnection zone can only be expelled along the lines of the 
(real, physical) magnetic field. As we have seen in section 3.1, the lateral wandering as one 
moves a distance $s$ along a field line is given by eqs.(\ref{B-rich}), (\ref{B-diff}), or 
\be       \langle y^2(s)\rangle \sim \cases{ (s^3/L_i)M_A^4  &if $s<L_i$ \cr     
                                                                       L_i s M_A^4,  &if $s>L_i$. \cr} \lb{B-wand} \ee
Since the plasma must move a distance $s\sim L_x$ along the field-line in order to 
escape the reconnection layer, we can estimate the thickness of the layer by $\Delta\sim 
\sqrt{\langle y^2(L_x)\rangle}.$ Using (\ref{B-wand}) we immediately recover (\ref{Delta-LV99}). 
The agreement of these two arguments demonstrates some basic dynamical consistency 
of the LV99 theory. 

We obtain from both arguments the LV99 estimate for the reconnection speed
\be  v_{rec} = v_A M_A^2 \min\left\{\left({{L_x}\over{L_i}}\right)^{1/2}, 
       \left({{L_i}\over{L_x}}\right)^{1/2}\right\}. \lb{vr-LV99} \ee
Note that $v_{trans}=v_A M_A^2$ is the amplitude of velocity fluctuations at the transition between 
weak and strong MHD regimes.  This can be a sizable fraction of the Alfv\'en velocity. Furthermore, 
{\it (\ref{vr-LV99}) is ``fast'' in the strictest sense of being independent of the microscopic resistivity.}  
The physical mechanism of this independence is spontaneous stochasticity, both of the Lagrangian 
particle trajectories and of the magnetic field lines. In both cases, initial separations by a microscopic 
resistive length are amplified to macroscopic distances as one traverses along the wandering curves 
by an amount (of time or of path-length, respectively) which is independent of the microscopic 
electrical resistivity.

The above discussion has approximated the reconnection layer as a zone of homogeneous
MHD turbulence, with a constant Alfv\'enic Mach number $M_A=u_L/v_A$ set by the reconnecting
$x$-component of the upstream magnetic field, $v_A=B_x/\sqrt{4\pi\rho}.$ In reality, there 
will be a ``magnetic shear layer'' with a continuous change of the mean magnetic field 
$\OL{B}_x(y)$ across the layer. Likewise, there are nontrivial profiles $\OL{u}_y(x,y)$ and 
$\OL{u}_x(x,y)$ of both incoming and outgoing velocities, which are known to have nontrivial 
effects on 2-particle dispersion \citep{ShenYeung97,Pope02}. An interesting 
refinement of the original  LV99 theory would be to consider the effect of variation of the mean 
velocity and magnetic fields across the reconnection layer. 
One plausible effect of magnetic shear will be to increase $M_A$ within the 
layer, permitting greater lateral wandering of field-lines and increasing the thickness
$\Delta$. Less easy to guess are the effects of the change in orientations of magnetic fields 
for flux-tubes reconnecting at an angle of less than $180^\circ$ \citep{Lintonetal01}.
The simulation results of \cite{Kowaletal09} on tubes crossed at various angles 
give support to the idea that the effects are minimal, but further theoretical and 
numerical study of this would be desirable. Another effect which 
has been neglected in our derivation above is the turbulent energy dissipation that exists 
in the reconnection layer. This can be estimated as $\rho\varepsilon \Delta L_x$ with 
$\varepsilon=u_L^4/v_AL_i.$ If this is added into the energy balance (\ref{energy}) 
one obtains a result for the outflow velocity 
\be v_{out}^2=v_A^2\left(1-2M_A^4\frac{L_x}{L_i}\right) \lb{outflow2} \ee
which can be substantially reduced from the Alfv\'en speed. By the mass conservation 
constraint (\ref{mass}), there is a similar reduction in the reconnection speed. However,
these effects are ignorable under the condition $M_A^4\ll L_i/L_x$ or, equivalently, 
$\Delta\ll L_i.$ This is generally well-satisfied. Other modifications and refinements of the 
LV99 model can be envisaged and are worth pursuing (see the opening paragraph of this 
section and also section 6.4).

It is important to emphasize that the original LV99 argument made no essential use of 
averaging over turbulent ensembles. The stochastic line-wandering which is 
the essence of their argument holds in every realization of the flow, at each instant of time. 
The ``spontaneous stochasticity" of field lines is not a statistical result in the usual sense of 
turbulence theory and does not arise from ensemble-averaging. The only use of ensembles 
in LV99 is to get a measure of the ``typical'' wandering distance $\Delta$ of the field lines (in 
an rms sense). If one looks at different ensemble members of the turbulent flow, or at different 
single-time snapshots of the steady reconnection state, then $\Delta$ will fluctuate. Thus, the 
reconnection rate will also fluctuate a considerable amount over ensemble members or over 
time.  E.g. see Figs.~12-14 in \cite{Kowaletal09}. But it will be  ``fast''  in each realization and 
at each instant of time, because the mass outflow constraint is lifted by the large wandering of field 
lines. 

Exactly the same remarks apply to the rederivation of LV99 using stochastic flux-freezing (SFF)
and Richardson diffusion. SFF is also not ``stochastic'' in the usual sense of turbulence theory. 
The new derivation uses ensemble-averaging just to get an estimate of the ``typical'' wandering 
of stochastic Lagrangian trajectories in an rms sense. Numerical studies of Richardson's $t^3$ law 
in hydrodynamic turbulence show that it is a very robust property. At most points in a turbulent flow, 
a ``cloud'' of stochastic particles spreads with a power-law very close to $t^3$. There is some 
intermittency  effect, with certain rarer, high-intensity points showing a much faster dispersion or 
low-intensity a slower dispersion, but it is not a huge effect.  Only the higher-order moments of the 
relative separation distance are strongly affected \citep{BoffettaSokolov02}.  

The LV99 theory, therefore, does not involve ``turbulent resistivity'' or ``turbulent 
magnetic diffusivity'' as this is usually understood. This is ordinarily meant to be an enhanced diffusivity 
experienced by the ensemble-averaged magnetic field $\langle\bB\rangle$. However, it is only if one 
assumes some scale-separation between the mean and the fluctuations that the effect of the fluctuations 
can be legitimately described as an enhanced diffusivity \citep{Moffatt83}. In realistic turbulent flows, 
with no scale-separation, this phenomenological description as an effective diffusivity can be 
wildly inaccurate. LV99 makes no appeal to such concepts and, indeed, never considers 
the ensemble-average field $\langle\bB\rangle$ at all. Neither does Nature. The Sun
does not average over an ensemble of coronal loops in order to get fast reconnection! 

\subsection{Small-Scale Reconnection and MHD Turbulence}

The focus of this paper has been on the reconnection of large-scale, oppositely-directed
magnetic flux-tubes in a turbulent plasma, since it is this phenomenon which has the most 
direct interest for astrophysics. On the other hand, a similar process of reconnection of local 
magnetic field-lines must take place at small-scales in homogeneous MHD turbulence, 
with or without a background magnetic field. It has been argued that such small-scale 
magnetic reconnection is an essential feature of MHD  turbulence \citep{MatthaeusLamkin86, 
LazarianVishniac99}. A substantial literature has since arisen on the relation between MHD turbulence 
and magnetic reconnection (e.g. see \cite{Lazarian05,EyinkAluie06,Servidioetal10,
MininniPouquet09}). We cannot review this entire subject here, but it is 
important to discuss briefly the implications of  ``spontaneous stochasticity'' for the problem.

It was observed already in LV99 that nearly parallel field-lines in adjacent turbulent 
eddies must frequently intersect and reconnect according to GS95 theory. The local
geometry is similar to that in large-scale reconnection ``writ small'',  with the local mean
magnetic field acting as a guide-field and the field-lines with anti-parallel transverse
components undergoing reconnection. The crossing-points and knots in the local
magnetic field-lines, if unresolved, would create topological obstructions to the plasma 
flow which would render hydrodynamic motion impossible. LV99 argued instead that 
line-intersections at perpendicular length-scale $\ell_\perp$ are resolved by magnetic
reconnection in the eddy-turnover time $\ell_\perp/\delta u(\ell),$ independent of 
the resistivity. Fast magnetic reconnection is thus crucial for the very existence of MHD 
turbulence of the kind supposed in GS95. 

The stochastic  ``line-wandering'' implied by GS95 theory (section 3.1) suggests that 
a very complex line-topology must exist in turbulent flows with magnetic fields that are
non-smooth on inertial-range scales. This complexity is made particularly evident by the 
``spontaneous stochasticity'' phenomenon (section 4.2), according to which there are 
infinitely-many field-lines passing though each point in the limit of zero resistivity! This result 
suggests that reconnection sites are distributed densely in space for MHD turbulence at 
very high conductivity. Turbulent plasma flows with rough magnetic and velocity fields 
have a very strange and intricate geometry when compared with the smooth, laminar 
flows that have often been the focus of theoretical and empirical studies of reconnection. 
Another indication of this complexity is the fractal distribution of clustered magnetic nulls 
for a magnetic field with a turbulent, power-law energy spectrum \citep{Albright99}.  
This latter fact is very intimately connected with the stochastic ``line-wandering'' 
phenomenon deduced in LV99, since the magnetic nulls form natural sites of separation
of  adjacent magnetic field-lines. See \cite{DavilaVassilicos03, GotoVassilicos04} 
for a discussion of the closely related hydrodynamic phenomenon associated with Lagrangian 
particle trajectories and stagnation points of the turbulent velocity field.    

The above facts strongly suggest that magnetic reconnection must be a process 
that is active always and  everywhere in homogeneous MHD turbulence. As one
measure of the local reconnection rate one may take the estimates 
(\ref{strong-2part-law}),(\ref{weak-2part-law}) of the lateral diffusion of magnetic 
field-lines (for strong and weak turbulence, resp.) For example, in the strong 
turbulence regime, field lines diffuse in an eddy-turnover-time $\tau_\ell=\varepsilon^{-1/3}
\ell_\perp^{2/3}$ at scale $\ell_\perp$ by a mean-square perpendicular distance $\epsilon 
\tau_\ell^3=\ell_\perp^2.$ Thus, reconnection rates are exactly large enough to resolve 
knots of field-lines at length-scale $\ell_\perp$ in the natural evolution time at that scale, as 
argued in LV99. Note that $d\ell_\perp/dt\simeq \delta u(\ell)$ and thus the reconnection 
speed of pairs of field-lines separated by inertial-scale separations $\ell_\perp$ is equal to 
the relative velocity $\delta u(\ell)$ between the two lines. 

Another way to quantify small-scale reconnection in MHD turbulence is by 
violations of Alfv\'en's Theorem at  length-scale $\ell.$ This was the approach adopted 
by \cite{EyinkAluie06}, who studied the MHD induction equation in a coarse-grained description 
at the (isotropic) length-scale $\ell$\footnote{The field-perpendicular variations therefore include  
contributions from eddies with $\ell_\perp$ down to $\ell$ but the field-parallel variations get  
contributions from somewhat smaller eddies with $\ell_\perp$ down to $M_A^2\ell^{3/2}/L_i^{1/2}.$}:
\be  \partial_t\OL{\bB}_\ell=\grad\btimes[\OL{\bu}_\ell\btimes\OL{\bB}_\ell+c\,\boepsilon_\ell
    -\lambda\grad\btimes\OL{\bB}_\ell]. \ee
Simple estimates show that the direct resistive contribution proportional to $\lambda$ is negligible 
at high magnetic Reynolds numbers whenever the magnetic energy stays finite in that limit. 
See \cite{Eyink05}. It is the turbulent subscale EMF $\boepsilon_\ell=\OL{(\bu\btimes\bB)}_\ell-\OL{\bu}_\ell
\btimes\OL{\bB}_\ell$ which provides the necessary electric field for reconnection at the scale $\ell.$ 
The magnetic flux through a material loop $\OL{C}_\ell(t)$ advected by the coarse-grained velocity 
$\OL{\bu}_\ell$ satisfies
\be  \frac{d}{dt}\oint_{\OL{C}_\ell(t)} \OL{\bA}_\ell\bdot d\bx = c
      \oint_{\OL{C}_\ell(t)} \OL{\boepsilon}_\ell\bdot d\bx.  \ee
\cite{EyinkAluie06} established necessary conditions for a non-vanishing line-voltage
on the righthand side of the above equation, independent of $\ell.$ These conditions are:
either (i) a non-rectifiable (fractal) loop $\OL{C}_\ell(t),$ or (ii) a blow-up of $\bu$ or $\bB$
along the loop, or (iii) tangential discontinuities of both $\bu$ and $\bB$ intersecting the 
loop $\OL{C}_\ell(t)$ over a finite length. One or more of these conditions is plausibly 
satisfied in turbulent flow, implying ubiquitous reconnection. 
An important remark is that MHD turbulence is dominated by scale-local interactions 
under very general assumptions. 
See \cite{Eyink05, AluieEyink10}. In particular,
the turbulent EMF $\boepsilon_\ell$ is dominated by contributions from velocity and 
magnetic field modes around the scale $\ell.$ It is the EMF from these modes which is 
responsible for reconnection of the lines of the coarse-grained magnetic fields at length-scale 
$\ell$ and microscopic plasma processes at smaller scales play no direct role. 

\subsection{2D MHD Turbulence and Reconnection}

While astrophysical reality is three-dimensional (3D), there is a scholarly and intellectual value
to considering MHD turbulence in two-dimensional (2D) space. Comparison of toy models with 
reality is often a good way to test and sharpen one's understanding. Numerical simulations can 
also be better resolved in 2D than in 3D, permitting study of higher kinetic and magnetic 
Reynolds-number flows. Furthermore, the conventional view is that MHD turbulence in 2D and in 3D 
are very similar, much more so than for hydrodynamics of neutral fluids. 3D MHD possesses 
the three ideal, quadratic invariants of total energy, cross-helicity and magnetic helicity and 2D 
MHD has a nearly identical set of ideal invariants, with magnetic helicity replaced by the mean-square 
magnetic potential. Absolute equilibrium spectra  in 3D suggest that energy and cross-helicity 
cascade to high-wavenumbers and magnetic helicity to low-wavenumbers \citep{Frischetal75},
while the same argument in 2D leads to an identical conclusion, with an inverse cascade of 
magnetic potential replacing that of magnetic helicity \citep{FyfeMontgomery76}. Numerical 
simulations have corroborated these predictions, with the most recent 2D MHD simulations giving 
energy spectra close to $k^{-5/3}$ or $k^{-3/2}$ in the forward cascade range \citep{Gomezetal99, 
BiskampSchwarz01, Ngetal03}. On the other hand, from the point of view of GS95 
theory, 2D MHD turbulence appears as a very degenerate case. In particular, shear-Alfv\'en 
waves that play the dominant role in 3D MHD turbulence according to GS95 are entirely lacking 
in 2D, where only pseudo-Alfv\'en wave modes exist. This fact suggests that the character of MHD 
turbulence in 2D must be quite different than in 3D and certainly not well-described 
by GS95.

Because of the numerical advantages, many turbulent reconnection studies have also been 
performed in 2D, beginning with the important work of \cite{MatthaeusLamkin85, MatthaeusLamkin86}. 
The numerical evidence is rather controversial, however. \cite{MatthaeusLamkin85} 
reported the reconnection of magnetic islands at rates nearly independent of 
resistivity in the early stages of their simulations. \cite{Servidioetal10} have more recently 
made a similar study of Ohmic electric fields at X points in homogeneous, decaying 2D MHD 
turbulence. This is really a case of small-scale magnetic reconnection, as considered in our
previous section 5.3, and not directly relevant to the issue of reconnection of large-scale 
flux tubes. 
Two other numerical studies have recently been made of large-scale 
magnetic reconnection in 2D, by \cite{Loureiroetal09} and \cite{Kulpa-Dubeletal10}, 
which reach different conclusions. On the one hand, \cite{Loureiroetal09} 
had a better resolution but used periodic boundary conditions, which strongly 
constrain the ability to do averaging of the reconnection rate and the attainment of the steady 
state for reconnection. They inferred from their data that the 2D turbulent reconnection rate 
may be independent of resistivity. On the other hand, \cite{Kulpa-Dubeletal10} used smaller 
data cubes but longer averaging, which is enabled by their outflow boundary conditions. They 
concluded that the reconnection does depend on resistivity and therefore is slow. The raw data
of the two groups are actually rather similar and higher Lundquist-number simulations are 
probably necessary to resolve the matter.

What does theory say? In particular, can LV99 or some LV99-like theory apply in 
2D\footnote{In a completely different class are recent approaches which 
attribute fast reconnection in 2D MHD to formation and merging of plasmoids {\it without} 
the presence of conventional turbulence in the current sheet
\citep{Samtaneyetal09,Cassaketal09,Bhattacharjeeetal09,HuangBhattacharjee10}.
These will be discussed later in section 6.3.}? At first 
thought, one might doubt that LV99 is relevant at all since its central element---stochastic
line-wandering---seems ruled out in 2D on the simple grounds that magnetic field-lines may 
not cross. However,  closer consideration shows that this argument is fallacious. Magnetic
field-lines {\it can} cross in 2D, at very special points, the magnetic nulls. Furthermore, in a 
rough magnetic field, the magnetic nulls and null-null lines  will form a very complex, fractal 
web \citep{Albright99} which may permit quite unconstrained wandering of field-lines. Indeed,
the phenomenon of spontaneous stochasticity of field-lines ought to occur in 2D as well as 
in 3D, if the magnetic field is rough. There are rigorous examples of this type in 2D, where 
integral curves of a non-smooth vector field are non-unique at every point in space (e.g.
see section II.5 and Fig.2 in \cite{Hartman02}.) From the Lagrangian point of view of Richardson 
diffusion and stochastic flux-freezing, there is also little difference between 2D and 3D. 
The Richardson $t^3$-law has been observed, for example, in the inverse-energy cascade
range of 2D hydrodynamic turbulence \citep{CelaniVergassola01, GotoVassilicos04} 
and its analogue could be expected in 2D MHD turbulence. Indeed, if velocity fields are sufficiently 
rough in 2D MHD turbulence, then spontaneous stochasticity of Lagrangian trajectories ought 
to hold in 2D just as in 3D, and an LV99-like theory ought to apply. 

The above arguments are, however, quite delicate. As a counterexample, we may consider
weak MHD turbulence in 2D. The most important difference from 3D is that only pseudo-Alfv\'en wave 
modes exist in 2D. Because the weak turbulence cascade preserves $k_\|=1/L_i$ while increasing 
$k_\perp,$ the total wavevector is nearly perpendicular and the pseudo-Alfv\'en polarization 
vector nearly parallel to the magnetic field. Our estimate from section 3.2 for the 2-particle 
diffusivity of weak MHD turbulence is thus reduced in the field-perpendicular direction by a factor
of $(\ell_\perp/\ell_\|)^2.$ The result is that 
\be  D_\perp(\ell) \sim \left(\frac{\ell_\perp}{\ell_\|}\right)^2 \frac{\varepsilon}{\omega_A^2}
     = \ell_\perp^2 \frac{\varepsilon}{v_A^2}. \ee
 Solving for the perpendicular particle separation from $d\ell_\perp^2/dt=D_\perp(\ell)$ 
 gives an exponential growth proportional to the initial separation
 \be  \ell^2_\perp(t) \simeq \ell^2_\perp(0)\exp( {\rm (const.)}\varepsilon t/v_A^2), \ee
 i.e., no spontaneous stochasticity! There will be no 2D fast reconnection if such a weak 
 turbulence cascade persists down to dissipation scales.  Naively, one would 
 expect this to be true in 2D from a simple time-scale argument. The interaction time of 
 a pair of pseudo-Alfv\'en waves with parallel length-scale $\ell_\|$ and fluctuation amplitude
 $\delta u_\ell$  is $\ell_\|/\delta u_\ell$, which will in general be longer than the Alfv\'en 
 wave-period $\ell_\|/v_A.$ This argument suggests that there is no strong MHD turbulence
 at all in 2D and thus no fast reconnection.  
 
 The only way to avoid this conclusion that we 
can see is that the 2D turbulence develops very singular structures or ``tangential discontinuities''
 with $\delta u_\ell\simeq v_A,$ independent of $\ell$ over the inertial range.  Note that the 
 rigorous result of \cite{EyinkAluie06} for the 2D case requires such discontinuities and/or 
 blowup in the magnitudes of the velocity or magnetic fields.  There seems also to 
 be support for this scenario from numerical simulations of 2D MHD turbulence, which show
 very intermittent current and vortex sheets in a background of weaker turbulence. It is 
 possible that this is one of those instances where the weak-turbulence cascade is  unstable
 to the formation of localized, singular structures (cf. \cite{Majdaetal97, Rumpfetal09}). 
 The crucial issue seems to be whether these intermittent structures occur sufficiently densely 
 in spacetime.  We shall not attempt to resolve this issue here, but only conclude that the 
 existence of strong MHD turbulence and fast turbulent reconnection 
 in 2D are quite open questions. 


\subsection{Alternative Ideas on the Role of Turbulence in Magnetic Reconnection}

As we mentioned in the introduction, various ideas how turbulence can increase the 
reconnection rate were discussed as far back as 40 years ago. However, these 
ideas fell short of solving the problem. For instance, some papers have concentrated 
on the effects that turbulence induces on the microphysical level. In particular, 
\cite{Speiser70} showed that in collisionless plasmas the electron collision time 
should be replaced with the electron retention time in the current sheet.  Also 
\cite{Jacobson84} proposed that the current diffusivity should be modified to include 
the diffusion of electrons across the mean field due to small scale stochasticity. All 
these effects can only marginally change the reconnection rates.

The closest predecessor to LV99 was the important work of \cite{MatthaeusLamkin85,MatthaeusLamkin86}
(ML85,86).  Those authors studied 2D magnetic reconnection in the presence of external turbulence,
both theoretically and numerically. They emphasized the very close analogies between 
the magnetic reconnection layer at high Lundquist numbers and homogeneous MHD 
turbulence. They also pointed out various turbulence mechanisms that would enhance reconnection 
rates, including multiple X-points as reconnection sites and motional EMF of magnetic bubbles
advecting out of the reconnection zone. However, ML85,86 did not understand the importance
of ``spontaneous stochasticity'' of field lines and of Lagrangian trajectories and they did not 
arrive at an analytical prediction for the reconnection speed.  An enhancement of the reconnection 
rate was reported in their numerical study, but the setup precluded the calculation of a long term 
average reconnection rate.  A more recent study along the approach in \cite{MatthaeusLamkin85} 
is one in \cite{Watsonetal07}, where the effects of small scale turbulence on 2D reconnection were 
studied and no significant effects of turbulence on reconnection were found. However, this 
new work studies flow-driven merging in the super-Alfv\'enic limit, with very strong flows ramming
magnetic field together. In this situation, turbulence-induced X-points that might provide additional 
reconnection sites are usually rapidly ejected from the sheet. As discussed in the previous 
section, it is still a wide open question whether fast reconnection can occur in 2D MHD.

Some other numerical simulation papers have claimed to see evidence for fast 
reconnection in resistive MHD simulations.   
An important study of tearing instability of current sheets in the presence 
of background 2D turbulence and the formation of large-scale, long-lived
magnetic islands has been performed in \cite{Politanoetal89}. They present 
evidence for ``fast energy dissipation'' in 2D MHD turbulence and show that
this result does not change as they change the resolution. A more recent work of 
of \cite{MininniPouquet09} provides evidence for ``fast dissipation'' also in 
3D MHD turbulence.  This phenomenon is consistent with the idea
of fast reconnection, but cannot be treated as a direct evidence of the process. 
Evidently, fast energy dissipation and fast magnetic reconnection are rather different  
physical processes. A series of papers by Galsgaard and Nordlund, 
in particular \cite{GalsgaardNordlund97b}, might also be interpreted as 
providing indirect evidence for fast reconnection.  The authors noted that in their 
simulations they could not produce highly twisted magnetic fields. One of the 
interpretations of this finding could be the relaxation of magnetic field via reconnection. 
In this case, this observations could be related to the numerical finding
of \cite{LapentaBettarini11} which shows that reconnecting magnetic configurations 
spontaneously get chaotic and dissipate, which in its turn may be related 
to the predictions in LV99. However, in view of many uncertainties of the
numerical studies, we are not confident of this connection. The highest resolution 
simulations of \cite{GalsgaardNordlund97b} were only $136^3$ and at modest 
Lundquist numbers that could not permit a turbulent inertial-range. 

A number of papers have attempted to treat turbulent magnetic reconnection 
by the formal methods of Mean-Field Electrodynamics (MFE), e.g. \cite{Strauss88, 
KimDiamond01}. Before we address these works in any detail, let us briefly
discuss the relationships of LV99 itself to MFE.  LV99 theory as originally formulated 
does not consider ensemble- or time-averaged magnetic fields, does not 
employ MFE and, in fact, makes only very limited use of probability and averaging. 
The key observation to derive LV99 predictions using the exact formula (\ref{path-int}) 
for the magnetic field---the ``stochastic Lundquist formula''---is that it does {\it not} reduce 
to the conventional Lundquist formula as $\lambda\rightarrow 0$ in a turbulent plasma. The 
typical lateral spread of Lagrangian trajectories that contribute in that limit is estimated 
from GS95 turbulence theory to be $\ell_\perp^2\sim \varepsilon |t-t_0|^3$ for $t<L_i/v_A$ 
and to be  
$\ell_\perp^2\sim u_L L_iM_A^3|t-t_0|$ for $t>L_i/v_A.$
This is a  turbulent mixing effect but not in the usual sense of an enhanced 
``eddy-diffusivity'' or ``$\beta$-effect'' experienced by a mean magnetic field. 
Within the latter concept (see \cite{Parker79}), magnetic fields are presumed 
to be mixed up passively by turbulence and the rate of reconnection thereby accelerated. 
This solution is, however, not tenable for any dynamical important magnetic field which 
would resist bending by turbulent motions at small scales. The concept in its original 
formulation is therefore not applicable to any astrophysical fields, apart from unrealistically 
weak fields, which are of marginal astrophysical importance, anyhow.

Of course, it is always possible to define in a purely formal manner an ``effective 
diffusivity'' which, when substituted into the MFE equations, will reproduce 
the predictions of {\it any} theory for the reconnection velocity $v_{rec}$ and 
width $\Delta$ of the reconnection layer. All that one needs to do is set 
\be \lambda_{turb}\equiv v_{rec}\Delta=\frac{L_x}{v_A}v_{rec}^2 \ee
for the desired $v_{rec}$ and for the corresponding layer width $\Delta\equiv L_x(v_{rec}/v_A)$
imposed by mass balance. In the case of LV99, this leads to a ``turbulent magnetic diffusivity'' 
\be \lambda_{\perp\,turb}= u_LL_iM_A^3
\min\left\{\left(\frac{L_x}{L_i}\right)^2,1\right\}, \lb{lambda-LV99} \ee
where the subscript ``$\perp$'' denotes that this diffusivity is in the direction 
transverse to the background field. There is not necessarily any real physics 
associated with this purely formal definition of a diffusivity.

In the weak turbulence regime for $L_x>L_i$, however, this ``turbulent magnetic diffusivity'' 
reduces to the turbulent thermal diffusivity derived by \cite{Lazarian06}. This is not
an accident. As we have discussed in deriving eq.(\ref{2part}), the nearly straight 
field-lines in the weak turbulence regime experience a Brownian motion in the plane 
perpendicular to their direction, with a diffusion constant $\lambda_{\perp\,turb}=u_LL_iM_A^3,$ 
as in (\ref{lambda-LV99}). This implies a term in the induction equation for the mean field 
$\overline{\bB}(\bx,t)$ of the form $\lambda_{\perp\,turb}\triangle_\perp\OL{\bB}.$
There is therefore a consistent MFE approach to recover the predictions of the LV99 
theory for reconnection in the weak turbulence regime, by means of an appropriate
{\it anisotropic} turbulent magnetic diffusivity. 

This is not the true for the strong turbulence regime with $L_x<L_i$. Recall from 
the discussion around eq.(\ref{2part}) that field lines in strong turbulence do {\it not} 
separate diffusively in the transverse direction, as $\ell_\perp^2(t)\sim \lambda_{\perp\,turb} t$ , 
but instead undergo a Richardson diffusion $\ell_\perp^2(t)\sim \varepsilon t^3.$ This 
superdiffusive line motion cannot be consistently described by a Laplacian term in the MFE
equations.  Note also that strict accuracy of the MFE description requires a slow variation of 
mean fields on the integral scale $L_i$ of the turbulence, whereas the width of the reconnection 
zone according to (\ref{Delta-LV99}) always satisfies $\Delta< L_i.$ 
One would therefore not expect MFE with a simple Laplacian 
diffusion term to consistently and accurately describe the reconnection layer in strong turbulence
predicted by LV99.   

With this discussion as backdrop, let us review some of the attempts to discuss magnetic 
reconnection using MFE, beginning with \cite{KimDiamond01} [KD01]. In the first place, 
it must be emphasized that they consider a rather unconventional set-up with a very strong 
guide field $\langle B_z\rangle$ and with fluctuating components $B_{rms}$ also much larger 
than the reconnecting component $\langle \bB_H\rangle,$ i.e. $\langle B_z\rangle\gg
B_{rms} \gg \langle |\bB_H|\rangle. $ Because of the first assumption, reconnecting lines 
of the mean field are almost exactly parallel, with only tiny components of opposing sign. 
Because of the second assumption $M_A\gg 1$ and there is no recognizable reconnection 
geometry for actual lines of the unaveraged magnetic field: the anti-parallel horizontal components 
are just a small mean effect. These conditions were adopted for analytical convenience and 
do not generally reflect astrophysical reality (particularly the second). Then KD01 
develop closure approximations for the turbulent magnetic diffusivity 
using the 3D RMHD equations. Assuming that the small-scale
fluctuations are statistically stationary and appealing to boundary conditions (e.g. periodic)
which  permit one to neglect surface flux terms, they derive a strongly quenched turbulent 
diffusivity with $\lambda_{turb}=O(\lambda),$ the plasma diffusivity. They therefore 
conclude that magnetic reconnection proceeds at the slow Sweet-Parker rate, in contradiction 
to LV99.  However, we find their conclusions of little relevance to LV99. In the first place, they
consider a quite different problem, with $M_A\gg 1.$ In the second place, their main finding   
of a quenched turbulent magnetic diffusivity is due to the conservation of magnetic potential 
by the nonlinear terms of the RMHD equations. This quenching is unlikely to persist 
with the open boundary conditions required for stationary reconnection or to be a feature 
of turbulent reconnection governed by the full 3D MHD equations. 

Finally, one should not mix up the concepts we discuss here with the so-called 
``hyper-resistivity'' \citep{Strauss86, BhattacharjeeHameiri86, HameiriBhattacharjee87,
DiamondMalkov03}, which is another attempt to derive fast reconnection from turbulence 
within the context of mean-field resistive MHD. The form of the parallel electric field 
can be derived from magnetic helicity conservation.  Integrating by parts one
obtains a term which looks like an effective resistivity proportional to the
magnetic helicity current.  There are several assumptions implicit in this
derivation. Fundamental to the hyper-resistive approach is the assumption that the 
magnetic helicity of mean fields and of small-scale, statistically-stationary turbulent fields 
are separately conserved, up to tiny resistivity effects.  However, this ignores magnetic 
helicity fluxes through open boundaries, essential for stationary reconnection, that 
vitiate the conservation constraint.

There is a general objection to all mean-field approaches which, to ``explain''  fast 
reconnection, make appeal to some effective dissipation (resistivity, hyper-resistivity, etc.) 
experienced by the fields once averaged over ensembles or over small space-time scales.  
The difficulty is that it is the lines of the {\it full} magnetic field that must be rapidly reconnected,
not just the lines of the mean-field. The former implies the latter, but not conversely. Nature 
does not average over ensembles of small-scale turbulence to get fast reconnection! No 
mean-field approach can claim to have explained the observed rapid pace of magnetic 
reconnection unless it is shown that the reconnection rates obtained in the theory are 
strictly independent of the length- and time-scales of the averaging. See \cite{EyinkAluie06} 
for further discussion. 

\subsection{Turbulent Versus Collisionless Reconnection}  

The most popular alternative to LV99 is currently not a competing 
turbulence theory of reconnection but instead non-MHD approaches based on the Hall 
effect in a two-fluid model \citep{Shayetal98, Shayetal99,Wangetal00,Birnetal01,Drake01,
Malakitetal09,Cassaketal10}, or on full kinetic Vlasov dynamics 
\citep{Daughtonetal06,Daughtonetal08,Daughtonetal11,Cheetal11}. We have argued 
in section 4.3 and at length in Appendix B that collisionless effects are, in most cases, 
irrelevant to determining reconnection speeds in the presence of turbulence.  We do 
believe that there are some situations in certain astrophysical systems  where kinetic 
effects will influence the rate of reconnection. However, formulating a criterion to 
determine if kinetic effects are important requires some care in the presence of turbulence. 
A popular criterion for ``collisionless reconnection'' is the condition that the Sweet-Parker 
thickness $\delta_{SP}$ be smaller than the ion-skin depth $\delta_i$ \citep{Yamadaetal06,
ZweibelYamada09}. This condition has been checked to correctly signal ``collisionless 
reconnection'' in laminar simulations and experiments. However, as we now
discuss, this condition in turbulent plasmas is not relevant to the validity of LV99.

There is even some question of how to define the ``Sweet-Parker thickness'' 
$\delta_{SP}$ in turbulent reconnection. In the original resistive MHD model of LV99, local, 
small-scale reconnection events were argued to have a Sweet-Parker Y-type structure 
and thickness 
 \be \ell_\eta^\perp \simeq L_i M_A^{-1}S_{L}^{-3/4}, \lb{A2} \ee 
with $\ell_\eta^\perp$ the resistive cut-off length of GS95 turbulence theory. See  LV99, 
eq.(14). However, this may not be the most relevant length-scale for determining the 
importance of collisionless effects. Within GS95 phenomenology one can 
estimate the pointwise ratio of the Hall electric field to the MHD motional field as 
\be \frac{J/en}{u} \simeq \frac{c\delta B(\ell_\eta^\perp)/4\pi ne\ell_\eta^\perp}{u_L} 
  \simeq \frac{\delta_i}{L_i} M_A S_L^{1/2}
\lb{A3} \ee
where $S_L=v_AL_i/\lambda$ is the Lundquist number based on the forcing length-scale
of the turbulence and $M_A=u_L/v_A$ is the Alfv\'enic Mach number. This suggests 
the definition of the ``Sweet-Parker thickness'' as 
\be \delta_{SP} =L_i M_A^{-1} S_L^{-1/2}, \ee
so that $(J/en)/u_L\simeq \delta_i/\delta_{SP}.$  The length-scale $\delta_{SP}$ is the 
analogue of the Taylor microscale in hydrodynamic turbulence, whereas the distance 
$\ell_\eta^\perp$ is the GS95 analogue of the Kolmogorov scale.  If the magnetic diffusivity 
in the definition of the Lundquist number is assumed to be that based on the Spitzer resistivity, 
given by $\lambda=\delta_e^2 v_{th,e}/\ell_{ei}$ where $\delta_e$ is the electron skin depth, 
$v_{th,e}$ is the electron thermal  velocity, and $\ell_{ei}$ is the electron mean-free-path length 
for collisions with ions, then 
$ S_L=\left(\frac{m_e}{m_i}\right)^{1/2} \beta^{-1/2} \left(\frac{\ell_{ei}}{\delta_e}\right)^2
                \left(\frac{L_i}{\ell_{ei}}\right), $ 
with $\beta=v_{th,i}^2/v_A^2$ the plasma beta. Substituting into (\ref{A3}) and defining
$M=v_{th,i}/u_*$ as the Mach number gives
\be \frac{\delta_i}{\delta_{SP}}\simeq \left(\frac{m_i}{m_e}\right)^{1/4} M
               \beta^{1/4}  \left(\frac{\ell_{ei}}{L_i}\right)^{1/2}, \lb{A6} \ee
which coincides precisely with the ratio defined by \cite{Yamadaetal06},eq.(6).

With this definition of $\delta_{SP},$ it is easy to see from the data in our Table 1 that the ratio
$\delta_i/\delta_{SP}\simeq 10^{-3}$ for the warm ISM, $\simeq 1$ for post-CME current sheets
and $\simeq 10^3$ for solar wind impinging on the magnetosphere. However, this has nothing 
to do with the validity of LV99 for those systems. What it does say is that Ohmic resistivity may 
not be the mechanism for line-breaking in the latter two cases and the original resistive model 
of LV99 is not an adequate description at sufficiently small scales. Thus, the structure of local, 
small-scale reconnection events should be strongly modified by Hall or other collisionless effects, 
with possibly an $X$-type structure, an ion layer thickness $\sim \delta_i,$ quadrupolar magnetic 
fields, etc. However, these local effects are irrelevant to determining the global rate of large-scale 
reconnection (see Appendix B).

What is the correct criterion to determine a breakdown of LV99 and the 
importance of collisionless effects? The LV99 model assumes that the thickness $\Delta$
of the reconnection layer is set by turbulent MHD dynamics (line-wandering and Richardson 
diffusion). Thus, self-consistency requires that the length-scale $\Delta$ must lie within the range 
of scales where shear-Alfv\'en modes are correctly described by incompressible MHD. 
This implies a criterion for ``turbulent collisionless reconnection''
\be  \rho_i\gtrsim \Delta  \lb{LV99-break} \ee
with $\Delta$ calculated from eq.(\ref{Delta-LV99}) and $\rho_i$ the ion cyclotron 
radius (see section 2). Since $\Delta \propto L_x,$ the large length-scale of the reconnecting 
flux structures, this criterion is far from being satisfied in most astrophysical settings. For example, 
in the three cases of Table 1, one finds using $\Delta = LM_A^2$ that $\rho_i/\Delta\simeq 10^{-13}$ 
for the warm ISM, $\simeq 10^{-6}$ for post-CME sheets, and $\simeq .1$ for the magnetosphere. 
Only in the latter case is the condition (\ref{LV99-break}) close to being satisfied. Reconnection 
events in the magnetosphere probably do typically involve collisionless effects in an 
essential way. It must be appreciated, however, that this is a highly nongeneric situation and turbulent 
fluid effects will generally predominate in astrophysical reconnection. 

Even in magnetospheric reconnection the presence of background turbulence
in the environment should not be ignored. Like nearly every cosmic plasma, the magnetosphere 
commonly exhibits turbulence, either statistically steady or in sporadic bursts \citep{Zimbardoetal10}.
Energy spectra are generally similar to those in the solar wind, with spectral exponents close 
to -5/3 and -7/3 at scales above and below the ion gyroradius/ion skin-depth (since $\beta\simeq
0.1-10,$ these are nearly the same), respectively. See Table 1 in \cite{Zimbardoetal10}. The original
LV99 theory does not apply in the situation that $\rho_i\simeq \Delta >\rho_e,$ but a version 
of LV99 based on EMHD might provide some insight. Of course, since $\rho_e$ is only 43 times 
smaller than $\rho_i$ for a hydrogen plasma,  any EMHD inertial range must be of limited extent. 
Kinetic effects enter at electron scales, not accurately described by a fluid model, and 
approaches based on solving the Vlasov equation (see section 6.3) become important.
Some insights of LV99 should still carry over, such as the importance of field-line 
wandering in enhancing reconnection rates. See section 6.3 for more discussion.

\subsection{Observational and Numerical Tests}

Let us discuss briefly the support for LV99 from observations and simulations.   


The solar corona is a system proximate to Earth where LV99 theory 
should be applicable.  Recent observations of a greatly thickened current sheet 
associated with coronal mass ejections give broad support to LV99 \citep{CiaravellaRaymond08, 
Bemporad08}. For example, the width $\Delta$ of the reconnection zone from the formula 
(\ref{Delta-LV99}) was found by \cite{CiaravellaRaymond08} to be just 10 times smaller 
than the observed thickness. This is quite good agreement, given that that formula has 
an unknown prefactor (which probably depends upon global geometry) and that some 
of the inputs (particularly the turbulence integral scale $L_i$) were unknown from 
observations and had to be crudely estimated.  LV99 predicted also the phenomenon 
of triggering of magnetic reconnection. Indeed, as the reconnection speed depends on the 
level of magnetic field stochasticity, the stochasticity induced by Alfv\'{e}n wave packages 
can enhance the reconnection speed. This is consistent with the recent observations by
\cite{Sychetal09} who reported a phenomenological relationship between 
oscillations in a sunspot and flaring energy  release above an active region
above the sunspot. The authors proposed that the pulsations in the flaring 
energy release are triggered by wave packages arising from sunspots. The
LV99 mechanism presents a very natural explanation for the phenomenon.

While the Sun presents one of the best studied examples of magnetic reconnection, 
interstellar medium has the advantage of testing LV99 in a collisional
environment (see \cite{Yamadaetal08}). Similarly, collisional reconnection 
was observed in the solar photosphere \citep{Parketal09}. For both types of environments 
effects of partial ionization may be important\footnote{For the interstellar media some 
phases are partially ionized (see \cite{DraineLazarian98} for a list of
idealized phases and their parameters).}. The generalization of the LV99 
model for a partially ionized gas is presented in \cite{Lazarianetal04}. 
The direct numerical testing of the regimes of reconnection described 
in that paper requires the use of a two-fluid code and has not been done so
far. Given the importance of magnetic fields in interstellar medium, 
such a testing is very appealing.

The most convincing tests of LV99 would be of its predicted scaling laws for the width
(\ref{Delta-LV99}) and for the reconnection velocity (\ref{vr-LV99}) in terms of the 
parameters $u_L,v_A,L_i,$ and $L_x.$ These scalings are most easily checked 
in numerical MHD studies, which achieve only moderate Reynolds numbers but which 
permit controlled experiments.  \cite{Kowaletal09} used the relation $\varepsilon \simeq 
u_L^4/v_AL_i$ to rewrite formula (\ref{vr-LV99}) as 
 \be  v_{{\rm rec}} = \left({{\varepsilon}\over {v_A L_x}}\right)^{1/2}\min\{L_x,L_i\}.
 \lb{vr-kowal} \ee
and tested the scalings in $\varepsilon$ and $L_i$ by means of numerical simulations. 
\cite{Kowaletal09} studied in particular the weak-turbulence regime with $L_x>L_i,$ 
investigating the predicted power-law scalings $v_{rec}\propto \varepsilon^{1/2}L_i.$ 
The dependence on $\varepsilon^{1/2}$ was confirmed, but the predicted linear 
relation $\propto L_i$ was not well verified and replaced with a weaker dependence 
closer to  $L_i^{3/4}$ for larger $L_i\lesssim L_x.$ This behavior is possibly associated 
with a crossover to the strong-turbulence regime with $L_i>L_x$. Note that the formula 
(\ref{vr-kowal}) for strong turbulence predicts independence of the reconnection rate from 
$L_i,$  except through the dependence on $L_i$ of the mean energy dissipation $\varepsilon$.

The independence of reconnection rate on $L_i$ for strong turbulence is quite convenient for 
observational tests of LV99, since $L_i$ is much harder to reliably estimate in natural reconnection 
phenomena than are $v_A,L_x$ and $\varepsilon.$ Indeed, the source 
of turbulence may not be so easily identified or localized in wavenumber. One approach is to 
obtain $u_L$ from Doppler line-broadening (e.g. \cite{Bemporad08}) and then to estimate $L_i=
u_L^4/v_A\varepsilon.$ Of course, if spectral information is available, then one can obtain 
$L_i$ directly from the peak of the energy spectrum.
The mean energy dissipation rate $\varepsilon$ is a source term for plasma heating, which can 
be estimated from observations of electromagnetic radiation. To make such an estimation reliably 
requires not only a radiation model but also some understanding of where the energy from the 
cascade is deposited at scales smaller than the ion gyroradius \citep{Schekochihinetal09}. 
It may be preferable to estimate $\varepsilon$ from coarse-grained measurements of 
the energy-flux rate $\Pi_\ell,$ which is equal, on average, to the dissipation rate.  
For example, the flux of magnetic energy is given by $\Pi_\ell^B=\OL{\bJ}_\ell\bdot\boepsilon_\ell$ 
where $\OL{\bJ}_\ell=(c/4\pi)\grad\btimes\OL{\bB}_\ell$ is the coarse-grained current and $\boepsilon_\ell
=\OL{(\bu\btimes\bB)_\ell}-\OL{\bu}_\ell\btimes\OL{\bB}_\ell$ is the subscale EMF. Because 
of the scale-locality of the MHD energy cascade it possible to develop very 
quantitatively accurate approximations to the subscale EMF which depend only
upon the coarse-grained gradients $\grad\OL{\bu}_\ell$ and $\grad\OL{\bB}_\ell.$ For discussion
of the approximation, see \cite{Eyink06}. 

 Additional testing of magnetic reconnection can be produced through indirect 
 means. Indeed, the process of reconnection may be responsible for a wide variety
of processes. For instance, the LV99 model of reconnection was invoked to explain 
gamma ray bursts \citep{Lazarianetal01, ZhangYan11}, acceleration of 
cosmic rays \citep{GouveiadalPinoLazarian03, Lazarian05, LazarianOpher09, 
LazarianDesiati10}, removal of magnetic field during star formation \citep{Lazarian05, 
SantosdeLimaetal10}. As the processes are quantified and elaborated, it is  
possible to get insight into the magnetic reconnection that is driving them.

\section{Discussion} 

Let us discuss here briefly the main results of this paper as well as 
some of their ramifications. 

\subsection{Flux-freezing in astrophysics}

Plasma conductivity is high for most astrophysical circumstances. It has therefore generally 
been assumed that  ``flux-freezing'' is an excellent approximation for astrophysical plasmas. 
The principle of ``frozen-in'' field-lines provides a powerful heuristic which allows simple,
back-of-the-envelope estimates in place of full solutions (analytical or numerical) of the MHD equations. 
As such, the ``flux-freezing'' principle has been applied to gain insight into diverse processes, 
such as star formation, stellar collapse, magnetic dynamo, solar wind-magnetospheric interactions, 
etc. The predictions of this simple principle often accord very well with observations. Flux-freezing 
is used to explain, for example, the magnetization of white dwarfs and neutron stars, the low 
angular momentum of stars, and the spiral structure of lines of force in the solar wind. 

However, for every success of the ``flux-freezing'' principle, there is an equally stark failure.
If  the standard ``flux-frozenness'' property held to good accuracy, then topology changes 
of magnetic field could not occur at the very fast rates observed in solar flares and coronal mass 
ejections. During star-formation, the magnetic pressure of in-falling field-lines would be so great 
as to prevent gravitational collapse altogether. The tangled line-structure in small-scale dynamos 
would quench the exponential growth of magnetic field.  The most common attempts to explain such 
contrary observations invoke additional physical effects that violate conventional ``flux-freezing'',  
e.g. additional terms in the generalized Ohm's law besides collisional resistivity. Unfortunately, 
most of these microscopic plasma effects seem to be too small to have much effect in astrophysical 
MHD plasmas at very high kinetic and magnetic Reynolds numbers. 

Turbulence is a natural suspect to explain apparent breakdowns of ``flux-freezing'' in 
astrophysical environments. Most attempts to use turbulence invoke it as just another 
effective non-ideal term in the generalized Ohm's law. Indeed, any averaging of Ohm's law 
(e.g. over ensembles, space or time) produces an additional ``turbulent EMF''  that leads to 
a possible resistivity-independent violation of ``flux-freezing''  for the {\it mean} magnetic field. 
This is a correct observation as far as it goes, but not a fundamental solution to the problem. 
``Flux-frozenness'' in the conventional sense must be violated for the full magnetic field
and not just for some averaged field. 

Our solution is quite different. We have argued that ``flux-freezing'' in the standard sense is violated 
in a turbulent inertial range at high kinetic and magnetic Reynolds numbers, but continues 
to hold in a novel stochastic sense. This {\it stochastic flux-freezing law} in turbulent plasmas
is a consequence of the remarkable turbulence phenomenon of ``spontaneous stochasticity'' of 
Lagrangian particle trajectories. As we have shown in detail, it is this stochastic form of ``flux-freezing'' 
which underlies the LV99 theory of fast magnetic reconnection. In any turbulent astrophysical 
system, ``stochastic flux-freezing'' is the natural replacement for conventional flux freezing. 
We believe that it  will provide a powerful heuristic tool to explain many astrophysical 
phenomena. To use this tool with confidence requires a sound understanding of MHD 
plasma turbulence, in particular Lagrangian particle statistics. This is a clear focus 
of further theoretical, experimental and numerical work. 

\subsection{Line-wandering and LV99} 

Spontaneous stochasticity is a property not only of Lagrangian trajectories, but also of magnetic 
field-lines themselves.  The identity of ``the'' field-line passing through a point becomes blurred
in a high magnetic-Reynolds-number turbulent plasma. 
Certainly, at any moment in time there are a unique set of field lines that describe the field, 
through each point. 
However, there are an infinite number of field lines that can be drawn through a small volume, and 
if we follow for some parallel distance along these field lines, then they diverge explosively away from 
one another.  In a turbulent plasma with a rough magnetic field, the distance that these lines diverge apart 
is independent of the dimensions of this initial volume, when the parallel distance travelled greatly 
exceeds this dimension. As a limiting situation, for vanishingly small initial volumes, there are   
infinitely many field-lines that  emerge and meander randomly out of one point! At sufficiently large 
scales, that is, above the dissipation scale, this line-wandering is independent of the value of resistivity, 
and, within limits, to the details of the microscopic physics (see, for example, Appendix B).  

This was the original insight of \cite{LazarianVishniac99}.  The important implication is that plasma, 
flowing a finite distance along magnetic fields lines, will diffuse away from the mean magnetic field 
direction. This effect can accelerate reconnection speeds not only in 
the presence of inertial-range turbulence, but whenever field-lines exhibit some disordered,   
random character. If we use line-wandering in a turbulent reconnection zone to calculate 
the width of the outflow region around the current sheet, then we find that the classic Sweet-Parker 
rate becomes the fast reconnection rate given in LV99,  independent of resistivity.
As we have shown in the present paper, this same reconnection rate can also be obtained 
by an alternative approach in which an ensemble of field lines move stochastically in time. 
This agreement follows from a formal equivalence between the two approaches.

The precise quantitative details of magnetic line-wandering depend upon the scaling laws of MHD turbulence.
We have adopted here the predictions of phenomenological GS95 theory, and these should surely be verified 
and possibly refined in future numerical and experimental studies. However, a parameter study in LV99 shows that
the conclusion of fast reconnection in a 3D turbulent flow holds for a range of spectral slopes and 
spectrum anisotropies that encompass the existing suggestions of modifying the GS95 model. 
Our work neglects also some features of large scale 
reconnection.  By treating the reconnection zone as a {\it typical} part of a turbulent plasma we are neglecting 
the effects of a strongly sheared large scale field and the fast outflow along magnetic fields that will accompany 
reconnection events.  These features are naturally included in numerical simulations \citep{Kowaletal09} and 
do not seem to result in any significant modification of the reconnection rate, but further work along this line 
is certainly desirable.

\subsection{Other recent work on reconnection}

Since our conclusions are consistent with the work of LV99 it is worth 
considering recent work on alternative approaches to calculating reconnection rates.
Over the last decade, more traditional approaches to reconnection have 
changed considerably. At the time of its introduction, the models competing with LV99 
were modifications of the single X-point collisionless reconnection scheme first introduced by \citet{petschek64}. 
Those models had 
point-wise localized reconnection regions which were stabilized via plasma effects so that the
outflow opened up on larger scales  (see Fig.~5). Such configurations would be difficult to 
realize in the presence of random forcing,  which would be expected to collapse the reconnection layer. 
Moreover, \citet{CiaravellaRaymond08} argued that observations of solar flares were inconsistent with single X-point 
reconnection.

In response to these objections, more recent models of collisionless reconnection have acquired several 
features in common with the LV99 model. In particular, they have moved to consideration of volume-filling 
reconnection, (although it is not clear how this volume filling is achieved in the presence of a single 
reconnection layer (see \cite{Drakeetal06})).  While much of the discussion still centers around magnetic 
islands produced by reconnection, in three dimensions these islands are expected to evolve into 
contracting 3D loops or ropes due to tearing-type instabilities in electron-current layers 
(\cite{Daughtonetal08,Daughtonetal11}). This is broadly similar to what is depicted in Fig.~5,  at least 
in the sense of introducing stochasticity to the reconnection zone. The 3D PIC simulation studies reported 
in these works should help, in particular, to interpret magnetospheric  reconnection phenomena. However, 
although the simulations are described as ``turbulent'', they do not exhibit the inertial-range power-law 
spectra observed in the magnetosphere and do not take into account either the pre-existing turbulence 
found in many of its regions (due to temperature anisotropy, velocity shear, Kelvin-Helmholtz instability, etc.)
or inertial-range turbulence generated as a consequence of reconnection itself  \citep{Zimbardoetal10}. 
An EMHD analogue of LV99 and \cite{Kowaletal09} studies may help to estimate such effects.

Similar remarks apply to the recent 3D PIC study of \cite{Cheetal11}, which 
observes micro-turbulence in the electron current layer during reconnection. The authors identify 
the source of this ``turbulence'' as a filamentation instability driven by current gradients, very
similar to a related instability in the EMHD model. The key aim of this work was to identify 
the term in the generalized Ohm's law which supplies the reconnection electric field  to break
the ``frozen-in'' condition. However, this study ignores the ambient inertial-range turbulence 
observed in the magnetosphere and other astrophysical plasmas, which may strongly modify 
laminar instabilities. Also, while there is interest in understanding the origin and character of 
reconnection electric fields (e.g. for particle acceleration), we have argued at length in section 4.3
that the precise mechanism of ``line-breaking'' is irrelevant for the rate of reconnection in the 
presence of high-Reynolds-number inertial-range turbulence.

Departure from the concept of laminar reconnection and the introduction of magnetic stochasticity 
is also apparent in a number of the recent papers appealing to the tearing mode instability to 
drive fast MHD reconnection \citep{Loureiroetal09, Bhattacharjeeetal09}.
\footnote{The idea of appealing to the tearing mode as a means of enhancing the reconnection 
speed can be traced back to \cite{Strauss88, Waelbroeck89, ShibataTanuma01}. 
LV99 showed that the linear growth of tearing modes is insufficient to obtain fast reconnection. 
More recent work is based on the idea that the non-linear growth of magnetic islands or plasmoids
due to mergers provides large scale growth rates larger than the tearing mode linear growth rates 
on these scales.  A situation where the non-linear growth is faster than the linear one is 
rather unusual and requires further investigation. See \cite{Diamondetal84}.}  Several 2D 
numerical studies \citep{Samtaneyetal09,Cassaketal09,Bhattacharjeeetal09,HuangBhattacharjee10}
have provided evidence that reconnection in the plasmoid-unstable region is independent of resistivity 
and  a simple, heuristic picture of a multi-level plasmoid hierarchy has been proposed for this regime 
by \cite{Uzdenskyetal10}. A recent very high-resolution study of \cite{NgRagunathan11}, 
however, has found that plasmoids do not form even when the Lundquist number exceeds the putative 
critical value, except  when insuffcient numerical resolution is used or when a small amount of noise 
is added externally. Thus, this alternative MHD mechanism of fast reconnection may require some
level of evironmental noise, similar to LV99. In any case, since tearing modes exist even in a collisional 
fluid, this may open another channel of reconnection in such fluids. As we discuss below, this reconnection 
should not be ``too fast'' to account for the observational data.

A fundamental consideration for all reconnection models is that they must explain fast reconnection in collisional
and collisionless plasmas.  At the same time it should be possible for reconnection to be delayed for significant
amounts of time.  Otherwise the accumulation of magnetic flux prior to a solar flare would be inexplicable.  Fast
reconnection as a feature of turbulence sets a minimum reconnection speed which 
allows the topology of the magnetic field to evolve on dynamical time scales. 
Thus the LV99 model can explain the accumulation of flux provided that 
there are no generically fast reconnection processes at work\footnote{To be specific, take typical 
coronal loop parameters $n=10^9\, cm^{-3},$ $T=100\, eV,$ $B=300\, G,$ $u_L=10^7\, cm/s$ and $L_i\simeq L_x
=10^9\, cm.$ Then $v_A=10^{10}\, cm/s$ and $\Delta =L_iM_A^2=10^3\, cm,$ comparable to the ion skin 
depth $\delta_i=700\, cm.$ Thus, the initial turbulent reconnection rate will be quite small and may be 
slower than collisionless mechanisms. However, the energy released in the slow reconnection process 
can make the region more turbulent, accelerating the reconnection and resulting in a flare.}. 
An alternative explanation based on collisionless reconnection in a laminar 
environment has been suggested by  \cite{Cassaketal06} and \cite{Uzdensky07}. However,
it is not clear whether this explanation will account also for the large observed thickness of the 
macroscopic current sheets \citep{CiaravellaRaymond08,Bemporad08}.


If local turbulence is driven by release of energy from the magnetic field, 
it may result in a runaway turbulent reconnection process which may be relevant 
to some numerical simulations\citep{Lapenta08, BettariniLapenta09}.
Alternatively, if tearing modes begin
by driving relatively slow reconnection, but by some process destabilize the current 
sheet, then a similar runaway might result. Turbulence looks like a natural candidate 
for such process, but one should be open to alternative suggestions. For instance, 
we note that while X-point reconnection is clearly unstable for an isolated current sheet in a collisional
fluid, interacting current sheets can produce bursts of fast X-point reconnection, 
separated by periods of slow evolution (\cite{pangetal2010}). Formally, in this case we might 
not need to invoke turbulent feedback\footnote{Energy release induces the feedback anyhow 
and the interaction of the resulting turbulence with the instabilities is an issue of further research.}, 
but simply rely on the relatively slow pace of stochastic reconnection to evolve the magnetic 
field from one outburst to the next.
 
In any case, in most astrophysical situations one has to deal with the {\it pre-existing turbulence}, 
which is the inevitable consequence of the high Reynolds number of astrophysical fluids 
and for which abundant empirical evidence exists. Such turbulence may modify or suppress instabilities, 
including the tearing mode instability. In this paper we have shown that it, by itself, induces fast reconnection 
on dynamical time scales. 

\subsection{Outstanding Issues for Turbulent Reconnection}

Much remains to be done to fully clarify the effects of MHD plasma 
turbulence on astrophysical reconnection. Without attempting to be exhaustive, let us 
mention some of the issues that we regard as most pressing.

Of the possible extensions and refinements of LV99 mentioned 
in section 5, we think one of the most important is including the effects of magnetic and velocity shear.  
A related issue not addressed in the current paper is the detailed structure of the turbulent 
reconnection zone. We have invoked here the GS95 phenomenology of MHD turbulence, but 
intermittency effects not included in GS95 could be relevant.  In principle, an independent  
theoretical derivation of the reconnection speed can be based on the line-voltage studied 
in \cite{Kowaletal09}. The arguments of LV99 suggest that the reconnection voltage should be
contributed primarily by the motional EMF of already reconnected field-lines, but a detailed analysis 
is required. Addressing these issues will be the subject of future work.

In its original formulation, LV99 is a model of non-relativistic reconnection. However, it is clear 
that the idea of enabling fast reconnection by extending the thickness of the outflow region through magnetic field wandering should be applicable to relativistic flows. This extension was implicitly used in some models of gamma ray bursts (see \cite{Lazarianetal02, ZhangYan11}). Formal and detailed studies of relativistic turbulent reconnection 
are therefore important.

Similarly, inertial-range turbulence with power-law spectra is an idealization adopted in LV99 to get analytically 
tractable results. In many cases, however, wandering of field-lines and of Lagrangian trajectories is significantly 
affected by larger or smaller scales outside the inertial-range. Deviations from pure power-law spectra may 
occur due to multiple turbulent energy injection scales or to the effects of Ohmic dissipation. These effects 
of inhomogeneity of turbulence driving are important subjects for further quantitative studies.

Last, but not least, effects of turbulence dissipation via viscosity is another issue where more studies are 
required. \cite{Lazarianetal04} extended LV99 to the case of partially ionized gas where the dissipation of 
turbulence arises mostly through viscosity. The concept of Richardson diffusion and its relation to turbulent 
reconnection in this case needs additional work.

\section{Summary}

The main points of our paper can be very briefly summarized as follows:

\begin{enumerate}
\item In turbulent plasmas with rough velocity fields, the lines of force are not ``frozen-in'' in the standard 
deterministic sense, but instead in a random sense associated to the ``spontaneous stochasticity''
of Lagrangian particle trajectories. 

\item Magnetic reconnection rates calculated based on ``stochastic flux-freezing'' and 
phenomenological GS95 turbulence theory recover the predictions of the LV99 theory 
of turbulent reconnection.

\item The LV99 predictions are independent of all non-ideal terms in the generalized 
Ohm's law. For example, they do not change if the Hall effect is included, whenever the ion skin depth is much 
smaller than the turbulent integral length or injection scale.
\end{enumerate}





\noindent {\bf Acknowledgements.} We thank A. Bhattacharjee, B. D. G. Chandran, 
P. H. Diamond, R.Kulsrud, A. Pouquet,  and M. Yamada for some useful correspondence 
and discussion. The criticisms and suggestions of an anonymous 
referee have helped to greatly improve the paper.
GE was partially supported by NSF grants AST 0428325 and CDI-II:  CMMI 0941530.  
AL thanks the NSF grant AST 0808118, NASA grant NNX09AH78G and the support of the 
Center for Magnetic Self Organization. The work of ETV is supported by the National Science 
and Engineering Research Council of Canada. \\






\appendix

\section{Formal Treatment of Turbulent Line-Separation}

Consider the problem of the growth in separation of a pair of magnetic 
field lines, starting at points displaced by vector $\Bell,$ as one moves a distance $s$ 
in arclength along the field lines passing through the two points (cf. \cite{Jokipii73}).  
The exact equation for the change in separation can be obtained from (\ref{line-ODE}) to be
\be {{d}\over{ds}} \Bell(s) =\hat{\bb}(\boxi'(s))-\hat{\bb}(\boxi(s)), \ee
where $s$ is arclength, $\Bell(s) =\boxi'(s)-\boxi(s),$ and $\hat{\bb}=\bB/|\bB|$ 
is the unit tangent vector along the magnetic field-line.  By analogy with Richardson diffusion 
for Lagrangian particle trajectories, one can define a ``2-line diffusivity''
\be D_{ij}^B(\Bell) = \int_{-\infty}^0 ds\,\langle\delta \hat{b}_i(\Bell,0)
\delta\hat{b}_j(\Bell,s)\rangle, \lb{DB} \ee
with units of length, where $\delta\hat{b}_i(\Bell,s)=\hat{b}_i(\boxi'(s))-\hat{b}_i(\boxi(s))$
with $\Bell=\boxi'(0)-\boxi(0),$ so that
\be {{d}\over{ds}}\langle \ell_i(s)\ell_j(s)\rangle=\langle D_{ij}^B(\ell)\rangle. \ee
The above equations are formally exact \citep{Batchelor50,Kraichnan66}. Note that (\ref{DB}) 
allows one to write 
\be D_{ij}^B(\Bell)\sim \delta \hat{b}_i(\Bell)\delta \hat{b}_j(\Bell)s_{int}(\Bell) \lb{DB2} \ee
where $s_{int}(\Bell)$ is an integral correlation length of the increment in the tangent 
vector moving along the lines.  (This is properly a tensor quantity, but here written as a scalar.) 
To determine the line wandering, one must have a model for this integral correlation length.

The LV99 theory is obtained by modelling the integrand in (\ref{DB}) for 
field-perpendicular increments  as 
\be  \langle\delta \hat{b}_\perp(\Bell,0)\delta\hat{b}_\perp(\Bell,s)\rangle \sim
    \frac{\delta u^2(\ell)}{v_A^2}{{\rm Re}}(e^{is/\ell_\|-|s|/\lambda(\ell)}).  \lb{dBdBs} \ee
Note that $|\delta \hat{b}_\perp(\ell)|\sim \delta B(\ell)/B_0\sim \delta u(\ell)/v_A.$ There 
are two effects in the $s$-dependent part. First, there is a correlation length $\lambda(\ell)$
of tangent-vector increments along the field line, which gives an exponential decay. 
Within LV99 theory, it is assumed that this length is the distance traveled by a random 
Alfv\'enic disturbance propagating along the field-line with velocity $v_A$ in one energy 
cascade time at scale $\ell_\perp.$ That is, 
\be  \lambda(\ell) = v_A\tau_\ell=v_A \frac{\delta u^2(\ell)}{\varepsilon}. \ee
The second effect arises from the periodic variation along the field-lines due to regular 
Alfv\'en wave trains. It is assumed that, at perpendicular scale $\ell_\perp,$ the Alfv\'en 
wave trains that are dominant are those with wavelength $\ell_\|,$ the parallel length-scale 
corresponding to perpendicular scale $\ell_\perp.$ Integrating (\ref{dBdBs}) in $s$ gives
the result
\be s_{int}(\ell) \sim \frac{1/\lambda(\ell)}{1/\lambda^2(\ell) + 1/\ell_\|^2}\sim 
        \frac{\ell_\|^2}{\lambda(\ell)}= \frac{\varepsilon}{v_A} 
        \frac{\ell_\|^2}{\delta u^2(\ell)} \lb{sint} \ee
for $\lambda(\ell)\geq \ell_\|.$ Substituting into (\ref{DB2}) gives
\be D^B_\perp(\ell) \sim \frac{\varepsilon\ell_\|^2}{v_A^3}=\frac{\ell_\|^2}{L_i}M_A^4, 
\lb{DB3} \ee
where notice that a factor of $\delta u^2(\ell)$ has cancelled between numerator and 
denominator. From this last formula, various specific cases can be considered.

In the strong GS95 turbulence regime, critical balance implies that  $\lambda(\ell)
\sim \ell_\|$ and thus, from (\ref{sint}), $s_{int}(\ell)\sim\ell_\|.$ This gives the intuitive result that 
\be D_\perp^B(\ell)\sim (\delta u_\ell/v_A)^2 \ell_\| \sim L_i\left({{\ell_\perp}\over{L_i}}\right)^{4/3}
M_A^{4/3}, \ee
where we have substituted from (\ref{Lambda}),(\ref{vl}) for $\ell_\|$ and $\delta u_\ell/v_A.$ 
Alternatively, one can just substitute for $\ell_\|$ from (\ref{Lambda}) into (\ref{DB3}).
The equation 
\be {{d}\over{ds}}\ell_\perp^2\sim D_\perp^B(\ell) \sim L_i\left({{\ell_\perp}\over{L_i}}\right)^{4/3}
M_A^{4/3}, \ee
is then easily solved to give eq.(\ref{B-rich}) in the text for $L_i>s\gg \ell_\perp^{(0)}.$ 
This result was already obtained for $M_A=1$ by \cite{Skillingetal74} in a discussion of cosmic-ray 
diffusion, assuming a simple isotropic Kolmogorov hydrodynamic model of magnetized turbulence.
In the weak turbulence regime one has instead 
$\ell_\|=L_i$ a constant, which substituted into (\ref{DB3}) gives 
\be  {{d}\over{ds}}\ell_\perp^2\sim D_\perp^B(\ell) \sim L_i M_A^4 \ee
and this is solved to give the diffusive result (\ref{B-diff}) in the text.

\section{Neglect of the Hall Term in Turbulent Reconnection}

All of the nonideal terms in the generalized Ohm's law (\ref{gen-ohm})---Ohmic resistivity, Hall term, 
pressure tensor, electron inertia, etc.---are expected to be insignificant in a long turbulent inertial range,
at sufficiently large length-scales. Thus, they do not alter the predictions of the LV99 theory for the 
reconnection rates of large-scale flux structures. We shall here present more detailed analytical arguments 
for this thesis,  in the specific example of the incompressible Hall MHD (HMHD) equations. As remarked
in the text, the HMHD dynamics is valid in a literal sense for almost no astrophysical plasmas. However, 
it has  been used in many reconnection studies  as a simple fluid model that exemplifies the Hall effect
\citep{Shayetal98, Shayetal99,Wangetal00,Birnetal01,Drake01,Malakitetal09,Cassaketal10}, so that 
it is a useful example to consider in this respect.

The incompressible Hall MHD equations can be written as
\be (\partial_t \bu+\bu\bdot\grad)\bu=-\grad p + \bj\btimes \bb + \nu \triangle \bu, \lb{B1} \ee
\be  \partial_t\bb = \grad\btimes [(\bu-\delta_i\bj)\btimes \bb] +\lambda \triangle \bb, \lb{A7} \ee
with $\grad\bdot\bu=\grad\bdot\bb=0$ and $\bj=\grad\btimes\bb$. Here we have introduced 
the Alfv\'en variable $\bb=\bB/\sqrt{4\pi\rho}$  with units of velocity and $\delta_i=c/\omega_{p,i}$
is the ion skin depth.  The above system has two inviscid constants of motion that cascade 
to small scales, the total energy
\be E=\frac{1}{2}\int d^3x\, [|\bu(\bx)|^2+|\bb(\bx)|^2] \lb{HMD-energy} \ee
and the cross-helicity 
\be H_C=\int d^3x\, \bu(\bx)\cdot[\bb(\bx)+\frac{1}{2}\delta_i \grad\btimes \bu(\bx)] \lb{HMD-HC}, \ee
as well as the magnetic helicity which cascades to large scales. It is not our purpose here to 
give an exhaustive account of the turbulent cascade phenomenology of the HMHD model. Instead, 
we just wish to establish the simple fact that the Hall term modifies turbulence properties only at 
length-scales $\ell\lesssim \delta_i,$ whereas length-scales $\ell\gtrsim \delta_i$ will behave as 
in standard MHD turbulence without the Hall term.

Our first argument is based on a spatial coarse-graining approach in an Eulerian formulation 
(e.g. see \cite{EyinkAluie06}). The HMHD induction equation coarse-grained (low-pass filtered) 
at length-scale $\ell$ becomes
\be \partial_t\OL{\bb}_\ell = \grad\btimes [(\OL{\bu}_\ell-\delta_i\obj_\ell)\btimes \OL{\bb}_\ell 
        +\boepsilon_\ell-\lambda\obj_\ell], \lb{A8} \ee 
where the total contribution to the turbulent subscale EMF is 
\be \boepsilon_\ell = \OL{(\bu_e\btimes\bb)_\ell}-\OL{\bu}_{e\,\ell}\btimes\OL{\bb}_\ell.
\lb{A9} \ee        
and $\bu_e\equiv \bu-\delta_i\bj$ is the electron fluid velocity. It is easy to see that 
\be \frac{\delta_i\obj_\ell}{\OL{\bu}_\ell}\simeq \frac{\delta_i \delta b(\ell)/\ell}{u_L} \simeq 
        \frac{\delta_i}{\ell} M_A^{1/3}\left(\frac{\ell}{L_i}\right)^{1/3}, \lb{A10} \ee
where the last expression uses GS95 theory to obtain the $1/3$ powers of $M_A$ and 
$\ell/L_i.$ However, a similar expression will hold in any theory of HMHD turbulence 
in which $\delta b(\ell)\sim v_A M_A^q (\ell/L_i)^p$ with $q\geq 1$ and $p>0.$ Hence
$\delta_i\OL{\j}_\ell\ll\OL{u}_\ell$ for $\delta_i\ll \ell\ll L_i,$ so that the coarse-grained dynamics 
in HMHD turbulence at those scales is identical in form to standard MHD. The only difference 
is that the EMF (\ref{A9}) has an additional contribution from the Hall term 
\be \boepsilon^{Hall}_\ell = -\delta_i [\OL{(\bj\btimes\bb)_\ell}-\obj_\ell \btimes\OL{\bb}_\ell]  
     = \frac{1}{2}\delta_i\grad[{\rm tr}(\botau_{M\,\ell})]- \delta_i\grad\bdot\botau_{M\,\ell}, 
     \lb{A11} \ee
where  
$\botau_{M\,\ell} = \OL{(\bb\bb)_\ell}-\OL{\bb}_\ell\OL{\bb}_\ell$      
is the turbulent Maxwell stress. (Notice incidentally that the gradient term in (\ref{A11}) 
does not contribute at  all when substituted into the curl in (\ref{A8}).) The total EMF 
in (\ref{A9}) is thus $\boepsilon_\ell=\boepsilon_\ell^{MHD}
+\boepsilon_\ell^{Hall}$ including the standard MHD contribution $\boepsilon^{MHD}_\ell
=\OL{(\bu\btimes\bb)_\ell}-\OL{\bu}_\ell\btimes\OL{\bb}_\ell$. It is easy to estimate 
the size of these various terms as 
\be  \boepsilon^{MHD}\sim \delta u(\ell)\delta b(\ell),\,\,\,\,
     \boepsilon^{Hall}\sim \delta_i\frac{\delta b^2(\ell)}{\ell}, \,\,\,\,
     \OL{\bE}^{Ohm}_\ell =\lambda\obj_\ell\sim \lambda \frac{\delta b(\ell)}{\ell} \ee
These estimates are rigorous upper bounds (cf. \cite{Eyink05}) but they must be 
good order of magnitude estimates of the terms as well, in order to allow constant fluxes of 
energy and cross helicity to small scales. In that case we must interpret $\ell$ in the upper 
bounds as $\ell_\perp,$ the scale perpendicular to the mean magnetic field\footnote{
Note that energy flux to small scales is given in HMHD by 
$$ \Pi^E_\ell=-\botau_\ell\bdots\grad\OL{\bu}_\ell-\boepsilon_\ell\bdot\obj_\ell $$
and cross-helicity flux by
$$ \Pi^C_\ell =-\botau_\ell\bdots[\grad\OL{\bb}_\ell+\delta_i\grad\OL{\bomega}_\ell]-
        \OL{\bomega}_\ell\bdot\boepsilon_\ell, $$
where $\bomega=\grad\btimes\bu$ is the vorticity and $\botau_\ell=\botau_{R\,\ell}-\botau_{M\,\ell}$ is 
the total stress, both Maxwell stress and Reynolds stress  $\botau_{R\,\ell} = \OL{(\bu\bu)_\ell}
-\OL{\bu}_\ell\OL{\bu}_\ell.$ At scales $\ell>\delta_i,$ the terms without $\delta_i$ dominate
and one gets constant energy and cross-helicity fluxes assuming GS95 scaling $\delta u(\ell)\sim
\delta b(\ell)\sim (\varepsilon \ell_\perp)^{1/3}.$} Instead at scales $\ell<\delta_i$ 
the terms proportional to $\delta_i$ dominate, so that 
$$ \Pi^E_\ell\sim -\obj_\ell\bdot\boepsilon_\ell^{Hall}\sim \delta_i\frac{\delta b^3(\ell)}{\ell_\perp^2}, \,\,\,\,\,
\Pi^C_\ell \sim -\delta_i \grad\OL{\bomega}_\ell\bdots\botau_\ell-
        \OL{\bomega}_\ell\bdot\boepsilon_\ell^{Hall}\sim \delta_i\frac{\delta u(\ell)}{\ell_\perp^2}
        [\delta u^2(\ell)+\delta b^2(\ell)]. $$
One thus gets constant fluxes with the scaling
$$ \delta u(\ell)\sim \delta b(\ell) \sim (\varepsilon \ell_\perp^2/\delta_i)^{1/3}, $$
in agreement with the numerical results of \cite{ShaikhShukla09}. 
As a matter of fact, the precise scaling at length-scales $\ell<\delta_i$ will not affect our conclusions.
As usual, the large-scale Ohmic electric field is negligible compared with the MHD turbulent 
EMF whenever $\ell\delta u(\ell)\gtrsim \lambda,$ or $\ell\gtrsim \ell_\eta^\perp=(\lambda^3/\varepsilon)^{1/4}$
assuming GS95\footnote{For completeness we remark that the Ohmic field is also negligible with
respect to $\boepsilon^{Hall}$ whenever $\delta_i\delta b(\ell)\gtrsim \lambda$. This defines 
another resistive scale $\ell^{Hall\,\perp}_\eta$ which is greater or smaller than $\delta_i$ according
as $\ell_\eta^\perp$ is greater or smaller than $\delta_i.$ In the latter case, $\ell^{Hall\,\perp}_\eta$ is
the true, physical resistive scale, not $\ell_\eta^\perp$}.
However, one can also see that $\boepsilon^{Hall}_\ell$ is negligible compared 
with $\boepsilon^{MHD}_\ell$ whenever $\ell\gtrsim\delta_i.$
Here we have assumed that 
$\delta b(\ell)\simeq\delta u(\ell),$ which is true in GS95 but also in any theory of MHD turbulence 
which predicts equipartition of kinetic and magnetic energies at small scales.  These considerations 
show that the coarse-grained dynamics in Hall  MHD turbulence reduce to those in standard MHD 
turbulence at length-scales $\ell\gtrsim \delta_i.$ This conclusion is in agreement with the 
numerical study of \cite{DmitrukMatthaeus06}. They compared results of Hall and standard 
MHD simulations at kinetic and magnetic Reynolds numbers of 1000, with $\delta_i$ for the 
Hall simulation chosen about 4 times greater than the resistive dissipation scale. They found 
virtually identical results in the two simulations at scales greater than $\delta_i$ for all of the 
various quantities, such as velocities, magnetic fields and electric fields.         

Now let us consider the Lagrangian description of HMHD turbulence.
Stochastic flux-freezing holds also in resistive Hall MHD, with magnetic field-lines 
stochastically frozen-in to the electron fluid velocity $\bu_e$ \citep{Eyink09}.  
Thus, the lines can be regarded to ``move'' according to the stochastic equation 
\be \frac{d\bx}{dt}=\bu(\bx,t)-\delta_i\bj(\bx,t)+\sqrt{2\lambda}\boeta(t) \ee 
At separations $\ell\gg \ell_\eta^\perp$ where the direct effects of  the resistive term
$\sqrt{2\lambda}\boeta(t)$ can be neglected,  the distance $\ell_\perp(t)=\sqrt{\langle y^2(t)\rangle}$ 
that field-lines advect apart in the field-perpendicular direction is obtained from
\be \frac{d}{dt}\ell^2_\perp  \sim D^e_\perp(\ell) \lb{A13} \ee
with 
\be D^e_\perp(\ell)=\int_{-\infty}^0 dt\,\langle \delta \bu_{e\,\perp}(\Bell)\bdot
\delta \bu_{e\,\perp}(\Bell,t)\rangle \lb{A14}  \ee
The electron velocity increment in (\ref{A14}) gets a contribution $\delta \bu(\Bell)$
from the bulk plasma velocity and another contribution $\delta_i\, \delta \bj(\Bell)$ from the Hall 
term. We would like to show that the Hall contribution is negligible whenever $\ell\gg\delta_i.$
This can be argued as follows:

First, we can estimate from energy balance alone without any other  assumptions 
that $\varepsilon\sim \lambda j_{rms}^2$  and thus $j\sim j_{rms}=(\varepsilon/\lambda)^{1/2}.$
This allows us compare the typical size of the Hall velocity with the flow velocity as
\be \frac{\delta_i j}{u}\sim \frac{\delta_i (\varepsilon/\lambda)^{1/2}}{u_L}=
       \frac{\delta_i}{\ell_\eta^\perp}\cdot S_L^{-1/4}, \lb{ratio} \ee
where we have used $\varepsilon=u_L^4/v_AL_i$ and we have {\it defined} $\ell_\eta^\perp
\equiv (\lambda^3/\varepsilon)^{1/4},$ without the assumption that it plays any dynamical role 
as a ``dissipation scale.'' Now let the Lundquist number $S_L$ increase by increasing $L_i$ 
with $\varepsilon$ fixed, and thus also $\lambda,\ell_\eta^\perp$ and the ratio 
\be \alpha\equiv \delta_i/\ell_\eta^\perp \lb{alpha} \ee
all fixed as well. It follows from (\ref{ratio}) that $\delta_i j\ll u$ always when $\alpha<1$ but 
even when $\alpha>1$ if $S_L\gg \alpha^4.$ We therefore expect that the Hall contribution
to the 2-particle diffusivity is negligible compared with the advective MHD contribution at high 
Lundquist numbers.  

Physically, the Hall term is very short-range correlated in both space and time, 
so that it acts on a pair of Lagrangian particles as two independent Brownian motions with 
an effective 1-particle diffusivity. This  can be shown more formally, by observing that 
\be \langle \delta\bj_\perp(\Bell)\bdot\delta\bj_\perp(\Bell,t)\rangle=
       2[ \langle\bj_\perp(\bzed)\bdot\bj_\perp(\bzed,t)\rangle- \langle\bj_\perp(\bzed)\bdot\bj_\perp(\Bell,t)\rangle]
       \lesssim 2 \langle\bj_\perp(\bzed)\bdot\bj_\perp(\bzed,t)\rangle, \ee
since the second term in the square brackets is a {\it decaying} term for increasing $\ell$
\footnote{Note that $\langle j_i(\br)j_k({\bf 0})\rangle
\sim \triangle \langle \delta b_i(\br)\delta b_k(\br)\rangle \sim r^{2(h_B-1)}$ decays for large $r$
whenever the magnetic-field increment $\delta \bb(\br)$ has scaling exponent $h_B<1.$}.  
This upper bound is a considerable overestimate for smaller values of $\ell,$  but it suffices 
for our purpose. We can then estimate
\be  \int_{-\infty}^0 dt\, \langle\bj_\perp(\bzed)\bdot\bj_\perp(\bzed,t)\rangle\sim j_{rms}^2\tau_\lambda \ee
where $\tau_\lambda=(\lambda/\varepsilon)^{1/2}=1/j_{rms}$ is the correlation time at the resistive 
dissipation scale. Multiplied by $\delta_i^2,$ this is the Hall contribution to the 1-particle diffusivity. 
We therefore obtain a conservative upper bound on the Hall contribution to the 2-particle 
diffusivity, independent of $\ell,$ as 
\be D^{Hall}_\perp(\ell) \lesssim \delta_i^2j_{rms}^2\tau_\lambda  =\alpha^2 \lambda, \lb{D-Hall} \ee
where we have used (\ref{ratio}) and (\ref{alpha}).

Finally, we compare with the advective 
contribution for $\ell>\delta_i.$ From the Eulerian argument presented earlier in the Appendix,
we expect that GS95 scaling works on those length-scales. We can thus estimate for 
$\ell_\perp>\delta_i$ that
\be D_\perp^{MHD}(\ell)\sim \ell_\perp\delta u(\ell)\sim (\varepsilon \ell_\perp^4)^{1/3}
\sim \alpha^{4/3}\lambda \left(\frac{\ell_\perp}{\delta_i}\right)^{4/3} \ee
where we have used $\lambda=(\varepsilon\ell_\eta^{\perp\,4})^{1/3}$ and the definition 
of $\alpha.$ We conclude that 
\be  D_\perp^{Hall}(\ell)/D_\perp^{MHD}(\ell)
\lesssim \alpha^{2/3}\left(\frac{\delta_i}{\ell_\perp}\right)^{4/3} \ll 1\ee
at least when $\ell_\perp\gg \sqrt{\alpha}\delta_i$.  Thus, Richardson diffusion of field-lines 
in HMHD turbulence is exactly the same as for MHD turbulence at perpendicular separations 
$\ell_\perp$ much greater than the ion skin depth $\delta_i.$
   
Now let us reconsider the problem of {\it large-scale} reconnection treated in section 5,
including the Hall effect within the HMHD model. Define $\Delta=L_iM_A^2$ taking $L_i\simeq L_x.$
We shall make the essential assumption that $\delta_i\ll \Delta.$ This assumption is almost always 
true in astrophysical  reconnection problems. For example, using $\delta_i=\rho_i/\sqrt{\beta_i}$ 
and the data in Table 1, we can see that $\delta_i/\Delta\sim 10^{-13}$ for the warm ISM, 
$\sim10^{-6}$ for post-CME current sheets and $\sim 10^{-1}$ for solar wind impinging 
on the magnetosphere. Only in the latter case is the condition not extremely well satisfied.  
Let us then repeat the arguments in section 5 taking into account the Hall term.  Consider first 
the derivation of LV99 based on the stochastic wandering of magnetic field-lines as one traces 
points of increasing arc-length $s$ along the lines.  At small initial separations, the Hall term is 
important and the rms transverse distance $\ell_\perp(s)=\sqrt{\langle y^2(s)\rangle}$ grows 
as $\ell_\perp(s)\sim (s^3/\delta_iL_i)M_A^4$ \footnote{
Obtained by solving $d\ell_\perp/ds\sim db(\ell)/v_A=(1/v_A)(\varepsilon\ell_\perp^2/\delta_i)^{1/3}$.}.
As soon as neighboring lines have separated to $\ell_\perp \gtrsim \delta_i,$ then their further separation 
is the same as in standard MHD turbulence without the Hall term. However, to reach the separation
$\sim \delta_i$ one must move along the field-lines a distance of only $s\sim L_i^*\equiv L_i\left(
\frac{\delta_i}{\Delta}\right)^{2/3}\ll L_i$ when $\delta_i\ll \Delta.$ In that case, the formula 
(\ref{B-wand}) in section 5.1 for $\langle y^2(s)\rangle$ still holds, if $s\gg L_i^*.$ It therefore follows 
that the LV99 estimate for the width of the reconnection layer and the formula (\ref{vr-LV99}) for the 
reconnection velocity will hold, as long as $\delta_i \ll \Delta,$ since most of the separation of lines 
will be achieved for $s\geq L_i^*.$ The same conclusion can be derived from the Lagrangian 
perspective. For small initial separations $\ell_\perp<\delta_i$ the lines will separate laterally 
as $\ell_\perp(t) \sim (\delta_i^2\varepsilon t^3)^{1/4}$ \footnote{Obtained 
by solving $d\ell_\perp/dt=\delta_i\delta b(\ell)/\ell_\perp=(\varepsilon \delta_i^2
/\ell_\perp)^{1/3}$}. However, after a time $t\sim t_c^*=(\delta_i^2/\varepsilon)^{1/3}=
\left(\frac{\delta_i}{\Delta}\right)^{2/3}(L_i/v_A)$ much shorter than $t_c=L_i/v_A$ 
the pairs of lines will reach the separation $\ell_\perp(t)\sim \delta_i$ and GS95 scaling 
will take over. The formula (\ref{2part}) for $\langle y^2(t)\rangle$ that follows from GS95 
will hold as soon as $t>t_c^*$ and the bulk of the separation of lines will occur after this 
time if $\delta_i\ll \Delta.$  Thus, the estimate (\ref{Delta-LV99}) for the width of the 
reconnection layer holds just as before, with only a negligible correction due to finite $\delta_i$

We thus reach the important conclusion that {\it the Hall term is irrelevant to determining 
the rate of global reconnection in the presence of high-Reynolds-number plasma turbulence.}
At most a modest enhancement of the reconnection rate is obtained, but vanishingly small 
when $\delta_i\ll \Delta.$  We should mention that a similar conclusion was reached by
\cite{Smithetal04} on the basis of numerical simulations. However, that study can be 
fairly criticized because the simulations were performed in 2D and at such low Reynolds 
numbers that most of the measured reconnection rates (in a transient regime) were 
smaller than the steady-state Sweet-Parker rates. Our theoretical arguments, on the 
other hand, are valid in 3D and at arbitrarily high Reynolds numbers.  This is not to say 
that the Hall term will have no effect whatsoever, but these will be confined to very small 
scales, e.g. modifying the structure of local reconnection events. However, these local 
effects are irrelevant to determining the global rate of large-scale reconnection. 

Similar conclusions can be reached not only for the 2-fluid Hall term, but also for other microscopic plasma 
mechanisms proposed to enhance reconnection speeds. For example, the MHD turbulent 
cascade may be terminated not by Spitzer resistivity but instead by some form of anomalous 
resistivity, e.g. due to ion acoustic instability. The latter produces a large effective magnetic diffusivity 
of order $(m_i/m_e)^{1/2}\delta_e^2\omega_{c,e}$ at scales smaller than $\delta_i/\sqrt{\beta}$ 
(e.g. see \cite{GaleevSagdeev84} and \cite{Kulsrud05}, section 14.7.) But larger scales will 
behave as standard MHD turbulence and the LV99 predictions will again hold as long as 
$\delta_i\ll \Delta.$ This was verified in the simulations of 
\cite{Kowaletal09}, who found no effect of anomalous resistivity on reconnection 
speed. Turbulence is such a powerful accelerator of magnetic line-separation that it completely 
dominates microscopic plasma processes of line diffusion, as soon as the magnetic and 
kinetic Reynolds numbers are sufficiently high.




\clearpage


\newpage

\begin{thebibliography}{}

\bibitem[Albright (1999)]{Albright99}
Albright, B.~J.\ 1999,
Phys. Plasmas, 6, 4222
\bibitem[Aluie \& Eyink (2010)]{AluieEyink10}
Aluie, H. \& Eyink, G.~L.\ 2010,
Phys. Rev. Lett., 104, 081101
\bibitem[Alfv\'en (1942)]{Alfven42}
Alfv\'{e}n, H.\ 1942, 
Ark. Mat., Astron. o. Fys., 29B, 1
\bibitem[Alfv\'en (1976)]{Alfven76}
Alfv\'en, H.\  1976, 
J. Geophys. Res., 81, 4019
\bibitem[Armstrong et al.(1995)]{Armstrongetal94} Armstrong, J.~W., 
Rickett, B.~J., \& Spangler, S.~R.\ 1995, \apj, 443, 209 
\bibitem[Axford (1984)]{Axford84}
Axford, W. I.\ 1984, 
in:  Magnetic Reconnection in Space and 
Laboratory Plasmas; Proceedings of the Chapman Conference on Magnetic Reconnection, 
Los Alamos, NM, October 3-7, 1983. American Geophysical Union, Washington, DC,
pp.~1
\bibitem[Balbus 
\& Hawley(1998)]{BalbusHawley98} 
Balbus, S.~A., \& Hawley, J.~F.\ 1998, Reviews of Modern Physics, 70, 1
\bibitem[Bale et al.(2005)]{Baleetal05} Bale, S.~D., Kellogg, 
P.~J., Mozer, F.~S., Horbury, T.~S., \& Reme, H.\ 2005, Physical Review Letters, 94, 215002
\bibitem[Balk (2002)]{Balk02}
Balk, A.\ 2002,  
J. Fluid Mech., 467, 163
\bibitem[Ballesteros-Paredes et al.(2007)]{Ballesteros-Paredesetal06} 
Ballesteros-Paredes, J., Klessen, R.~S., Mac Low, M.-M., 
\& Vazquez-Semadeni, E.\ 2007, Protostars and Planets V, 63 
\bibitem[Batchelor (1950)]{Batchelor50} 
Batchelor, G. K.\ 1950, 
Q. J. R. Meteorol. Soc., 76, 133
\bibitem[Bemporad et al. (2008)]{Bemporad08}
Bemporad, A.\ 2008,
\apj, 689, 572
\bibitem[Beresnyak (2011)]{Beresnyak11} 
Beresnyak, A.\ 2011, 
\prl, 106,  075001
\bibitem[Beresnyak 
\& Lazarian(2006)]{BeresnyakLazarian06} Beresnyak, A., \& Lazarian, A.\ 2006, \apjl, 640, L175 
\bibitem[Beresnyak 
\& Lazarian(2008)]{BeresnyakLazarian08} Beresnyak, A., \& Lazarian, A.\ 2008, \apj, 682, 1070 
\bibitem[Beresnyak 
\& Lazarian(2009)]{BeresnyakLazarian09} Beresnyak, A., \& Lazarian, A.\ 2009, \apj, 702, 1190 
\bibitem[Beresnyak 
\& Lazarian(2010)]{BeresnyakLazarian10} Beresnyak, A., \& Lazarian, A.\ 2010, \apjl, 722, L110 
\bibitem[Bernard et al. (1998)]{Bernardetal98} 
Bernard, D., Gaw\c{e}dzki, K.  \& Kupiainen,  A.\ 1998, 
J. Stat. Phys., 90, 519
\bibitem[Bettarini \& Lapenta (2009)]{BettariniLapenta09}
Bettarini, L. \&  Lapenta, G.\ 2010,
A\&A 518, A57 
\bibitem[Bhattacharjee 
\& Hameiri(1986)]{BhattacharjeeHameiri86} Bhattacharjee, A., \& Hameiri, E.\ 1986, 
Physical Review Letters, 57, 206 
\bibitem[Bhattacharjee et al. (1999)]{Bhattacharjeeetal99}
Bhattacharjee, A., Ma, Z.~W. \& Wang, X.\ 1999,
\jgr, 104, 14543
\bibitem[Bhattacharjee et al. (2009)]{Bhattacharjeeetal09}
Bhattacharjee, A.,  Huang, Y.-M.,  Yang, H.  \& Rogers, B.\ 2009,
Phys. Plasmas, 16, 112102
\bibitem[Bieber et al.(1994)]{Bieberetal94} Bieber, J.~W., 
Matthaeus, W.~H., Smith, C.~W., Wanner, W., Kallenrode, M.-B., 
\& Wibberenz, G.\ 1994, \apj, 420, 294
\bibitem[Birn et al.(2001)]{Birnetal01} Birn, J., et al.\ 2001, 
\jgr, 106, 3715 
\bibitem[Biskamp \& Schwarz (2001)]{BiskampSchwarz01}
Biskamp, D. \& Schwarz, E.\ 2001,
Physics of Plasmas, 8, 3282
\bibitem[Biskamp(2003)]{Biskamp03} Biskamp, D.\ 2003, 
Magnetohydrodynamic Turbulence, by Dieter Biskamp, pp.~310.~ISBN 
0521810116.~Cambridge, UK: Cambridge University Press, September 2003.,
\bibitem[Boffetta \& Sokolov (2002)]{BoffettaSokolov02}
Boffetta, G. \& Sokolov, I.~M.\ 2002,
Phys. Rev. Lett., 88, 094501
\bibitem[Boldyrev (2005)]{Boldyrev05} 
Boldyrev, S.\ 2005,  
\apj, 626, L37
\bibitem[Boldyrev (2006)]{Boldyrev06}
Boldyrev, S.  2006,
Phys. Rev. Lett., 96, 115002 
\bibitem[Braginsky (1965)]{Braginsky65}
Braginsky, S.~I.\ 1965. Rev. Plasma Phys. 1, 205.
\bibitem[Busse \& M\"uller (2008)]{BusseMuller08}
Busse, A. \& M\"uller, W.-C.\ 2008,
Astron. Nachr., 329, 714
\bibitem[Caflisch et al.(1997)]{Caflischetal97} Caflisch, R.~E., Klapper, I., 
\& Steele, G.\ 1997, Communications in Mathematical Physics, 184, 443 
\bibitem[Cassak et al. (2006)]{Cassaketal06}
Cassak, P.~A., Drake, J.~F. and Shay, M.~A.\ 2006. 
\apj,  644, L145.
\bibitem[Cassak \& Shay (2007)]{CassakShay07}
Cassak, P.~A. \& Shay, M.~A.\ 2007,
Phys. Plasmas, 14, 102114
\bibitem[Cassak et al. (2009)]{Cassaketal09}
Cassak, P.~A., Shay, M.~A. and Drake, J.~F.\ 2009.
Phys. Plasmas 16, 120702.
\bibitem[Cassak et al.(2010)]{Cassaketal10} Cassak, P.~A., Shay, 
M.~A., \& Drake, J.~F.\ 2010, Physics of Plasmas, 17, 062105
\bibitem[Castaing et al.(1990)]{1990PhyD...46..177C} Castaing, B., Gagne, 
Y., \& Hopfinger, E.~J.\ 1990, Physica D Nonlinear Phenomena, 46, 177 
\bibitem[Celani \& Vergassola (2001)]{CelaniVergassola01}
Celani, A. \& Vergassola, M.\ 2001,
Phys. Rev. Lett., 86, 424
\bibitem[Chandran(2005)]{Chandran05} Chandran, B.~D.~G.\ 2005, 
\apj, 632, 809
\bibitem[Chandran(2008)]{Chandran08} Chandran, B.~D.~G.\ 2008, 
\apj, 685, 646 
\bibitem[Chaves et al. (2003)]{Chavesetal03}
Chaves, M., Gaw\c{e}dzki, K., Horvai, P., Kupiainen, A.  \& Vergassola, M. 2003,
J. Stat. Phys., 113, 643
\bibitem[Che et al. (2011)]{Cheetal11}
Che, H., Drake, J.~F., \& Swisdak, M.\ 2011,  Nature, to appear
\bibitem[Chepurnov(1998)]{Chepurnov98} Chepurnov, A.~V.\ 1998, Astronomical and Astrophysical Transactions, 17, 281
\bibitem[Chepurnov 
\& Lazarian(2010)]{ChepurnovLazarian10} Chepurnov, A., \& Lazarian, A.\ 2010, \apj, 710, 853
\bibitem[Cho 
\& Lazarian(2002)]{ChoLazarian02} Cho, J., \& Lazarian, A.\ 2002, Physical Review Letters, 88, 245001 
\bibitem[Cho 
\& Lazarian(2003)]{ChoLazarian03} Cho, J., \& Lazarian, A.\ 2003, \mnras, 345, 325 
\bibitem[Cho et al.(2003)]{Choetal03} Cho, J., Lazarian, A., 
\& Vishniac, E.~T.\ 2003, Turbulence and Magnetic Fields in Astrophysics, 614, 56 
\bibitem[Cho \& Lazarian(2004)]{ChoLazarian04} Cho, J., \& Lazarian, A.\ 2004, 
\apjl, 615, L41
\bibitem[Cho 
\& Lazarian(2005)]{ChoLazarian05} Cho, J., \& Lazarian, A.\ 2005, Theoretical and Computational Fluid Dynamics, 19, 127 
\bibitem[Cho et al.(2002)]{Choetal02} Cho, J., Lazarian, A., 
\& Vishniac, E.~T.\ 2002, \apj, 564, 291 
\bibitem[Cho 
\& Vishniac(2000)]{ChoVishniac00} Cho, J., \& Vishniac, E.~T.\ 2000, \apj, 539, 273 
\bibitem[Ciaravella 
\& Raymond(2008)]{CiaravellaRaymond08} Ciaravella, A., \& Raymond, J.~C.\ 2008, \apj, 686, 1372 
\bibitem[Daughton et al.(2006)]{Daughtonetal06} Daughton, W., Scudder, 
J., \& Karimabadi, H.\ 2006, Physics of Plasmas, 13, 072101
\bibitem[Daughton et al.(2008)]{Daughtonetal08} Daughton, W., 
Roytershteyn, V., Albright, B.~J., Bowers, K., Yin, L., 
\& Karimabadi, H.\ 2008, AGU Fall Meeting Abstracts, A1705 
\bibitem[Daughton et al.(2011)]{Daughtonetal11} Daughton, W., 
Roytershteyn, V., Karimabadi, H., Yin, L., Albright, B.~J., Bergen, B., \& 
Bowers, K., \ 2011, Nature Physics, in press.
\bibitem[Davila \& Vassilicos (2003)]{DavilaVassilicos03}
D{\'a}vila, J. \& Vassilicos, J.~C.\ 2003, 
Phys. Rev. Lett., 91, 144501
\bibitem[de Gouveia Dal Pino 
\& Lazarian(2003)]{GouveiadalPinoLazarian03} de Gouveia Dal Pino, E.~M., \&
Lazarian, A.\ 2003, arXiv:astro-ph/0307054 
\bibitem[de Gouveia dal Pino 
\& Lazarian(2005)]{deGouveiadalPinoLazarian05} de Gouveia dal Pino, E.~M., \& 
Lazarian, A.\ 2005, \aap, 441, 845 
\bibitem[Diamond et al. (1984)]{Diamondetal84}
Diamond, P.~H., Hazeltine, R.~D., An, Z.~G., Carreras, B.~A. 
\& Hicks, H.~R.\ 1984, 
Physics of Fluids, 27, 1449
\bibitem[Diamond 
\& Malkov(2003)]{DiamondMalkov03} Diamond, P.~H., \& Malkov, M.\ 2003, 
Physics of Plasmas, 10, 2322 
\bibitem[Dmitruk \& Matthaeus (2006)]{DmitrukMatthaeus06}
Dmitruk, P. \& Matthaeus, W. H. 2006,
Phys. Plasmas, 13, 042307
\bibitem[Draine 
\& Lazarian(1998)]{DraineLazarian98} Draine, B.~T., \& Lazarian, A.\ 1998, \apj, 508, 157 
\bibitem[Drake(2001)]{Drake01} Drake, J.~F.\ 2001, \nat, 410, 
525 
\bibitem[Drake et al.(2006)]{Drakeetal06} Drake, J.~F., Swisdak, 
M., Che, H., \& Shay, M.~A.\ 2006, \nat, 443, 553 
\bibitem[E \& vanden Eijnden (2000)]{EvandenEijnden00}
E, W. \& vanden-Eijnden, E.\ 2000, 
Proc. Natl. Acad. Sci., 97,  8200
\bibitem[E \& vanden Eijnden (2001)]{EvandenEijnden01}
E, W. \& vanden-Eijnden, E.\ 2001, 
Physica D, 152-153, 636
\bibitem[Elmegreen
\& Scalo(2004)]{ElmegreenScalo04} Elmegreen, B.~G., \& Scalo, J.\ 2004, \araa, 42, 211 
\bibitem[En{\ss}lin 
\& Vogt(2006)]{EnsslinVogt06} En{\ss}lin, T.~A., \& Vogt, C.\ 2006, \aap, 453, 44
\bibitem[Esquivel 
\& Lazarian(2005)]{EsquivelLazarian05} Esquivel, A., \& Lazarian, A.\ 2005, \apj, 631, 320
\bibitem[Eyink (2005)]{Eyink05}
Eyink, G.~L.\ 2005, 
Physica D, 207, 91
\bibitem[Eyink(2006)]{Eyink06}
Eyink, G.~L.\ 2006, 
J. Fluid Mech., 549, 159
\bibitem[Eyink (2007)]{Eyink07}
Eyink, G. L.\ 2007,  
Phys. Lett. A, 368, 486
\bibitem[Eyink(2008)]{Eyink08} Eyink, G.~L.\ 2008, Physica D 
Nonlinear Phenomena, 237, 1956
\bibitem[Eyink (2009)]{Eyink09}
Eyink, G. L.\ 2009, 
J. Math. Phys., 50,  083102 
\bibitem[Eyink (2010)]{Eyink10}
Eyink, G. L.\ 2010,
Phys. Rev. E, 82,  046314
\bibitem[Eyink (2011)]{Eyink11}
Eyink, G. L.\ 2011,
Phys. Rev. E 83, 056405
\bibitem[Eyink \& Aluie (2006)]{EyinkAluie06}
Eyink, G. L.  \& Aluie, H.\ 2006,   
Physica D, 223, 82
\bibitem[Eyink \& Neto (2010)]{EyinkNeto10}
Eyink, G. L. \& Neto, A. F.\ 2010,
New J. Phys., 12, 023021
\bibitem[Falkovich et al. (2001)]{Falkovichetal01}
Falkovich, G., Gaw{\c e}dzki, K. \& Vergassola, M.\ 2001, 
Rev. Mod. Phys., 73, 913
\bibitem[Ferri{\`e}re(2001)]{Ferriere01} Ferri{\`e}re, K.~M.\ 
2001, Reviews of Modern Physics, 73, 1031
\bibitem[Fitzpatrick (2011)]{Fitzpatrick11}
Fitzpatrick, R.\ 2011, ``Introduction to Plasma Physics'', online lecture notes, URL:
{\tt http://farside.ph.utexas.edu/teaching/plasma/plasma.html}
\bibitem[Fox et al.(2005)]{Foxetal05} Fox, D.~B., et al.\ 2005, 
\nat, 437, 845 
\bibitem[Freidlin \& Wentzell (1984)]{FreidlinWentzell84}
Freidlin, M.~I. \& Wentzell, A.~D.\ 1984, Random Perturbations of Dynamical Systems. 
Springer, New York.
\bibitem[Frisch et al. (1975)]{Frischetal75}
Frisch, U., Pouquet, A., Leorat, J. \& Mazure, A.\ 1975,
J. Fluid Mech., 68, 769
\bibitem[Fyfe \& Montgomery (1976)]{FyfeMontgomery76}
Fyfe, D. \& Montgomery, D.\ 1976, 
Journal of Plasma Physics, 16, 181
\bibitem[Galama et al.(1998)]{Galamaetal98} Galama, T.~J., et al.\ 
1998, \nat, 395, 670 
\bibitem[Galeev \& Sagdeev (1984)]{GaleevSagdeev84}
Galeev, A.~A. \& Sagdeev, R.~Z.\ 1984. 
in: Basic Plasma Physics: Selected Chapters, Handbook of Plasma Physics, Vol. 1,
eds. A.~A.~Galeev \& R.~N.~Sudan, North-Holland, Amsterdam. pp.271-303
\bibitem[Galsgaard 
\& Nordlund(1997a)]{GalsgaardNordlund97a} Galsgaard, K., \& Nordlund, 
{\AA}.\ 1997, \jgr, 102, 219
\bibitem[Galsgaard \& Nordlund (1997b)]{GalsgaardNordlund97b}
Galsgaard, K. \& Nordlund, \AA.\ 1997,
\jgr, 102, 231
\bibitem[Galtier et al. (2000)]{Galtieretal00}
Galtier, S., Nazarenko, S. V., Newell, A. C. \& Pouquet, A.\ 2000,
J. Plasma Phys., 63, 447
\bibitem[Galtier et al. (2002)]{Galtieretal02}
Galtier, S., Nazarenko, S. V., Newell, A. C.  \&  Pouquet, A.\ 2002,
\apjl , 564, L49
\bibitem[Gaw\c{e}dzki \& Vergassola (2000)]{GawedzkiVergassola00} 
Gaw\c{e}dzki, K. \& Vergassola, M.\ 2000, 
Physica D, 138, 63
\bibitem[Gerrard 
\& Hood(2003)]{GerrardHood03} Gerrard, C.~L., \& Hood, A.~W.\ 2003, 
\solphys, 214, 151
\bibitem[Goldreich \& Sridhar (1995)]{GoldreichSridhar95}
Goldreich, P. \& Sridhar, S.\ 1995
\apj, 438, 763
\bibitem[Goldreich \& Sridhar (1997)]{GoldreichSridhar97}
Goldreich, P.  \& Sridhar, S.\ 1997,
\apj , 485, 680
\bibitem[Gogoberidze(2007)]{Gogoberidze07} Gogoberidze, G.\ 2007, 
Physics of Plasmas, 14, 022304 
\bibitem[Gomez et al. (1999)]{Gomezetal99} 
Gomez, T., Politano, H. \& Pouquet, A.\ 1999,
Physics of Fluids, 11, 2298
\bibitem[Goto \& Vassilicos (2004)]{GotoVassilicos04}
Goto, S. \& Vassilicos, J.~C.\ 2004,
New J. Phys., 6, 65
\bibitem[Greene (1993)]{Greene93}
Greene, J.~M.\ 1993,
Phys. Fluids B, 5, 2355
\bibitem[Hameiri 
\& Bhattacharjee(1987)]{HameiriBhattacharjee87} 
Hameiri, E., \& Bhattacharjee, A.\ 1987, Physics of Fluids, 30, 1743 
\bibitem[Hartman (2002)]{Hartman02}
Hartman, P.\, 2002,  
Ordinary Differential Equations, 2nd Ed., 
Soc. Indust. Appl. Math. 
\bibitem[Higdon(1984)]{Higdon84} Higdon, J.~C.\ 1984, \apj, 285, 
109
\bibitem[Hornig \& Schindler (1996)]{HornigSchindler96}
Hornig, G. \& Schindler, K.\ 1996,
Phys. Plasmas, 3, 781
\bibitem[Huang \& Bhattacharjee (2010)]{HuangBhattacharjee10}
Huang, Y.-M. and Bhattacharjee, A.\ 2010.
Phys. Plasmas 17, 062104.
\bibitem[Innes et al.(1997)]{Innesetal97} Innes, D.~E., Inhester, 
B., Axford, W.~I., \& Wilhelm, K.\ 1997, \nat, 386, 811 
\bibitem[Iroshnikov (1964)]{Iroshnikov64}
Iroshnikov, R. S.\ 1964,
Soviet Astron., 7, 566
\bibitem[Jacobson 
\& Moses(1984)]{Jacobson84} Jacobson, A.~R., \& Moses, R.~W.\ 1984, \pra, 29, 3335
\bibitem[Jokipii (1973)]{Jokipii73}
Jokipii, J.~R.\ 1973, 
\apj, 183, 1029. 
\bibitem[Kim 
\& Diamond(2001)]{KimDiamond01} 
Kim, E.-j., \& Diamond, P.~H.\ 2001, \apj, 556, 1052 
\bibitem[Kowal 
\& Lazarian(2010)]{KowalLazarian10} Kowal, G., \& Lazarian, A.\ 2010, \apj, 720, 742 
\bibitem[Kowal et al.(2009)]{Kowaletal09} Kowal, G., Lazarian, A., 
Vishniac, E.~T., \& Otmianowska-Mazur, K.\ 2009, \apj, 700, 63 
\bibitem[Kraichnan (1965)]{Kraichnan65}
Kraichnan, R. H.\ 1965, 
Phys. Fluids, 8, 1385
\bibitem[Kraichnan (1966)]{Kraichnan66}
Kraichnan, R. H.\ 1966,
Phys. Fluids, 9, 1937
\bibitem[Kulpa-Dybe{\l} et 
al.(2010)]{Kulpa-Dubeletal10} Kulpa-Dybe{\l}, K., Kowal, G., 
Otmianowska-Mazur, K., Lazarian, A., \& Vishniac, E.\ 2010, \aap, 514, A26
\bibitem[Kulsrud (1983)]{Kulsrud83}
Kulsrud, R.\ 1983, 
Handbook of Plasma Physics, eds. M. N. 
Rosenbluth \& R. Z. Sagdeev (North Holland, New York) 
\bibitem[Kulsrud (2005)]{Kulsrud05}
Kulsrud, R.\ 2005 
Princeton University Press,  Princeton, NJ
\bibitem[Kupiainen (2003)]{Kupiainen03}
Kupiainen, A.\ 2003,
Ann. Henri Poincar\'{e}, 4, Suppl. 2, S713
\bibitem[Lapenta (2008)]{Lapenta08}
Lapenta, G.\ 2008, 
Phys. Rev. Lett., 100, 235001
\bibitem[Lapenta 
\& Bettarini(2011)]{LapentaBettarini11} Lapenta, G., \& Bettarini, L.\ 2011, arXiv:1102.4791 
\bibitem[Lazarian et al.(2002)]{Lazarianetal01} Lazarian, A., 
Pogosyan, D., 
\& Esquivel, A.\ 2002, Seeing Through the Dust: The Detection of HI 
and the Exploration of the ISM in Galaxies, 276, 182 
\bibitem[Lazarian et al.(2004)]{Lazarianetal04} Lazarian, A., 
Vishniac, E.~T., \& Cho, J.\ 2004, \apj, 603, 180 
\bibitem[Lazarian(2005)]{Lazarian05} Lazarian, A.\ 2005, Magnetic 
Fields in the Universe: From Laboratory and Stars to Primordial 
Structures., 784, 42 
\bibitem[Lazarian(2006)]{Lazarian06} Lazarian, A.\ 2006, \apjl, 
645, L25 
\bibitem[Lazarian 
\& Desiati(2010)]{LazarianDesiati10} Lazarian, A., \& Desiati, P.\ 2010, \apj, 722, 188 
\bibitem[Lazarian et al. (2010)]{Lazarianetal10}
Lazarian, A., Kowal, G. \& de Gouveia dal Pino, E. 2010, 
Fast Magnetic Reconnection and Energetic Particle Acceleration, Planetary and Space Science, doi:10.1016/j.pss.2010.07.020
\bibitem[Lazarian et al.(2002)]{Lazarianetal02} Lazarian, A., 
Petrosian, V., Yan, H., 
\& Cho, J.\ 2002, Beaming and Jets in Gamma Ray Bursts, 45 
\bibitem[Lazarian 
\& Opher(2009)]{LazarianOpher09} Lazarian, A., \& Opher, M.\ 2009, \apj, 703, 8 
\bibitem[Lazarian 
\& Vishniac(1999)]{LazarianVishniac99} Lazarian, A., \& Vishniac, E.~T.\ 1999, \apj, 517, 700 
\bibitem[Lazarian\& Vishniac (2000)]{LazarianVishniac00}
Lazarian, A.  \& Vishniac, E.\ 2000, 
Rev. Mex. Astron. AstroÞs. Ser. Conf., 9, 55
{\tt astro-ph/0002067}
\bibitem[Leamon et al.(1998)]{Leamonetal98} 
Leamon, R.~J., Smith, 
C.~W., Ness, N.~F., Matthaeus, W.~H., \& Wong, H.~K.\ 1998, \jgr, 103, 4775
\bibitem[Linton et al. (2001)]{Lintonetal01}
Linton, M.~G., Dahlburg, R.~B. \& Antiochos, S.~K.\ 2001, 
\apj, 553, 905
\bibitem[Lithwick 
\& Goldreich(2001)]{LithwickGoldreich01} Lithwick, Y., \& Goldreich, P.\ 2001, \apj, 562, 279 
\bibitem[Lithwick et al.(2007)]{LithwickGoldreich07} Lithwick, Y., 
Goldreich, P., \& Sridhar, S.\ 2007, \apj, 655, 269 
\bibitem[Loureiro et al. (2009)]{Loureiroetal09}
Loureiro, N. F.,  Uzdensky, D. A. , Schekochihin, A. A.,  
Cowley, S. C.  \& Yousef, T. A.\ 2009, 
Mon. Not. R. Astron. Soc., 399, L146
\bibitem[Lovelace(1976)]{Lovelace76} Lovelace, R.~V.~E.\ 1976, 
\nat, 262, 649 
\bibitem[Majda et al. (1997)]{Majdaetal97}
Majda, A. J., McLaughlin, D. W. \& Tabak, E. G.\ 1997,
J. Nonlinear Sci., 6, 9
\bibitem[Malakit et al.(2009)]{Malakitetal09} Malakit, K., Cassak, 
P.~A., Shay, M.~A., \& Drake, J.~F.\ 2009, \grl, 36, 7107
\bibitem[Maron \& Goldreich(2001)]{MaronGoldreich01} 
Maron, J., \& Goldreich, P.\ 2001, \apj, 554, 1175 
\bibitem[Mason et al. (2006)]{Masonetal06} 
Mason, J., Cattaneo, F.  \& Boldyrev, S.\ 2006,  
Phys. Rev. Lett., 97, 255002 
\bibitem[Masuda et al. (1994)]{Masudaetal94}
Masuda, S., Kosugi, T., Hara, H., Tsuneta, S. \& Ogawara, Y.\ 1994,
\nat, 371, 495
\bibitem[Matthaeus et al.(1983)]{Matthaeusetal83} Matthaeus, W.~H., 
Montgomery, D.~C., 
\& Goldstein, M.~L.\ 1983, Physical Review Letters, 51, 1484
\bibitem[Matthaeus \& Lamkin (1985)]{MatthaeusLamkin85}
Matthaeus, W.~H. \& Lamkin, S.~L.\ 1985, 
Phys. Fluids, 28, 303
\bibitem[Matthaeus \& Lamkin (1986)]{MatthaeusLamkin86}
Matthaeus, W.~H. \& Lamkin, S.~L.\ 1986, 
Phys. Fluids, 29, 2513
\bibitem[Matthaeus et al.(1990)]{Matthaeusetal90} Matthaeus, W.~H., 
Goldstein, M.~L., \& Roberts, D.~A.\ 1990, \jgr, 95, 20673 
\bibitem[McKee 
\& Ostriker(2007)]{McKeeOstriker07} McKee, C.~F., \& Ostriker, E.~C.\ 2007, \araa, 45, 565 
\bibitem[Mininni et al. (2007)]{Mininnietal07}
Mininni, P.-D., Alexakis, A. \& Pouquet, A.\ 2007,
J. Plasma Physics, 73, 377
\bibitem[Mininni \& Pouquet (2009)]{MininniPouquet09}
Mininni, P.~D. \& Pouquet, A.\ 2009,
\pre, 80, 025401
\bibitem[Moffatt (1983)]{Moffatt83}
Moffatt, H.~K.\ 1983, 
Rep. Prog. Phys., 46, 621
\bibitem[Monin 
\& Yaglom(1975)]{MoninYaglom74} Monin, A.~S., \& Yaglom, A.~M.\ 1975, 
Cambridge, Mass., MIT Press, 1975.~882 p.
\bibitem[Montgomery \& Matthaeus (1995)]{MontgomeryMatthaeus95}
Montgomery, D. \& Matthaeus, W.~H.\ 1995, 
\apj, 447, 706
\bibitem[Montgomery 
\& Turner(1981)]{MontgomeryTurner81} Montgomery, D., \& Turner, L.\ 1981, Physics of Fluids, 24, 825
\bibitem[Mordant et al. (2001)]{Mordantetal01}
Mordant, N., Metz, P., Michel, O. \& Pinton, J.-F.\ 2001,
Phys. Rev. Lett., 87, 214501
\bibitem[Narayan \& Medvedev (2001)]{NarayanMedvedev01}
Narayan, R. \& Medvedev, M.~V.\ 2001, 
\apjl, 562, L129
\bibitem[Newcomb (1958)]{Newcomb58}
Newcomb, W. A.\ 1958, 
Ann. Phys. (N.Y.) 3, 347
\bibitem[Ng \& Bhattacharjee (1996)]{NgBhattacharjee96}
Ng, C.~S. \& Bhattacharjee, A.\ 1996,
\apj, 465, 845
\bibitem[Ng et al. (2003)]{Ngetal03}
Ng, C.~S., Bhattacharjee, A., Germaschewski, K. \& Galtier, S.\ 2003,
Physics of Plasmas, 10, 1954
\bibitem[Ng \& Ragunathan (2011)]{NgRagunathan11}
Ng, C.~S.\&  Ragunathan, S.\ 2011. ASP Conference Series, submitted. 
\bibitem[Norman 
\& Ferrara(1996)]{NormanFerrara96} Norman, C.~A., \& Ferrara, A.\ 
1996, \apj, 467, 280
\bibitem[Obukhov (1941)]{Obukhov41} 
Obukhov, A. M.\ 1941, 
Izv. Akad. Nauk SSSR, Ser. Geogr. GeoÞz., 5, 453
\bibitem[Ossendrijver(2003)]{Ossendrijver03} Ossendrijver, M.\ 2003, \aapr, 11, 287 
\bibitem[Oughton et al.(2003)]{Oughtonetal03} Oughton, S., Dmitruk, 
P., 
\& Matthaeus, W.~H.\ 2003, Turbulence and Magnetic Fields in Astrophysics, 614, 28 
\bibitem[Padoan et al.(2004)]{Padoanetal04} Padoan, P., Jimenez, R., 
Juvela, M., \& Nordlund, {\AA}.\ 2004, \apjl, 604, L49 
\bibitem[Pang, Pen, \& Vishniac (2010)]{pangetal2010} Pang, B., Pen, U.-L., \& Vishniac, E.T.\ 2010,
{Phys. Plasmas}, 17, 102302
\bibitem[Park et al.(2009)]{Parketal09} Park, S., Chae, J., 
\& Litvinenko, Y.~E.\ 2009, \apjl, 704, L71 
\bibitem[Parker(1970)]{Parker70} Parker, E.~N.\ 1970, \apj, 162, 
665 
\bibitem[Parker(1979)]{Parker79} Parker, E.~N.\ 1979, Oxford, 
Clarendon Press; New York, Oxford University Press, 1979, 858 p.,
\bibitem[Parker(1993)]{Parker93} Parker, E.~N.\ 1993, \apj, 408, 
707
\bibitem[Perez 
\& Boldyrev(2009)]{PerezBoldyrev09} Perez, J.~C., \& Boldyrev, S.\ 2009, 
Physical Review Letters, 102, 025003 
\bibitem[Petschek (1964)]{petschek64} Petschek, H.E.\ 1964, The Physics of Solar Flares, AAS-NASA Symposium (NASA SP-50), 
ed. W. H. Hess (Greenbelt, MD: NASA), 425
\bibitem[Podesta (2010)]{Podesta10}
Podesta, J.~J.\ 2010, 
Twelfth International Solar Wind Conference, 1216, 128
\bibitem[Politano et al. (1989)]{Politanoetal89}
Politano, H., Pouquet, A. \& Sulem, P.~L.\ 1989, 
Physics of Fluids B, 1, 2330
\bibitem[Pope(2002)]{Pope02} Pope, S.~B.\ 2002, Physics of 
Fluids, 14, 1696
\bibitem[Priest \& Forbes (2000)]{PriestForbes00}
Priest, E. \& Forbes, T.\ 2000,
in:  Magnetic Reconnection: MHD theory and applications, Eds. 
E. Priest \& T. Forbes, pp.~612, Cambridge, UK: Cambridge University Press.
\bibitem[Priest 
\& Forbes(2002)]{PriestForbes02} Priest, E.~R., \& Forbes, T.~G.\ 2002, \aapr, 10, 313 
\bibitem[Richardson (1926)]{Richardson26}
Richardson, L. F.\ 1926,   
Proc. R. Soc. London, Ser. A, 110, 709
\bibitem[Rumpf et al. (2009)]{Rumpfetal09}
Rumpf, B., Newell, A.~C. \& Zakharov, V.~E.\ 2009,
Phys. Rev. Lett., 103, 074502
\bibitem[Salazar \& Collins (2009)]{SalazarCollins09}
Salazar, J. P. L. C. \& Collins, L. R.\ 2009,  
Annu. Rev. Fluid Mech., 41, 405
\bibitem[Samtaney et al. (2009)]{Samtaneyetal09}
Samtaney, R., Loureiro, N.~F., Uzdensky, D.~A., Schekochihin, A.~A.  
and Cowley, S.~C.\ 2009.
\prl 103, 105004.
\bibitem[Sano \& Stone (2002)]{SanoStone02}
Sano, T. \& Stone, J. M.\ 2002,
\apj, 570, 314
\bibitem[Santos-Lima et al.(2010)]{SantosdeLimaetal10} Santos-Lima, R., 
Lazarian, A., de Gouveia Dal Pino, E.~M., \& Cho, J.\ 2010, \apj, 714, 442 
\bibitem[Schekochihin et al.(2007)]{Schekochihinetal07} Schekochihin, 
A.~A., Cowley, S.~C., \& Dorland, W.\ 2007, Plasma Physics and Controlled 
Fusion, 49, 195 
\bibitem[Schekochihin et al. (2009)]{Schekochihinetal09} Schekochihin, 
A.~A., Cowley, S.~C., Dorland, W., Hammett, G.~W., Howes, G.~G., Quataert, 
E., \& Tatsuno, T.\ 2009, \apjs, 182, 310
\bibitem[Schuecker et 
al.(2004)]{Schueckeretal04} 
Schuecker, P., Finoguenov, A., Miniati, F., B{\"o}hringer, H., \& Briel, U.~G.\ 2004, \aap, 426, 387
\bibitem[Servidio et al.(2010)]{Servidioetal10} Servidio, S., 
Matthaeus, W.~H., Shay, M.~A., Dmitruk, P., Cassak, P.~A., 
\& Wan, M.\ 2010, Physics of Plasmas, 17, 032315
\bibitem[Shaikh 
\& Shukla(2009)]{ShaikhShukla09} Shaikh, D., \& Shukla, P.~K.\ 2009, 
Physical Review Letters, 102, 045004
\bibitem[Shay 
\& Drake(1998)]{ShayDrake98} Shay, M.~A., \& Drake, J.~F.\ 1998, \grl, 25, 3759 
\bibitem[Shay et al.(1998)]{Shayetal98} Shay, M.~A., Drake, J.~F., 
Denton, R.~E., \& Biskamp, D.\ 1998, \jgr, 103, 9165 
\bibitem[Shay et al.(1999)]{Shayetal99} Shay, M.~A., Drake, J.~F., 
Rogers, B.~N., \& Denton, R.~E.\ 1999, \grl, 26, 2163
\bibitem[Shebalin et al.(1983)]{Shebalinetal83} Shebalin, J.~V., 
Matthaeus, W.~H., 
\& Montgomery, D.\ 1983, Journal of Plasma Physics, 29, 525 
\bibitem[Shen \& Yeung (1997)]{ShenYeung97}
Shen, P. \& Yeung, P.~K.\ 1997,
Phys. Fluids, 9, 3472
\bibitem[Shibata \& Tanuma (2001)]{ShibataTanuma01}
Shibata, K.  \& Tanuma, S. 2001,
Earth Planets Space, 53, 473
\bibitem[Skilling et al.(1974)]{Skillingetal74} Skilling, J., McIvor, 
I., \& Holmes, J.~A.\ 1974, \mnras, 167, 87P
\bibitem[Smith et al. (2004)]{Smithetal04}
Smith, D., Ghosh, S., Dmitruk, P. \& Matthaeus, W. H.\ 2004,
Geophys. Res. Lett. 31, L02805
\bibitem[Speiser(1970)]{Speiser70} Speiser, T.~W.\ 1970, \planss, 18, 613
\bibitem[Sridhar \& Goldreich (1994)]{SridharGoldreich94}
Sridhar, S. \& Goldreich, P.\ 1994, 
\apj, 432, 612-621
\bibitem[Sturrock(1966)]{Sturrock66} Sturrock, P.~A.\ 1966, \nat, 
211, 695 
\bibitem[Strauss(1986)]{Strauss86} Strauss, H.~R.\ 1986, Physics 
of Fluids, 29, 3668
\bibitem[Strauss (1988)]{Strauss88}
Strauss, H.~R.\ 1988,
\apj, 326, 412
\bibitem[Subramanian et al.(2006)]{Subramanianetal06} 
Subramanian, K., 
Shukurov, A., \& Haugen, N.~E.~L.\ 2006, \mnras, 366, 1437
\bibitem[Sych et 
al.(2009)]{Sychetal09} Sych, R., Nakariakov, V.~M., Karlicky, M., 
\& Anfinogentov, S.\ 2009, \aap, 505, 791 
\bibitem[Taylor (1921)]{Taylor21}
Taylor, G. I.\ 1921,
Proc. Roy. Soc. London A, 20, 196
\bibitem[Tennekes \& Lumley (1972)]{TennekesLumley72}
Tennekes, H. \& Lumley, J.~L.\ 1972, 
First Course in Turbulence, Cambridge, MIT Press
\bibitem[Toschi \& Bodenschatz (2009)]{ToschiBodenschatz09}
Toschi, F.  \& Bodenschatz, E.\ 2009,  
Annu. Rev. Fluid Mech., 41, 375
\bibitem[Uzdensky 
\& Kulsrud(2006)]{UzdenskyKuslrud06} Uzdensky, D.~A. 
\& Kulsrud, R.~M.\ 2006,  Physics of Plasmas, 13, 062305
\bibitem[Uzdensky(2007)]{Uzdensky07} 
Uzdensky, D.~A.\ 2007, 
\apj, 671, 2139. 
\bibitem[Uzdensky et al. (2010)]{Uzdenskyetal10}
Uzdensky, D.~A., Loureiro, N.~F. and Schekochihin, A.~A.\ 2010. 
\prl 105, 235002.
\bibitem[Vasyliunas (1972)]{Vasyliunas72}
Vasyliunas, V.\ 1972, 
J. Geophys. Res., 77, 6271
\bibitem[Vasyliunas (1975)]{Vasyliunas75}
Vasyliunas, V.~M.\ 1975, 
Reviews of Geophysics and Space Physics, 13, 303
\bibitem[Vishniac \& Lazarian (1999)]{VishniacLazarian99}
Vishniac, E. \& Lazarian, A.\ 1999, in:
Plasma Turbulence and Energetic Particles in Astrophysics; 
Proceedings of the International Conference, Cracow, Poland, 5-10 September, 1999 
eds. M. Ostrowski \& R. Schlickeiser. 
Obserwatorium Astronomiczne, Uniwersytet Jagiello\'{n}ski, Krak\'{o}w. 
\bibitem[Vogt 
\& En{\ss}lin(2005)]{VogtEnsslin05} 
Vogt, C., \& En{\ss}lin, T.~A.\ 2005, \aap, 434, 67
\bibitem[Waelbroeck(1989)]{Waelbroeck89} Waelbroeck, F.~L.\ 1989, 
Physics of Fluids B, 1, 2372 
\bibitem[Wang et al.(2000)]{Wangetal00} Wang, X., Bhattacharjee, 
A., \& Ma, Z.~W.\ 2000, \jgr, 105, 27633
\bibitem[Watson et al. (2007)]{Watsonetal07}
Watson, P.~G., Oughton, S. \& Craig, I.~J.~D.\ 2007, 
Physic. Plasmas, 14, 032301
\bibitem[Wicks et al. (2010)]{Wicksetal10}
Wicks, R.~T., Horbury, T.~S., Chen, C.~H.~K. \&  Schekochihin, A.~A.\ 2010, 
\mnras, 407, L31
\bibitem[Wicks et al. (2011)]{Wicksetal11}
Wicks, R.~T., Horbury, T.~S., Chen, C.~H.~K. \&  Schekochihin, A.~A.\  2011,
Physical Review Letters, 106, 045001
\bibitem[Wilmot-Smith et al. (2005)]{WilmotSmithetal05}
Wilmot-Smith, A. L., Priest, E. R.  \& Hornig, G.\ 2005,
Geophys. Astrophys. Fluid Dyn., 99, 177
\bibitem[Yamada(1999)]{Yamada99} Yamada, M.\ 1999, \jgr, 104, 
14529
\bibitem[Yamada et al.(2006)]{Yamadaetal06} Yamada, M., Ren, Y., Ji, 
H., Breslau, J., Gerhardt, S., Kulsrud, R., 
\& Kuritsyn, A.\ 2006, Physics of Plasmas, 13, 052119 
\bibitem[Yamada(2007)]{Yamadaetal08} Yamada, M.\ 2007, Physics of 
Plasmas, 14, 058102
\bibitem[Yamada et al.(2010)]{Yamadaetal10} Yamada, M., Kulsrud, R., 
\& Ji, H.\ 2010, Reviews of Modern Physics, 82, 603
\bibitem[Yokoyama 
\& Shibata(1995)]{YokoyamaShibata95} Yokoyama, T., \& Shibata, K.\ 1995, \nat, 375, 42 
\bibitem[Zhang 
\& Yan(2011)]{ZhangYan11} Zhang, B., \& Yan, H.\ 2011, \apj, 726, 90 
\bibitem[Zimbardo et al.(2010)]{Zimbardoetal10} Zimbardo, G., Greco, 
A., Sorriso-Valvo, L., Perri, S., V{\"o}r{\"o}s, Z., Aburjania, G., 
Chargazia, K., \& Alexandrova, O.\ 2010, \ssr, 156, 89
\bibitem[Zweibel \& Rhoads (1995)]{ZweibelRhoads95}
Zweibel, E.~G. \& Rhoads, J.~E.\ 1995,
\apj, 440, 407
\bibitem[Zweibel 
\& Yamada(2009)]{ZweibelYamada09} Zweibel, E.~G., \& Yamada, M.\ 2009, \araa, 47, 291
\end{thebibliography}
\end{document}